\documentclass[journal]{IEEEtran}
\IEEEoverridecommandlockouts
\usepackage{amsfonts}
\usepackage{amssymb}
\usepackage[cmex10]{amsmath}
\usepackage{subfigure}
\usepackage{verbatim}
\usepackage{cite}
\usepackage{tikz}
\usepackage{pgflibraryarrows}
\usepackage{pgflibrarysnakes}
\usepackage{diagbox}
\usepackage{tabularx,booktabs}
\usepackage{caption}
 \usepackage{caption}
\usepackage{amsthm}
\usepackage{booktabs}
\usepackage{enumerate}
\usepackage{graphicx}
\usepackage{multicol}
\usepackage[ruled,lined]{algorithm2e}
\usepackage{setspace}
\usepackage{stfloats}
\usepackage{fancyhdr}
\usepackage{algorithmic}
\usepackage[tikz]{bclogo}
\usepackage{multicol}
\usepackage{tikz}
\usetikzlibrary{positioning}
\tikzset{>=stealth}

\newtheorem{Def}{Definition}
\newtheorem{Pro}{Proposition}

\newtheorem{Exa}{Example}

\newtheorem{Rem}{Remark}

\allowdisplaybreaks

\ifCLASSINFOpdf
\else
\fi
\IEEEoverridecommandlockouts
\begin{document}
\title{Principled information fusion  for multi-view multi-agent surveillance systems}
\author{ Bailu Wang, Suqi Li*, Giorgio Battistelli, Luigi Chisci, Wei Yi
\thanks{

B. Wang and S. Li  are with  Chongqing University, Chongqing, 400044, China (Email: w\_b\_l3020@163.com; qi\_qi\_zhu1210@163.com;).
G. Battistelli and L. Chisci are  with Dipartimento di Ingegneria dell'Informazione, Universit$\grave{\mbox{a}}$ degli Studi di Firenze, Via Santa Marta 3, 50139, Firenze, Italy (Email: giorgio.battistelli@unifi.it; luigi.chisci@unifi.it).

W. Yi is with University of Electronic Science and Technology of China, Chengdu, Sichuan, 611731, China (Email: kussoyi@gmail.com)

Corresponding author: Suqi Li.


}}
\maketitle

\IEEEpeerreviewmaketitle

\begin{abstract}
A key objective of multi-agent surveillance systems is
to monitor a much larger region than the limited \textit{field-of-view} (FoV) of any individual agent by successfully exploiting cooperation among multi-view agents.
Whenever either a centralized or a distributed approach is pursued, this goal cannot be achieved unless an appropriately designed fusion strategy is adopted.
This paper presents a novel principled information fusion approach for dealing with multi-view multi-agent case,  on the basis of \textit{Generalized Covariance Intersection} (GCI).
The proposed method can be used to perform multi-object tracking
on both a centralized and a distributed peer-to-peer sensor network.
Simulation experiments on realistic multi-object tracking scenarios demonstrate effectiveness of the proposed solution.
\end{abstract}
\section{Introduction}
The recent breakthrough of \textit{wireless sensor network} (WSN) technology has made it possible to develop efficient surveillance systems
that consist of multiple radio-interconnected, low-cost and low energy consumption devices (agents) with sensing, communication and processing capabilities.
 In this respect, it has become of paramount importance to redesign multi-object tracking algorithms tailored to the new features of  emerging surveillance systems.
 In particular, a fundamental issue to be suitably addressed is how to consistently fuse information from different agents \cite{Battistelli}.
As well known, the computational cost of extracting the common information can make the optimal fusion \cite{CY-Chong} intractable in
most practical cases, so that a suboptimal solution with demonstrated tractability has been proposed relying on  the
\textit{generalized covariance intersection} (GCI)~\cite{Mahler-1,Hurley}, also called \textit{exponential mixture density} (EMD) in \cite{EMD-Julier,Clark}.
From an information-theoretic viewpoint, GCI fusion admits a meaningful interpretation in that the fused density is the left centroid of the local posteriors with
Kullback-Leibler divergence (from local to fused density) considered as distance;
 accordingly, GCI fusion is also called  \textit{Kullback-Leibler average}
(KLA) in \cite{Battistelli,double-counting}.
 It has been mathematically proved that the KLA intrinsically avoids double counting of common information \cite{double-counting}.
Implementation of GCI fusion combined with a variety of multi-object trackers including PHD, CPHD, multi-Bernoulli as well as certain labeled \textit{random finite set} (RFS) filters has been investigated in
 \cite{Battistelli,GCI-MB,Fantacci-BT,GCI-GMB}.
Another relevant issue, that has recently received growing attention in the literature, is that, due to the limited sensing ability, individual sensor nodes can typically detect objects within a limited range and/or angle.
In a multi-agent surveillance system, agent nodes are  deployed in different locations of the
surveillance area, thus having  partially/non-overlapping \textit{fields of view} (FoVs).
A potential benefit of having a multi-view multi-agent surveillance system is actually  to extend the FoV of an individual sensor node to a much larger one, with the ultimate goal to make the latter equal to the union of the FoVs of all individual sensor nodes in the network (called global FoV).
To achieve this ultimate goal,  information fusion among multiple agents needs to  achieve sufficient and valid information gain within the global FoV.

Standard GCI fusion provides unsatisfactory performance for the multi-view case in that it tends to preserve only the objects that are in the intersection of the FoVs of all nodes involved in the fusion (common FoV).
In order to overcome this mis-behavior of  standard GCI fusion, some improved fusion strategies aiming at  certain classes of multi-object densities have been presented in the literature
\cite{PHD-DFOV,DFoV-LMB,GCI-PHD-guchong,DFOV-LMB-Reza,Fusion-CS,Fusion-LKLA,AA-li,DFOV-PHD-WeiYi,DFOV-LMB-Gostar,DFOV-CPHD-TianchengLi} for multi-view multi-object estimation problems.
In particular, \cite{PHD-DFOV} concerns multi-robot PHD-based distributed map construction with different robot FoVs, and proposes to
initialize the local density of each agent to an uninformative (flat) prior over the global FoV.
Further, multi-sensor fusion with different FoVs for the LMB filter is addressed in  \cite{DFoV-LMB}, wherein
 a suitable compensation strategy for the exclusive FoV to be applied before GCI fusion is proposed. The use of a compensation strategy is further investigated in \cite{GCI-PHD-guchong} where distributed fusion with a multi-view sensor network using PHD filters is considered. In \cite{DFOV-LMB-Reza}, an intuitive approach which tunes the weights of LMB posteriors automatically according to amount of information is proposed to overcome the drawback of GCI fusion for centralized fusion.
A further proposal has been to replace GCI fusion, penalized by its multiplicative nature, with
\textit{additive mixture density} (AMD)  fusion \cite{Fusion-CS,Fusion-LKLA, AA-li,DFOV-PHD-WeiYi,DFOV-LMB-Gostar,DFOV-CPHD-TianchengLi}.   

However, the analysis in this paper will show how the mis-behaviour of standard GCI fusion is mainly due to the inconsistency of the local multi-object density outside the FoV, rather than to the multiplicative nature of GCI.
For an independently-working agent, multi-object filtering usually  truncates the density outside the FoV for  the sake of computational feasiblity, leading to a multi-object posterior with nearly null \textit{yes-probability}  outside the individual FoV.
This feature of the multi-object posterior has clearly no influence on the performance of independently-working agents, since the region outside the FoV is not of interest.
However, our analysis shows that, for multi-agent fusion with partially/non-overlapping FoVs, whenever the local posterior is fused with other agent densities according to the GCI rule, the fused density will have null \textit{yes-probability} outside the common FoV (which means that no object existence can be declared outside the common FoV).  As a direct consequence for object tracking, only tracks within the common FoV can
 be preserved after fusion, while other tracks are lost.
Such null \textit{yes-probability} indicates a zero  information entropy, conveying specific information that objects cannot exist outside the common FoV, which is obviously
inconsistent with the fact that each individual sensor cannot
receive measurements of objects outside the FoV.

To avoid null \textit{yes-probability} outside the common FoV, a consistent propagation of the multi-object posterior is of crucial importance.
A first possibility is to restrict measurement update to the local FoV, and only apply prediction outside the FoV, which is the natural result of the local filtering  by setting to zero the probability of detection outside the FoV for the multi-object likelihood.
However, prediction of the multi-object density can induce a significant bias with respect to the true multi-object state, especially when some object moves in the exclusive FoV of a certain sensor for a long time.

Conversely, in the present paper, we provide  a rigorous mathematical definition of multi-object uninformative density.
Such a density reflects the complete unavailability of multi-object information and, therefore, represents a consistent form for the posterior outside the FoV of each agent.

The contributions of this paper can be summarized as follows.
\begin{enumerate}
\item First, we propose a novel principled information fusion rule for  a multi-view multi-agent surveillance system, called \textit{Bayesian-operation InvaRiance on Difference-sets (BIRD), on the basis of the GCI} which can satisfactorily cope with both centralized and distributed fusion.
\item  Then, the proposed BIRD fusion is applied to  multi-object Poisson processes and, accordingly, a multi-view multi-agent fusion algorithm under the PHD filtering framework is developed.
\item Finally, it is shown how the proposed BIRD PHD fusion  for  multi-object tracking can be implemented over a peer-to-peer sensor network in a fully distributed manner via a \textit{Gaussian mixture} (GM) approach.
\end{enumerate}

In addition to the proposed multi-view fusion rule, we also provide two basic mathematical tools that fill some blanks in the \textit{FInite Set STatistics} (FISST) \cite{mahler_book}.
\begin{enumerate}
\item The marginal and conditional densities with respect to  disjoint subspaces of the RFS, which further enable a general and accurate decomposition of RFS densities.
\item An RFS-based multi-object uninformative density, to be used as multi-object prior whenever there is no available prior knowledge.  Note that, to the best of the authors' knowledge, this is the first time that an RFS-based uninformative density is proposed in a rigorous way.
\end{enumerate}

\begin{figure}[tb]
\begin{minipage}[htbp]{0.49\linewidth}
  \centering
  \centerline{\includegraphics[width=4.9cm]{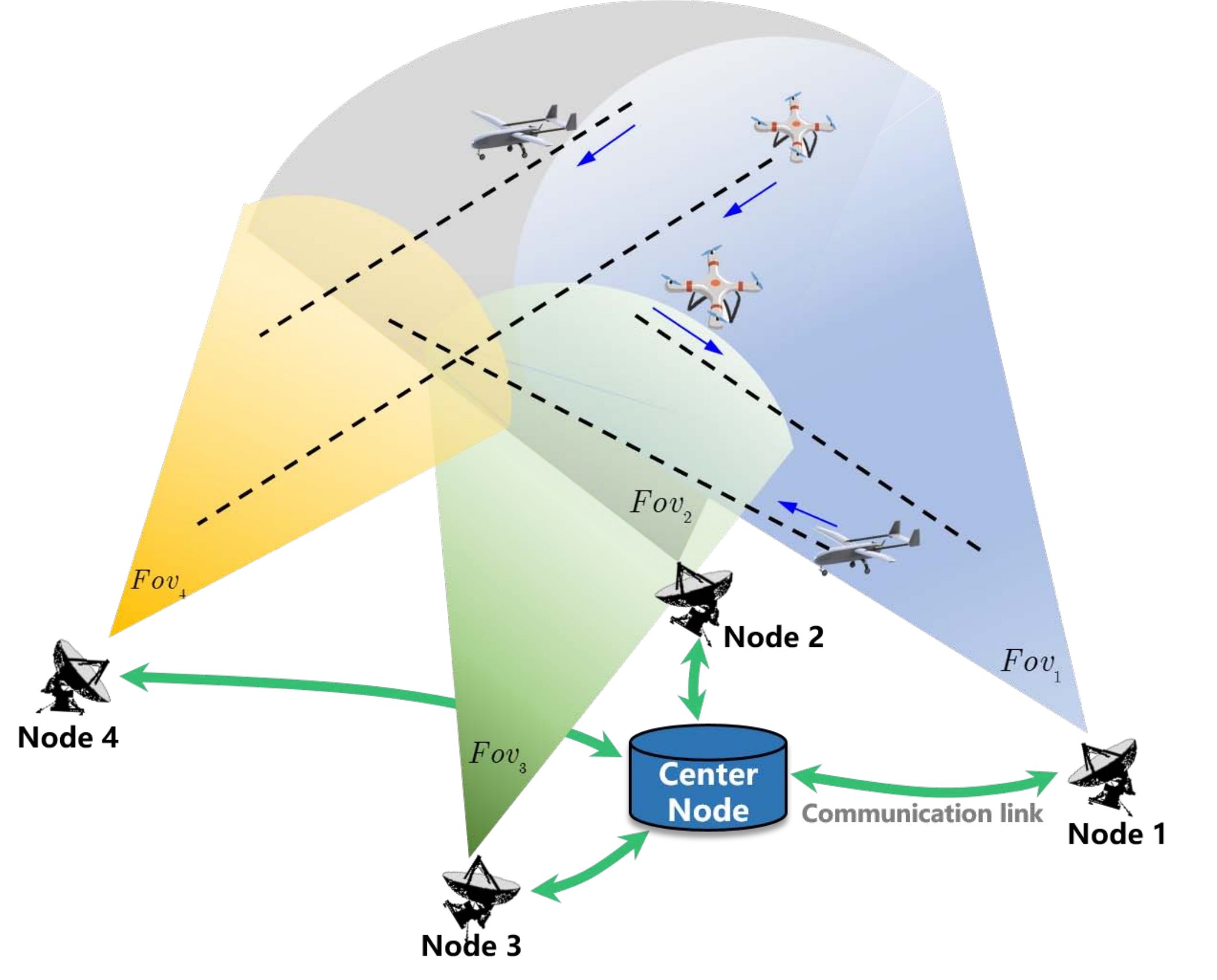}}

  \centerline{\small{\small{(a)}}}\medskip
\end{minipage}
  \hfill
  \hfill
\begin{minipage}[htbp]{0.49\linewidth}
  \centering
  \centerline{\includegraphics[width=4.9cm]{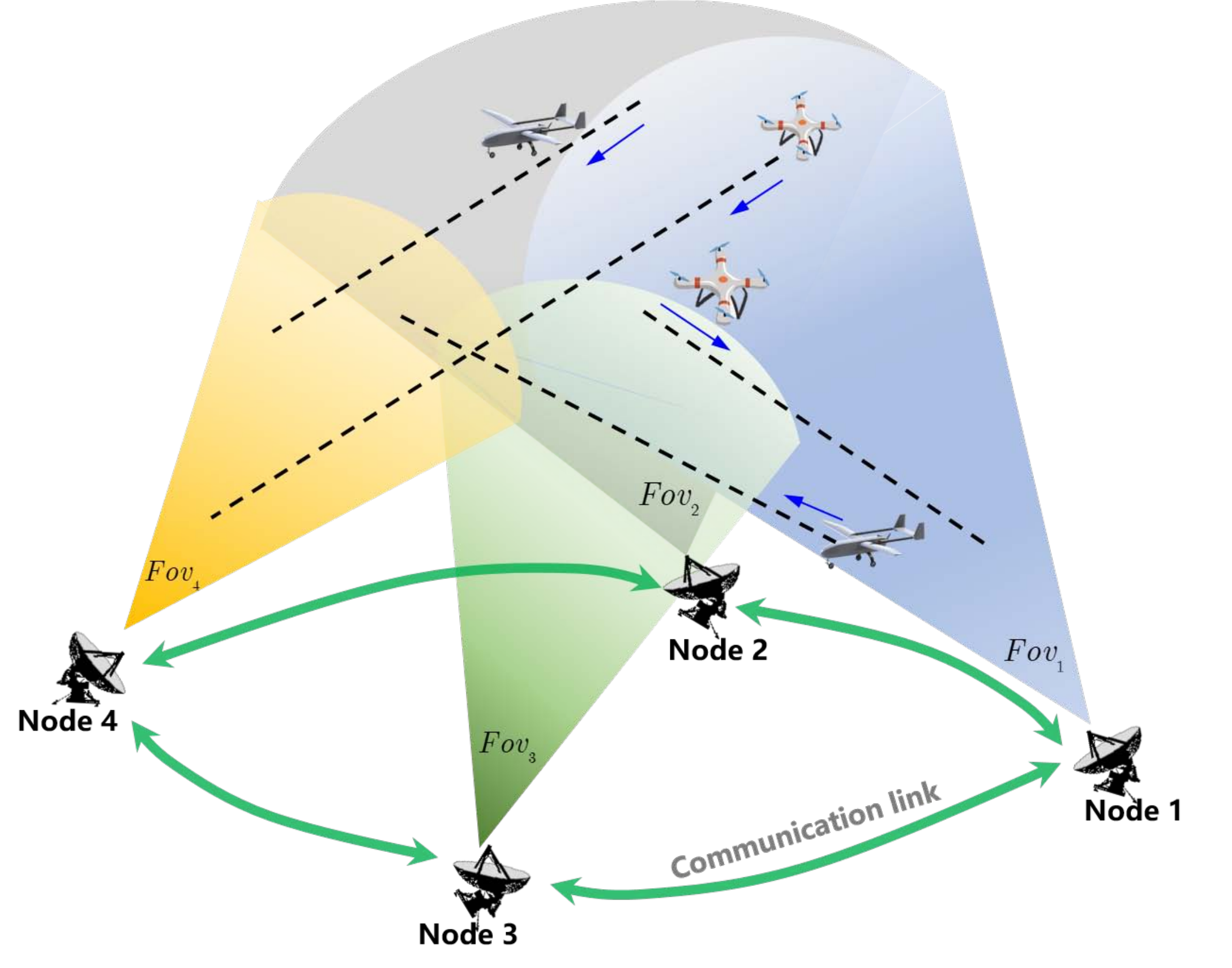}}
  \centerline{\small{\small{(b)}}}\medskip
\end{minipage}

\caption{Examples of multi-view sensor network : (a) centralized  network; (b) distributed peer-to-peer network.}
  \label{sensor-network}\end{figure}
\section{Problem formulation and mathematical tools}
\subsection{Network model}
This work considers two types of  networks as depicted in Figs. \ref{sensor-network} (a) and (b), namely the centralized and distributed peer-to-peer network.
Each sensor node (agent) $i\in\mathcal{N}$ has a limited \textit{Field-of-View} (FoV), denoted by $Fov_i$, where   $\mathcal{N}=\{1,\cdots,N_s\}$ is the set of nodes. Let $\mathbb{X}$ denotes the object state space. Herein  each $Fov_i$ is  defined as a finite subset of $\mathbb{X}$, i.e. $Fov_i\in\mathcal{F}(\mathbb{X})$. Specifically, each sensor node $i$ can only receive measurements  of kinematic variables
(e.g., angles, distances, Doppler shifts, etc.) relative to objects  states within $Fov_i$. The global FoV of the network, denoted
by $Fov_G$, is the union of the FoVs of all  sensor nodes, i.e.  $Fov_G =\displaystyle{\cup_{i\in \mathcal{N}}} \, Fov_i$. Note that to facilitate easy understanding and plotting,  the FoVs shown in figures of the paper are the projections of FoVs on the measurement space.

1) \emph{Centralized sensor network:} The  network consists of heterogenous and geographically dispersed nodes with processing, communication and sensing capabilities,
as well as of a fusion center. Each node transmits its local posterior to the fusion center, wherein the fusion of all local posteriors is carried out.

2) \emph{Distributed sensor network:} The distributed network consists only of heterogenous and geographically dispersed nodes with processing, communication and sensing capabilities.
Further features are that: 1) there is no central fusion node; 2) nodes are unaware of the network topology, i.e., the number of nodes
and their links.
The network is represented by a directed graph $\mathcal{G}=(\mathcal{N},\mathcal{A})$ , where  $\mathcal{A}\subseteq \mathcal{N}\times \mathcal{N}$ the set of edges, such that
$(i,j)\in\mathcal{A}$ if and only if node $j$ is able to receive data from node  $i$.
For each node $j\in\mathcal{N}$, $\mathcal{N}_j=\{ i \in\mathcal{N}: (i,j)\in \mathcal{A}\}$ denotes its set of  in-neighbors, i.e.,
the set of nodes from which node $j$ can receive data. By definition, $(j,j)\in \mathcal{A}$ and, hence, $j\in\mathcal{N}_j$ for all $j$.
Each sensor node  can process local data as well as exchange information with its neighbours.
\subsection{ GCI Fusion}
GCI fusion has been first proposed by Mahler \cite{Mahler-1} to  develop multi-object tracking in a multi-agent  setting.
The name GCI stems from the fact that the fusion  is the multi-object counterpart of the analogous fusion rule for (single-object) probability densities \cite{EMD-Julier} which, in turn, is a generalization of \textit{covariance intersection} originally conceived  for fusion of Gaussian probability densities \cite{Uhlmann}.

Suppose that, in  each node $ i \in\mathcal{N}$ of the network, an RFS
density $\pi_i$, defined over a suitable state space $\mathbb{X}$ and computed on the basis of local information, is available.
GCI fusion amounts to computing the geometric mean, or the exponential mixture of the set of the local multi-object  densities, i.e.,
\begin{align}\label{G-CI}
\begin{split}
\overline\pi(X)=\frac{\prod_{i\in\mathcal{N}}[\pi_i(X)]^{\omega_i} }                                             {\int\prod_{i\in\mathcal{N}}[\pi_{i}(X)]^{\omega_i} \delta X}
\end{split}
\end{align}
 where $\int ~\cdot ~\delta X$ denotes  set integration \cite{mahler_book} and fusion weights
$\omega_i\geqslant 0$ satisfy $\sum_{i\in\mathcal{N}}\omega_i=1$.

In \cite{Heskes,Battistelli} it has been shown that the GCI fusion in (\ref{G-CI}) essentially minimizes the weighted sum of the \textit{Kullback-Leibler divergences} (KLDs) with respect to the local densities, i.e.,
\begin{equation}\label{KLA}
  \overline\pi=\arg \min_\pi \sum_{i\in\mathcal{N}}\omega_i D_{\text{KL}}(\pi||\pi_i)\end{equation}
where $D_{\text{KL}}(\pi||\pi_{i})$ denotes the KLD of $\pi_{i}$ from $\pi$, i.e.,
\begin{equation}\label{KLD}
\begin{split}
 D_{\text{KL}}(\pi||\pi_{i})\triangleq\int \pi(X)\log{\frac{\pi(X)}{\pi_{i}(X)}}\delta X.
  \end{split}
\end{equation}

In view of (\ref{KLA}), GCI fusion is also called  \textit{Kullback-Leibler average} (KLA)  \cite{Battistelli}.

For notational simplicity, let us  introduce operators $\oplus$ and $\odot$ as
\begin{align}
\pi\oplus \pi'& \triangleq \frac{\pi(X)\, \pi'(X)}{\int \pi(X) \,  \pi'(X) \delta X}\\
\omega\odot \pi &\triangleq \frac{\pi(X)^\omega}{\int \pi(X)^\omega \, \delta X}.
\end{align}
Since $\oplus$ is consistent with the Bayes rule \cite{mahler_book},  taking $\pi'(\cdot)$ as prior density and $\pi(\cdot)$ as likelihood, we call $\oplus$  Bayesian operator.
Further, $\odot$ is called exponentation operator.

In terms of the  operators $\oplus$ and $\odot$, GCI fusion (\ref{G-CI}) can be re-expressed as
\begin{equation}\label{GCI-operation}
\overline\pi=\bigoplus_{i\in\mathcal{N}} \, \left( \omega_i \odot\pi_i \right).
\end{equation}
\subsection{Multi-object uninformative  density}
In the Bayes' theorem, the uninformative prior is typically used when there is no available prior
information, thus implying that the posterior turns out to be equal to the normalized likelihood. In an early stage,  originally postulated that the uninformative prior should be constant. Later,
Jeffreys, using heuristic arguments, suggested that a better uninformative prior is the $1/\sigma$
function \cite{uninformative-1}. Then, Jaynes provided a more formal derivation of the  $1/\sigma$ prior for certain classes of
probability functions \cite{uninformative-2}. However, for the RFS case, to the best of the authors' knowledge, no proper definition of uninformative prior has been given.

In this paper, we  first provide the rigorous notion of  multi-object uninformative density, which is a key concept for the subsequent developments.
\begin{Def}
Consider a family $\digamma$ of RFS densities  over a state space $\mathbb{X}$, then $\pi_{ui}$ is uninformative if
\begin{align}
\label{property-1}\pi\oplus \pi_{ui} &=\pi, & \forall\pi\in\digamma\\
\label{property-2}\omega \odot \pi_{ui} &=\pi_{ui},  & \forall \omega \in [0,1].
\end{align}
\end{Def}
\begin{Rem}
A multi-object uninformative density is the   zero-element of the algebra defined by the Bayesian operator $\odot$ and the exponentiation operator $\oplus$.
\end{Rem}

According to (\ref{property-1}),  the Bayesian operation between an arbitrary  multi-object density $\pi$ and a multi-object uninformative density $\pi_{ui}$,
always provides $\pi$ as a result, and hence this property is called as \emph{the Bayesian-operator invariance}.
Furthermore, according to (\ref{GCI-operation}) and Definition 1, it is also easy to see that GCI fusion between an arbitrary  multi-object density $\pi$ and an uninformative density $\pi_{ui}$,
always provides $\pi$ as a result.


\subsection{Decomposition of a multi-object density with respect to disjoint sub-spaces}
For the subsequent developments, we  propose now the marginal and conditional densities with respect to  disjoint subspaces of the RFS, which enable a general decomposition of RFS densities.
\begin{Pro}\label{pro:marginal-density}
Given the RFS $X$  defined on $\mathbb{X}$ with the multi-object density $\pi(X)$, the marginal density of $X_W=X\cap\mathbb{W}$ with $\mathbb{W}\subseteq\mathbb{X}$ is given by
\begin{equation}\label{marginal-density}
\begin{split}
\pi_{W}(X)=&[1_{\mathbb{W}}]^{X}\int_{\mathbb{V}}\pi(X\cup X') \,\delta X'\\
\end{split}
\end{equation}
where: $\mathbb{V}$ is the complementary set of $\mathbb{W}$, i.e. $\mathbb{V}=\mathbb{X}\backslash\mathbb{W}$; the
 indicator function $1_\mathbb{W}$  is equal to $1$ if  $x\in \mathbb{W}$ and to zero otherwise; and
\begin{equation}\label{conditioned-density}
\begin{split}
&\int_{\mathbb{V}}\pi(X\cup X')\delta X'\\
=&\sum_{n'=0}^{+\infty}\frac{1}{n'!}\int_{\mathbb{V}}\prod_{i=1}^{n'} \pi(X \cup\{x'_{1},\cdots,x'_{n'}\})  d(x'_{1},\cdots,x'_{n'}).
\end{split}
\end{equation}
\end{Pro}
\noindent
\textit{Proof:} see  Section I of the supplemental materials. \qed \\

For any $\pi_{W}(X)\neq 0$, the density of $X_V=X\cap\mathbb{V}$ conditioned on $X_W=X\backslash X_V$, denoted by $\pi_{V}(X'|X)$ is
\begin{equation}\label{conditioned-density-V}
\pi_{V}(X'|X)=\frac{[1_{\mathbb{W}}]^{X}[1_{\mathbb{V}}]^{X'}\pi(X'\cup X)}{\pi_{W}(X)}.
\end{equation}
\begin{Rem}
To check that $\pi_{V}(X'|X)$ is an RFS density on $\mathbb{V}$, we further compute its set integral on  $\mathbb{V}$ as
\begin{equation}
\begin{split}
&\int_{\mathbb{V}} \frac{[1_{\mathbb{W}}]^{X}[1_{\mathbb{V}}]^{X'}\pi(X'\cup X)}{\pi_{W}(X)} \delta X'\\
=&\frac{[1_{\mathbb{W}}]^{X}\int_{\mathbb{V}} \pi(X'\cup X)\delta X'}{\pi_{W}(X)}=1.
\end{split}
\end{equation}
\end{Rem}
For  consistency, let us further define
\begin{equation}\label{definition_zero}
\pi_{V}(X'|X)\triangleq 1, \,\,\, \mbox{if } \pi_W(X)=0.
\end{equation}

\begin{Pro} \label{decomposition-disjoint-subsets}
Any RFS density $\pi(X)$ on $\mathbb{X}$ can be decomposed as follows:
\begin{equation}\label{decomposition-pi}
\pi(X)=\pi_{V}(X\cap \mathbb{V}|X\cap\mathbb{W}) \, \pi_{W}(X\cap\mathbb{W})
\end{equation}
where $\mathbb{W}$ and $\mathbb{V}$ are any two disjoint sets satisfying $\mathbb{W}\uplus\mathbb{V}=\mathbb{X}$.
\end{Pro}
\noindent
\textit{Proof:} see  Section II of the supplemental materials. \qed \\
\section{Problem analysis: GCI fusion with multi-view multi-agent system}
\label{section3}
This section provides a systematic analysis of the failure of standard GCI fusion in the multi-view multi-agent system.
To this end, let us only consider two agents, indexed by $a$ and $b$, with fields-of-view $Fov_a$ and respectively $Fov_b$, performing multi-object filtering recursively in time.

\begin{figure*}[bp]
  \centering
\includegraphics[width=15cm]{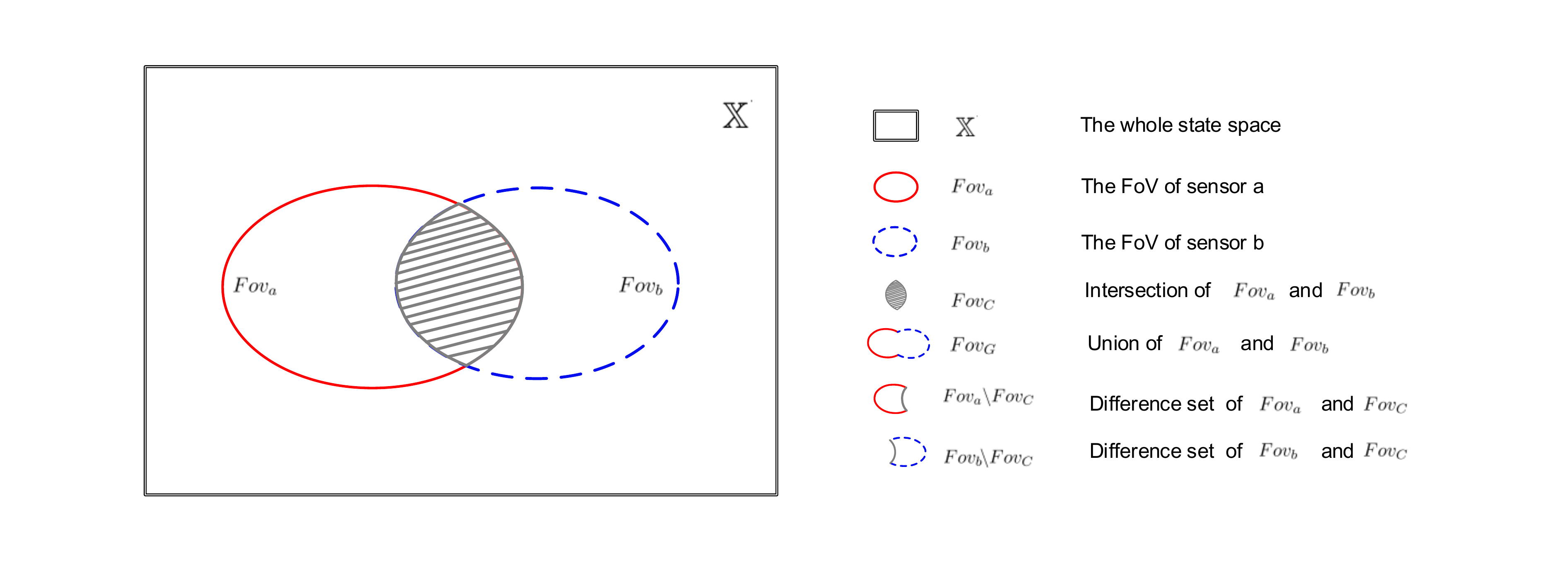}
\caption{The relationship between  FoVs involved in Subsection II-D}
\label{fig:proof_of_21_and_22}
\end{figure*}

For the subsequent analysis, we decompose the local multi-object posterior of each agent $i \in \{a,b \}$ according to (\ref{decomposition-pi}), i.e.,
\begin{equation}\label{decomposition-piab}
\pi_i(X)\!  =  \!\pi_{i,in}(X\cap Fov_i) \, \pi_{i,ou}(X\cap (\mathbb{X}\backslash Fov_i)| X\cap Fov_i)
\end{equation}
where: we have set $\mathbb{W}= Fov_i$ and $\mathbb{V} = \mathbb{X} \backslash Fov_i$  in (\ref{decomposition-pi});
$\pi_{i,in}$ denotes the marginal density inside the FoV of agent $i$, while $\pi_{i,ou}$ is the conditional density outside the FoV of agent $i$. To help understanding the technical content of this section, the involved FoVs are illustrated in Fig. \ref{fig:proof_of_21_and_22}.

For agents working independently, the multi-object density $\pi_{i,ou}(X\cap (\mathbb{X}\backslash Fov_i)| Fov_i)$ is  irrelevant since, usually, the region outside the FoV is not of interest and object tracks outside the FoV will not be displayed.

However, the objective of a sensor network is actually to surveil a much larger region of interest than the one that can be  covered by an individual agent, through the cooperation of multiple agents with partially/non-overlapping FoVs.
To this end, the fusion of multiple posteriors of individual agents is carried out at each time step, in order to get an average multi-object posterior
that hopefully integrates information of all objects within the union of the individual FoVs.
Employing the GCI rule and applying (\ref{decomposition-piab}), the fusion of multi-object posteriors of sensors $a$ and $b$ yields:
\begin{equation}\label{fusion-all}
\begin{split}
\overline \pi(X)
&\propto\pi_a(X)^{\omega_a} \, \pi_b(X)^{\omega_a}\\
&=\pi_{a,in}(X\cap Fov_a)^{\omega_a}\pi_{a,ou}(X\cap (\mathbb{X}\backslash Fov_a)| X \cap Fov_a)^{\omega_a}\\
&\times\pi_{b,in}(X\cap Fov_b)^{\omega_b}\pi_{b,ou}(X\cap (\mathbb{X}\backslash Fov_b)| X \cap Fov_b)^{\omega_b}\\
\end{split}
\end{equation}
 As can be seen from (\ref{fusion-all}), the posteriors $\pi_{i,ou}(X\cap (\mathbb{X}\backslash Fov_i)| X \cap Fov_i)$ outside the FoV of each individual agent  are  involved in the fusion, and hence have a substantial impact on the fusion result.

Next in this section,  we will analyse two usual implementations of GCI fusion relative to two different forms of $\pi_{i,ou}(X\cap (\mathbb{X}\backslash Fov_i)| X \cap Fov_i)$, and the corresponding impact on the fusion result.

\subsubsection{Form-I - Missing objects outside the common FoV} The posterior of each individual agent has a tiny \textit{yes-probability} \cite{mahler_book} outside the FoV
(corresponding to a no-object probability close to $1$), i.e.,
\begin{equation}\label{desntiy-form-1}
\pi_{i,ou}(X\cap \left(\mathbb{X}\backslash Fov_i\right)=\emptyset|X\cap Fov_i)\approx 1,~~ i=a,b.
\end{equation}
\begin{Rem}
Note that the yes-probability of a given region  represents the probability that the number of objects existing within the region is non-zero.
 \end{Rem}

This usually occurs in the local filtering when :
 \begin{itemize}
\item the survival probability is null outside the local FoV, i.e., $$P_S(x)=0, \forall x\in\mathbb{X}\backslash Fov_i;$$
\item and/or the observation model  does not properly take into account the fact that sensors cannot receive object measurements outside the FoV. For example, a sufficiently large  detection probability outside the FoV is  adopted, i.e., $P_D(x)=P_D, \forall x\in\mathbb{X}\backslash Fov_i$.
\end{itemize}
\begin{Pro}\label{limitation-non-common-FoV}
Consider the  GCI fusion of $\pi_a(\cdot)$ and $\pi_b(\cdot)$, i.e., $\overline\pi(\cdot)$ and its decomposition according to Proposition \ref{decomposition-disjoint-subsets} by setting $\mathbb{V}=Fov_C$ and $\mathbb{W}=\mathbb{X}\backslash Fov_C$, i.e.,
\begin{equation}\label{fusion-form-1-1}
\begin{split}
&\overline\pi(X)= \overline \pi_{co}(X\cap Fov_C) \, \overline\pi_{nc}(X\cap\left(\mathbb{X}\backslash Fov_C\right)|X\cap Fov_C)
\end{split}
\end{equation}
where $Fov_C=Fov_a\cap Fov_b$ is the common FoV of agents $a$ and $b$. Then if
\begin{equation}\label{desntiy-form-1}
\pi_{i,ou}(X\cap \left(\mathbb{X}\backslash Fov_i\right)=\emptyset|X\cap Fov_i)\to 1,~~ i=a,b,
\end{equation} the following holds:
\begin{equation}\label{fusion-form-1-2}
\overline\pi_{nc}(X\cap\left(\mathbb{X}\backslash Fov_C\right)=\emptyset|X\cap Fov_C) \to 1.
\end{equation}
\end{Pro}
\textit{Proof:} see  Section III of the supplemental materials. \qed \\

In particular, Proposition \ref{limitation-non-common-FoV} indicates that, if $\pi_{i,ou}(X\cap \left(\mathbb{X}\backslash Fov_i\right))$ takes Form I, i.e., (\ref{desntiy-form-1}), then after fusion, only the existence of objects within the common FoV of sensors can be declared, while the objects outside $Fov_C$ are lost, since the corresponding no-object probability is close to 1. 

This behavior is due to the fact that, even if each agent  has no sensing ability for objects outside its FoV, it becomes too much overconfident to infer that ``the object does not exist''.
As a result,  (\ref{desntiy-form-1}) is unreasonable and represents the main reason why  standard GCI fusion can lose objects outside the common FoV.
\subsubsection{Form-II -  Biased estimation of objects outside the common FoV} The posterior of each agent outside the FoV is the prediction of the prior density.
This occurs if
 \begin{itemize}
\item $P_S(x)=P_S, \forall x\in\mathbb{X}\backslash Fov_i$ (where $P_S$ is a value close to 1);
\item the FoV is properly modelled, i.e. $P_D(x)=0$ for all $ x\in\mathbb{X}\backslash Fov_i$.
\end{itemize}


Notice that $P_D(x) =0, \forall x\in\mathbb{X}\backslash Fov_i$, can be used in combination with any observation model, meaning that the sensor has null probability to receive observations of  objects outside $Fov_i$.  With this respect,  the multi-object likelihood can be further represented by
\begin{equation}\label{ture-likelihood}
\begin{split}
g(Z_i|X)
=&g(Z_i|X\cap Fov_i, X\cap(\mathbb{X}\backslash Fov_i)\\
=& g(Z_i|X\cap Fov_i).
\end{split}
\end{equation}
where the second equality in  (\ref{ture-likelihood}) is obtained since observations  are independent of objects outside $Fov_i$.

Suppose that the prediction density on the space $\mathbb{X}$ at the current time step is as follows,
\begin{equation}
\pi_{i}^{-}(X)=\pi_{i,in}^-(X\cap Fov_i)\pi_{i,ou}^-(X\cap (\mathbb{X}\backslash Fov_i)|X\cap Fov_i).
\end{equation}
According to the Bayes rule \cite{mahler_book}, the posterior is then computed by
\begin{equation}\label{update-all}
\begin{split}
&\pi_i(X|Z_i)\propto\pi^-_{i,in}(X)g(Z_i|X)=\\
&\pi^-_{i,in}(X\!\cap\!Fov_i)g(Z_i|X\cap Fov_i)\pi^-_{i,ou}(X\!\cap\! (\mathbb{X}\backslash Fov_i)|X\cap\!Fov_i).
\end{split}
\end{equation}

Eq. (\ref{update-all}) suggests that the update procedure is only performed with the posterior within the FoV, while the posterior outside the FoV is just the prediction from the previous time step.
Moreover, since  the survival probability outside the FoV is close to 1, predicted kinematic states of objects  outside the FoV
are just obtained from the prior kinematic states exploiting the dynamic motion model,  and such objects inherit the existence probabilities from the prior density.
In fact, a common situation is that an object moves outside the FoV of a given agent for a long time so that the corresponding posterior is predicted for multiple steps ahead
 which, in turn, produces larger and larger estimation biases over time.
 Then, after fusion with posteriors from other agents, the resulting fused density also inherits the estimation biases, thus implying deteriorated fusion performance.

\section{Principled fusion rule for multi-view multi-agent  system}
As we have analyzed in Section III,  the misbehaviour of  standard GCI fusion is  due to the inconsistency of the local multi-object densities outside the FoV. In this section, a consistent multi-object posterior is  constructed, so that the density within the FoV follows the marginal posterior provided by local filtering,
while the density outside the FoV adopts a multi-object uninformative prior.
Then, on the basis of this decomposition, we further propose a principled rule to handle multi-agent multi-view fusion.
\subsection{Principled fusion rule}
\subsubsection{Form III - Consistent marginal density outside the local FoV}
\begin{Pro}
  \label{Multi-object uninformative density} \textbf{Multi-object uninformative density over a bounded state space} --
Let $\digamma$ be the family of multi-object densities over a state space $\zeta$ of finite volume $V(\zeta)$.
Then, a multi-object uninformative density over $\digamma$ can be defined as follows
\begin{equation}\label{uninformative-density-1}
\pi_{ui}(X)=
\left\{\begin{array}{ll}
\ \dfrac{1}{\sum_{m=0}^{n_{\text{max}}} \frac{[V(\zeta)]^m}{m!}}, & \text{if } X\in\mathcal{F}(\zeta) \vspace{2mm} \\
\ 0, & \text{otherwise}
\end{array}\right.
\end{equation}
where $n_{\text{max}}$ is the maximum cardinality, and if $n_{\text{max}}\rightarrow +\infty$,
\begin{equation}\label{uninformative-density-2}
\pi_{ui}(X)=
\left\{\begin{array}{ll}
\ e^{-V(\zeta)}, & \text{if } X\in\mathcal{F}(\zeta)\\
\ 0, & \text{otherwise.}
\end{array}\right.
\end{equation}
\end{Pro}
\noindent\textit{Proof:} see  Section IV of the supplemental materials. \qed \\
\begin{Rem}
The multi-object uninformative density  (\ref{uninformative-density-1}) defines an i.i.d. cluster process
\begin{equation}
\pi_{ui}(X)=n! \, \rho(n) \, [p]^X,
\end{equation}
with cardinality distribution
\begin{equation}
\rho(n)=\frac{\frac{[V(\zeta)]^n}{{n!}}}{\sum_{m=0}^{n_{\text{max}}} \frac{[V(\zeta)]^m}{m!}},
\end{equation}
and location density
\begin{equation}
p(x)=\frac{1}{V(\zeta)}.
\end{equation}
\end{Rem}

\subsubsection{Principled fusion for  the two-agent case}
Let us again consider, without loss of generality, a network with two agents $a$ and $b$.
Let $\pi_{a,ui}$ and $\pi_{b,ui}$ denote multi-object uninformative densities on the state spaces $Fov_a\backslash Fov_C$ and $Fov_b\backslash Fov_C$, respectively.
Adopting  $\pi_{a,ui}$ and $\pi_{b,ui}$ as densities outside the local FoVs of agents $a$ and $b$, the posteriors of such agents both defined on the global FoV $Fov_{G}=Fov_a\cup Fov_b$
turn out to be
 \begin{equation}\label{posterior-form-III-a}
  \pi_a(X)=\pi_a(X\cap Fov_{a}) \, \pi_{a,ui}(X\cap (Fov_b\backslash Fov_C))
 \end{equation}
 and
 \begin{equation}\label{posterior-form-III-b}
 \pi_b(X)=\pi_b(X\cap Fov_{b})\, \pi_{b,ui}(X\cap (Fov_a\backslash Fov_C)).
 \end{equation}
\begin{Rem}
Please note that $\pi_{i,ui}$ is not a posterior, but actually a diffuse prior.
However, considering that $\pi_{i,in}(X\cap Fov_i)$ is the marginal density of the posterior, with a little abuse of terminology, we also call (\ref{posterior-form-III-a}) and (\ref{posterior-form-III-b}) as posteriors of agents $a$ and $b$.
\end{Rem}
\begin{Rem}
 $\pi_a$ and $\pi_b$ both are considered to be defined on $Fov_G$. The densities outside the global FoV, $Fov_G$, i.e., $\pi_i(X\cap\mathbb{X}\backslash Fov_G|X\cap Fov_G)$ are neglected,  since the corresponding region $\mathbb{X}\backslash Fov_G$ is outside the FoVs of both $a$ and $b$, wherein both $a$ and $b$ cannot get valid measurements.
\end{Rem}
Further, based on (\ref{decomposition-pi}), $\pi_i(X\cap Fov_{i})$  can be decomposed as
\begin{equation}  \begin{split}
&\pi_i(X\cap Fov_{i})\!=\pi_{i,co}(X\cap Fov_{C})\times\\
&\,\,\,\,\,\,\,\pi_{i,nc}(X\cap (Fov_{i}\backslash Fov_{C})|X\!\cap\! Fov_{C}),~~ i=a,b
\end{split}\end{equation}
where $\pi_{i,co}(X\cap Fov_{C})$ and $\pi_{i,nc}(X\cap (Fov_{i}\backslash Fov_{C})|X\!\cap\! Fov_{C})$ describe the statistics of objects within the common FoV and, respectively,  non-common (exclusive) FoV of agent $i$.

According to the GCI rule, the fusion of $\pi_a(X)$ and $\pi_b(X)$ yields:
\begin{equation}\label{GCI}
\begin{split}
\overline\pi(X)\propto &{\pi_{a,co}(X\cap Fov_{C})}^{\omega_a}{\pi_{b,co}(X\!\cap\! Fov_{C})}^{\omega_b}\times\\
&\pi_{a,nc}(X\!\cap\! (Fov_{a}\backslash Fov_{C})|X\!\cap Fov_{C})^{\omega_a}\times\\
&{\pi_{b,ui}(X\cap (Fov_a\backslash Fov_C))}^{\omega_b}\times\\
&{\pi_{a,ui}(X\!\cap\! Fov_b\backslash Fov_C)}^{\omega_a}\times\\
&{\pi_{b,nc}(X\!\cap\!(Fov_{b}\backslash Fov_{C})|X\!\cap\!Fov_{C})}^{\omega_b},
\end{split}
\end{equation}
where $\omega_i \in [0,1]$, for $i \in \{a,b\}$, and $\omega_a+\omega_b=1$.

Recall that the normalization of the fusion weights (i.e., $\omega_a+\omega_b=1$) is to avoid the double counting of common information of different agents \cite{Battistelli,double-counting}.
Looking at (\ref{GCI}), the common information between agents $a$ and $b$ can only be on $X\cap Fov_{C}$ which are the object states within the common FoV of agents $a$ and $b$.
As for the exclusive object states $X\cap \left( Fov_{a}\backslash Fov_C \right)$ and $X\cap \left( Fov_{b}\backslash Fov_C \right)$, they are  only observed by a single agent so that the corresponding density of the other agent used for fusion is a multi-object uninformative density, implying no common information between the two agents.
Hence, if fusion weights $\omega_a < 1$ and/or $\omega_b <1$  are applied also to the exclusive densities $\pi_{a,nc}, \pi_{a,ui}$ and/or $\pi_{b,nc}, \pi_{b,ui}$ in
(\ref{GCI}), an under-confidence problem occurs.

Hence, to avoid this problem, a better alternative for the fusion with respect to $X\cap \left( Fov_{a}\backslash Fov_C \right)$ or $X\cap \left( Fov_{b}\backslash Fov_C \right)$ is to perform Bayesian rule \cite{mahler_book}, that amounts to power-raising  posteriors of different agents with unit weights
$ \omega_a= \omega_b=1$.  The rationale is that,
as a result, the corresponding fusion  becomes (\ref{GCI-fixed}).
\begin{figure*}[bph]
\hrulefill
\begin{equation}\label{GCI-fixed}
\begin{split}
\overline\pi(X)&\propto \underbrace{{\pi_{a,co}(X\cap Fov_{C})}^{\omega_a}{\pi_{b,co}(X\cap Fov_{C})}^{\omega_b}}_{\text{Fusion on } Fov_C}\times\underbrace{{\pi_{a,ui}(X\cap Fov_b\backslash Fov_C)}{\pi_{b,nc}(X\cap(Fov_{b}\backslash Fov_{C})|X\!\cap\!Fov_{C})}}_{\text{Fusion on the difference set }  Fov_b\backslash Fov_C}\\
&\,\,\,\,\,\,\,\,\,\,\,\,\,\,\,\,\,\,\,\,\,\,\,\,\,\,\,\,\,\,\,\,\,\,\,\,\,\,\,\,\,\times\underbrace{\pi_{a,nc}(X\cap(Fov_{a}\backslash Fov_{C})|X\!\cap\ Fov_{C}){\pi_{b,ui}(X\cap (Fov_a\backslash Fov_C))}}_{\text{Fusion on the difference set } Fov_a\backslash Fov_C}
\end{split}
\end{equation}
\end{figure*}

Exploiting the Bayesian-operator invariance of multi-object uninformative densities, after normalization (\ref{GCI-fixed}) can be further expressed as (\ref{GCI-fixed-1}),
\begin{figure*}[bp]
\begin{equation}\label{GCI-fixed-1}
\!\!\!\!\overline\pi(X)= \underbrace{\overline\pi_{co}(X\cap Fov_C)}_{\text{GCI}}\times\underbrace{{\pi_{a,nc}(X\cap  (Fov_{a}\backslash Fov_{C})|X\cap Fov_{C})}}_{\text{Bayesian-operator invariance}}\times\underbrace{{\pi_{b,nc}(X\cap  (Fov_{b}\backslash Fov_{C})|X\cap Fov_{C})}}_{\text{Bayesian-operator invariance}}
\end{equation}
\end{figure*}
where,  $\forall X\in \mathcal{F}(Fov_C)$,
\begin{equation}\label{partial-GCI}
\overline\pi_{co}(X)=\frac{{\pi_{a,co}(X)}^{\omega_a} \, {\pi_{b,co}(X)}^{\omega_b}}{\int {\pi_{a,co}(X)}^{\omega_a} \, {\pi_{b,co}(X)}^{\omega_b} \delta X}.\end{equation}
The proposed fusion rule (\ref{GCI-fixed-1})-(\ref{partial-GCI})  is referred to hereafter as \textit{Bayesian-operator InvaRiance on Difference-sets} (BIRD)  where difference-sets refer to $Fov_{b}\backslash Fov_{C}$ and  $Fov_{b}\backslash Fov_{C}$.
\begin{Rem}
The proposed BIRD fusion  is in the same spirit of the  split CI  fusion  \cite{CU-Fusion} that performs CI fusion for the unknown correlated components, while applies the Kalman filter to the known independent components.
\end{Rem}

For a single agent $i$, which is not to get object observations outside its individual FoV, only the posterior $\pi_i(X\cap Fov_i)$ is informative and reliable;
 accordingly, only detections/estimates of objects moving within $Fov_i$ can be provided.
 However, by employing the proposed BIRD fusion in (\ref{GCI-fixed-1})-(\ref{partial-GCI}), a reliable and informative density $\overline\pi(X)$ over the global FoV is obtained, so that
 detections/estimates of all objects within the union of individual FoVs can be provided.
 Furthermore, the fused density achieves maximum information gain from the local posteriors within the common FoV, and also preserves, in the exclusive FoV, sufficient and valid information provided by the observing agent.

\subsubsection{Multi-agent case}
The BIRD fusion approach, presented above for $N_s=2$ agents,
can be easily extended to  the case of $N_s>2$ agents by sequentially applying the pairwise fusion (\ref{GCI-fixed})-(\ref{GCI-fixed-1}) $N_s-1$ times.
The same sequential fusion strategy has already been adopted for standard GCI fusion
with CPHD \cite{Battistelli} and, respectively, MB filters \cite{GCI-MB}.

The pseudocode of sequential BIRD fusion is given in Algorithm \ref{algorithm:sequential_fusion}, for which the following comments are in order.
\begin{algorithm}[tb]
\caption{Sequential BIRD fusion}
\small
{\underline{INPUT:} Local posterior $\pi_i$ and field-of-view $Fov_i$ of any agent $i \in \{ 1, 2, \dots, N_s \}$ involved in the fusion\;

\underline{OUTPUT:} Fused density $\overline\pi$ and domain of definition $\overline{Fov}$\;

\textbf{function} $(\overline\pi, \overline{Fov})=\mathfrak{F}$($\{(\pi_1,Fov_1),\cdots,(\pi_{N_s},Fov_{N_s})\}$\\
\For{$j=2:N_s$\\}
{Calculate the common FoV: $Fov_C=Fov_1\cap Fov_j$\;
Calculate the marginal densities $ \pi_{1,co}(X\cap Fov_C)$ and $\pi_{j,co}(X\cap Fov_C)$ according to (\ref{marginal-density})\;
Calculate the conditional densities $\pi_{1,nc}(X\cap  (Fov_1\backslash Fov_{C})|X\cap Fov_{C})$ and $\pi_{j,nc}(X\cap  (Fov_{i}\backslash Fov_{C})|X\cap Fov_{C})$
according to (\ref{conditioned-density-V})\;
Compute the fused density according to (\ref{GCI-fixed-1}) and (\ref{partial-GCI}):
\begin{equation}\notag
\begin{split}\overline\pi(X):=&\overline\pi_{co}(X\cap Fov_{C})\\
&\pi_{1,nc}(X\cap  (Fov_1\backslash Fov_{C})|X\cap Fov_{C})\times\\
&{\pi_{i,nc}(X\cap  (Fov_{j}\backslash Fov_{C})|X\cap Fov_{C})};
\end{split}\end{equation}

$\overline{Fov}:=Fov_1 \cup Fov_j$\;
$\pi_1:=\overline\pi$\;
$Fov_1:=\overline{Fov}$\;}
\textbf{return}: $\overline\pi$ and $\overline{Fov}$.}
\label{algorithm:sequential_fusion}
\end{algorithm}
 \begin{itemize}
 \item
After performing fusion $N_s-1$ times sequentially, the final fused density $\overline\pi(X)$ will be informative over  the global Fov of the sensor network, i.e., $\overline{Fov}=Fov_G$, integrating object information of all agents.

 \item Let $\{A_1,\cdots, A_N\}$   provide a partition of the global FoV $Fov_G$, where each $A_j$ is the intersection of the FoVs of all agents $i$ that can illuminate the region $A_j$, i.e.,
 \begin{equation}
A_j=\bigcap_{i: A_j\subseteq Fov_i} Fov_i.
 \end{equation}
 After the $N_s-1$ pairwise fusion steps, for each $A_j$, the fusion is only performed with all agents that can illuminate $A_j$.
 As for other sensors which cannot illuminate $A_j$, since multi-object uninformative densities are considered in the fusion, there is no impact on the fusion result.
\end{itemize}
The above two points ensure that the fused density integrates the most appropriate amount of information from all agents over the global FoV.
To further clarify the aforementioned observations, we give hereafter an example relative to three agents.
\begin{Exa}
Let us consider fusion of three densities $\pi_1$, $\pi_2$, $\pi_3$ provided by three agents with respective fields-of-view $Fov_1$, $Fov_2$, $Fov_3$.
The global FoV is partitioned into $6$ regions denoted by $A_i$ ($i=1,\cdots 6$) as shown in Fig. \ref{fig:Division of the global FoV}.
\begin{figure}[h]
  \centering
\includegraphics[width=8.5cm]{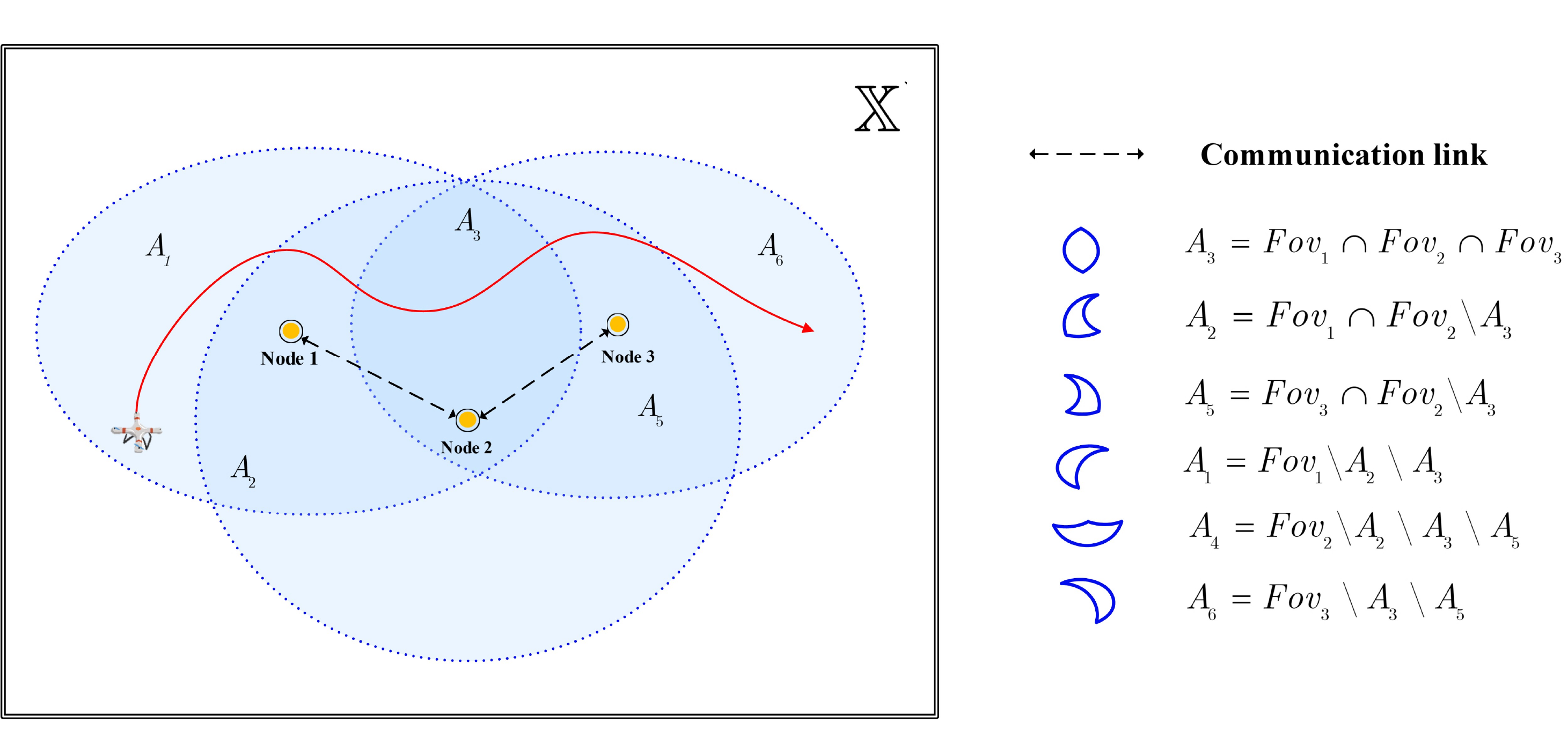}
\caption{{\bf } Division of the global FoV of the sensor network.}
\label{fig:Division of the global FoV}
\end{figure}

By adopting the sequential pairwise strategy, fusion is first performed between $\pi_1$ and $\pi_2$ according to (\ref{GCI-fixed-1}), resulting into
\begin{equation}\label{fusion-1-2}
{\small{\begin{split}
\!\!&\overline\pi_{1,2}(X)\\
\!\!\propto&\pi_{1,co}(X\cap\left(A_2\cup A_3\right))^{\omega}\pi_{2,co}(X\cap \left(A_2\cup A_3\right))^{1-\omega}\times\\
&\pi_{1,nc}(X\cap A_1|\left(A_2\cup A_3\right))\pi_{2,nc}(X\cap\left(A_4\cup A_5\right)|\left(A_2\cup A_3\right))
\end{split}}}
\end{equation}
After normalization, we get a fused density $\overline\pi_{1,2}(X)$ defined on $$Fov_1\cup Fov_2=A_1\cup A_2 \cup A_3 \cup A_4\cup A_5.$$
Then, fusion is further performed between $\overline\pi_{1,2}(X)$ and $\pi_3(X)$.
According  to (\ref{GCI-fixed-1}), the final fused density yields:
\begin{equation}\label{fusion-1-2-3}
{\small{\begin{split}
&\overline\pi_{1,2,3}(X)\\
\propto&\overline\pi_{1,2,c}(X\cap\left(A_3\cup A_5\right))^{\omega}\pi_{3,c}(X\cap \left(A_3\cup A_5\right))^{(1-\omega)}\times\\
&\overline\pi_{1,2,nc}(X\cap \left(A_1\cup A_2 \cup A_4\right)|X\cap\left(A_3\cup A_5\right))\times\\
&\pi_{3,nc}(X\cap A_6|X\cap\left(A_3\cup A_5\right))
\end{split}}}
\end{equation}
After normalization of (\ref{fusion-1-2-3}), the final fused density $\overline\pi_{1,2,3}(X)$ is defined on the global Fov, i.e.
$\overline {Fov}=\cup_{i=1}^6 A_i=Fov_G$, integrating object information of all three agents.

Specifically, observing (\ref{fusion-1-2}) and (\ref{fusion-1-2-3}):
\begin{itemize}
\item  for the region $A_3$ illuminated by all three agents, the fusion actually involves all the three densities;
\item  for the regions $A_2$ and $A_5$ illuminated by only two agents, the fusion involves the two densities of such agents;
\item for the regions $A_1$, $A_4$ and $A_6$ illuminated by a single agent, no fusion is actually performed but the corresponding density of the illuminating agent is preserved.
\end{itemize}
\end{Exa}
\subsection{Implementation of BIRD fusion}
This subsection provides the specific implementation of BIRD fusion, from the perspective of local filtering  and consensus-based distributed fusion.
\subsubsection{Local filtering for BIRD fusion}
Observing the proposed  fusion formula (\ref{GCI-fixed-1}), only the marginal density with respect to $Fov_i$ of each agent's posterior, i.e. $\pi_i(X\cap Fov_i)$, is used for fusion. This is consistent with the fact that the local filter can only provide valid updates to the prior within the local FoV. As a result, before BIRD fusion, the marginal posterior $\pi_i(X\cap Fov_i)$ is calculated  first. Hence, it does not matter how the probabilities of survival and detection are chosen outside the local FoV for  local filtering, which  further demonstrates the robustness of the proposed fusion.

\subsubsection{Consensus-based distributed fusion}
The fusion formula (\ref{GCI-fixed-1}) can be readily applied to the centralized case by adopting the sequential fusion given in Algorithm \ref{algorithm:sequential_fusion}. However, how to utilize the fusion formula in a fully  distributed way is not obvious.  In this subsection, we  discuss the
application of (\ref{GCI-fixed-1}) to the distributed peer-to-peer setting.

In a distributed sensor network, each agent has also limited communication capability and hence, at each time step, can only receive the posteriors of its neighbours.
In order to make each agent share the global information of the whole network, a consensus approach \cite{Xiao, Battistelli} is usually adopted.
The core idea of consensus-based distributed fusion is that, at each time instant $k$,
the collective fusion can be approximated by iterating regional fusions, called consensus steps, among neighboring nodes.
More precisely, at each sampling interval $k$, a certain number of consensus steps, say $L$, is performed.
Let us denote by $\pi_{i}^{\ell} (X)$ the density of agent  $i$ at (time $k$, not indicated for the sake of simplicity,  and) consensus step $\ell$.
Then, by applying consensus to the proposed enhanced fusion rule (\ref{GCI-fixed}),
we obtain the recursive algorithm, for each $\ell=1,\cdots,L$,
\begin{align}\label{consensus:iterate}
\begin{split}
(\pi_i^{(\ell)},  Fov^\ell)=\mathfrak{F}(\{(\pi_j^{(\ell-1)},Fov^{(\ell-1)}\}_{j\in\mathcal{N}_i}), \forall i\in\mathcal{N},
\end{split}
\end{align}
where $\pi_i^{(0)}=\pi_i$, $Fov^{(0)}=Fov_i$, $i\in\mathcal{N}$,  $\mathfrak{F}(\cdot)$ is given in Algorithm I.

According to the proposed fusion rule, the actual FoV of each agent is extended with the progression of consensus steps.
After a certain number of steps (which is at least equal to the network diameter), each agent can share the global FoV and all information of the other agents.

The pseudocode of the  consensus-based distributed BIRD fusion is given in Algorithm \ref{algorithm:consensus_fusion}.
\begin{algorithm}[tb]
\caption{Consensus based BIRD fusion performed by agent $i\in\mathcal{N}$.}
\small
{\underline{INPUT:} Local posterior $\pi_j$ and field-of-view $Fov_j$ of each agent $j \in \mathcal{N}_i$; number of consensus steps $L$\;

\underline{OUTPUT:} Consensus-based fused density $\pi_i^{(L)}$ and domain of definition $Fov_i^{(L)}$\;

\textbf{function} $\mathfrak{F}_{con}$($\{ \left(\pi_j,Fov_j \right) \}_{j \in \mathcal{N}_i} $)\\
Perform the initialization:
\begin{align}
\notag\pi_i^{(0)}:=&\pi_i, \\
\notag Fov_i^{(0)}:=&Fov_i
\end{align}
\For {$\ell=1:L$}
{Exchange posteriors with neighbours\;
Perform the sequential BIRD fusion  given in Algorithm I
\centerline{$(\pi_i^{(\ell)}, Fov_i^{(\ell)})=\mathfrak{F} ( \{ (\pi_j^{(\ell-1)}, Fov_j^{(\ell-1)}) \}_{j\in\mathcal{N}_i} )$}
}
\textbf{return}: $ \left( \pi_i^{(L)},Fov_i^{(L)} \right)$.}
\label{algorithm:consensus_fusion}
\end{algorithm}

%
\section{BIRD fusion of Poisson RFSs}
For the subsequent developments, the attention is devoted to specific RFSs called \textit{Poisson}, more precisely  \textit{Poisson point processes} \cite{stone-streit} or \textit{multi-object Poisson processes}\cite{mahler_book}, that play a fundamental role in PHD filtering \cite{mahler_book, PHD-Vo}.
\subsection{Decomposition and union of Poisson RFSs}
We will first derive  in Proposition \ref{marginal density} the marginal density of a Poisson RFS, based on which  the decomposition of a Poisson RFS with respect to disjoint subspaces in  the form of (\ref{decomposition-pi}) will be given in Proposition \ref{decomposition-Poisson}.
Finally, Proposition \ref{Combination-Poisson} will show that for any two Poisson RFSs defined on disjoint spaces, their union is another Poisson RFS defined on the union of such spaces.
\begin{Pro}
\label{marginal density}
Let us consider a Poisson RFS on $\mathbb{X}$ with density
\begin{equation}
\pi(X)=[\lambda]^{|X|}\, \exp(-\lambda) \, p^{X},\end{equation}
where $p^{X}\triangleq\prod_{x\in X}p(x)$ with $p^\emptyset\triangleq1$ by convention.
Then, $X_W = X\cap\mathbb{W}$, with $\mathbb{W}\subseteq\mathbb{X}$, is also a Poisson RFS with marginal density
\begin{equation}
\pi_{W}(X)=[\lambda K_{W}]^{|X|}\exp(-\lambda K_{W})p_{W}^{X}
\end{equation}
where
\begin{align}
\label{para-p-w}p_{W}(x)\triangleq&\frac{1_{\mathbb{W}}(x)p(x)}{K_{W}}\\
\label{para-K-w}K_{W}\triangleq&\int_{\mathbb{W}}p(x) dx
\end{align}
Further, $X_V=X\backslash X_W$  defined on $\mathbb{V}=\mathbb{X}\backslash\mathbb{W} $ and conditioned on $X_W$, is also a Poisson RFS,
with the conditional density
\begin{equation}\label{conditioned-density-V-Poisson}
\pi_{V}(X'|X)=\pi_V(X')=[\lambda K_{V}]^{|X'|} \, \exp(-\lambda K_V) \,p_{V}^{X'}
\end{equation}
where
\begin{align}
\label{para-p-v}p_{V}(x)\triangleq&\frac{1_{\mathbb{V}}(x)p(x)}{K_V}\\
\label{para-K-v}K_{V}\triangleq&\int_{\mathbb{V}}p(x) dx=1-K_W
\end{align}
\end{Pro}
\noindent
\textit{Proof:} see Subsection V of the supplemental materials . \qed \\
\begin{Pro}
\label{decomposition-Poisson}
(Decomposition) Let us consider a Poisson RFS on $\mathbb{X}$ with density
\begin{equation}
\pi(X)=[\lambda]^{|X|} \exp(-\lambda)p^{X}\end{equation}
and an arbitrary partition $\mathbb{X}=\mathbb{W}\uplus\mathbb{V}$.
Then, the density can be factored as follows:
\begin{equation}
\pi(X)=\pi_{W}(X\cap\mathbb{W}) \, \pi_{V}(X\cap\mathbb{V})
\end{equation}
where: $\pi_{W}(X)$ is a Poisson RFS density on $\mathbb{W}$ with parameter pair $(\lambda K_{W}, p_{W}(x))$;
$\pi_{V}(X)$ is a Poisson RFS density on $\mathbb{V}$ with parameter pair $(\lambda K_{V}, p_{V}(x))$;
$p_W, K_W, p_V, K_V$ are given by (\ref{para-p-w}), (\ref{para-K-w}), (\ref{para-p-v}) and (\ref{para-K-v}), respectively. \end{Pro}
\noindent\textit{Proof:}
Proposition \ref{decomposition-Poisson} is a direct consequence of Proposition \ref{marginal density}, considering that, by dividing the state space into disjoint subspaces, one can always write a Poisson
RFS as union of independent Poisson RFSs on such disjoint subspaces. It is, therefore,  a special property of  the Poisson RFS family. \qed \\
\begin{Pro} (Union)
\label{Combination-Poisson}
Let us consider  Poisson RFSs $X_W$ and $X_V$ defined on disjoint state spaces $\mathbb{W}$ and $\mathbb{V}$ (i.e., $\mathbb{W}\cap\mathbb{V}=\emptyset$), with  densities $\pi_W(\cdot)$ and $\pi_V(\cdot)$ given by
\begin{align}\label{density-w}
\pi_W(X)=&[\lambda_W]^{|X|} \exp(-\lambda_W)p_W^{X}\\
\label{density-v}\pi_V(X)=&[\lambda_V]^{|X|} \exp(-\lambda_V)p_V^{X}.
\end{align}
Then, the union of $X_W$ and $X_V$, i.e. $X=X_W\cup X_V$, turns out to be a Poisson RFS on state space $\mathbb{X}=\mathbb{W}\cup\mathbb{V}$, with density
\begin{equation}
\begin{split}
\pi(X)=&[\lambda]^{|X|} \exp(-\lambda)p^{X}
\end{split}
\end{equation}
where
\begin{align}
\lambda=&\lambda_W+\lambda_V\\
p(x)=&\frac{\lambda_W}{\lambda}p_{W}(x)+\frac{\lambda_V}{\lambda}p_{V}(x)
\end{align}
\end{Pro}
\noindent
\textit{Proof:} see Subsection VI of the supplemental materials. \qed \\
\subsection{Fusion formulas of Poisson RFSs}
Now, the fusion of Poisson posteriors using the BIRD rule (\ref{GCI-fixed-1}) is detailed.
For the sake of simplicity and without loss of generality, the case of two agents will be dealt with, recalling that more agents can be anyway handled by sequential pairwise fusion steps according to Algorithm \ref{algorithm:sequential_fusion}, or the consensus based distributed fusion according to Algorithm \ref{algorithm:consensus_fusion}.
Hence, let us assume that agents $a$ and $b$ have Poisson posteriors, defined on $Fov_{a}$ and $Fov_{b}$, of the form
\begin{equation}\label{posterior-poisson}
\pi_{i}(X)=[\lambda_{i}]^{|X|} \, \exp(-\lambda_{i}) \, p_{i}^{X}, ~~~i=a,b
\end{equation}
Notice that the Poisson RFS (\ref{posterior-poisson}) is completely characterized by the parameter pair $(\lambda_i, p_i)$ \cite{mahler_book}.

The BIRD fusion of $\pi_a$ and $\pi_b$ can, therefore, be performed via the following three-step procedure.

\textbf{STEP 1:}
 Perform the decomposition of the posteriors $(\lambda_i,p_i(x))$ on $Fov_i$ of agents $i=a,b$ by dividing the  state space into disjoint subsets, $Fov_C$  and $Fov_i \backslash Fov_C$, i.e.
\begin{equation}
\pi_i(x)=\pi_{i,co}(X\cap Fov_C) \, \pi_{i,nc}(X\cap( Fov_i\backslash  Fov_C)).
\end{equation}
According to Proposition 4 and proposition 5,  $\pi_{i,co}$ and $\pi_{i,nc}$   are both Poisson parameterized by $(\lambda_{i,co},p_{i,co}(x))$ and $(\lambda_{i,nc},p_{i,nc}(x))$, respectively, with
 \begin{align}
\label{decomposition-lambda-C} \lambda_{i,co}=&\lambda_{i}K_{i,co}
 \\
\label{decomposition-K-C}  K_{i,co}=&\int_{Fov_C} p_{i}(x) dx
 \\
\label{decomposition-Poisson-C} p_{i,co}(x)=&\frac{1_{ Fov_C}(x)p_{i}(x)}{K_{i,co}}\\
  \lambda_{i,nc}=&\lambda_{i}K_{i,nc}
 \\
K_{i,nc}=&\int_{Fov_i\backslash {Fov_C}}p_{i,nc}(x) dx
 \\
 \label{decomposition-Poisson-NC}p_{i,nc}(x)=&\frac{1_{Fov_i\backslash  Fov_C}(x)p_{i,nc}(x)}{K_{i,nc}}
 \end{align}

\textbf{{STEP 2:}}
Compute the fused density over the common FoV, $Fov_C$, namely the GCI fusion of $\pi_{a,co}(X)$ and $\pi_{b,co}(X)$ (both defined on
the common field-of-view $Fov_C$.

Since $\pi_{a,co}(X)$ and $\pi_{b,co}(X)$ are both Poisson, as proved in \cite{Clark}, the fused density $\overline\pi_{co}$ is also Poisson, parameterized by $(\overline\lambda_{co}, \overline p_{co})$ where
\begin{align}
\overline \lambda_{co}=&\lambda_{a,co}^{\omega_{a}} \, \lambda_{b,co}^{\omega_{b}}  \, \overline K_{co}\label{fusion-lambda-w1}\\
\overline p_{co}=&\frac{1_{Fov_{C}}(x) \, p_{a,co}(x)^{\omega_{a}} \, p_{b,co}(x)^{\omega_{b}}}{\overline K_{co}}
\label{fusion-p-w1}\\
 \overline K_{co}=&\int_{Fov_{C}} p_{a,co}^{\omega_{a}}(x) \, p_{b,co}^{\omega_{b}}(x) \, dx \label{fusion-C-w1}
 \end{align}

\textbf{STEP 3:}
According to (\ref{GCI-fixed-1}), compute the global fused density $\overline \pi(X)$ over the global FoV, i.e.,
\begin{equation}
\begin{split}
Fov_G=&(Fov_a\backslash Fov_C)\cup Fov_C \cup (Fov_b\backslash Fov_C)\\
=&Fov_a\cup Fov_b.
\end{split}
\end{equation}
This step amounts to performing disjoint union of three Poisson RFSs with densities $\pi_{i,nc}$ (on state space $Fov_i\backslash Fov_C$), $i \in \{a,b \}$, and
$\overline\pi_{co}(X)$ (on state space $Fov_C$).
Applying Proposition \ref{Combination-Poisson} twice, the global fused density $\overline\pi$  yields another Poisson RFS parameterized by $(\overline\lambda,\overline p)$ with
\begin{align}
\label{fused-location-density} \overline p(x) \triangleq &
\frac{\overline\lambda_{co}}{\overline\lambda} \, \overline p_{c}(x) +
\frac{\lambda_{a,nc}}{\overline\lambda} \, p_{a,nc}(x) +
\frac{\lambda_{b,nc}}{\overline\lambda} \, p_{b,nc}(x) \\
\label{fused-cardinality} \overline\lambda=&\overline\lambda_{co}+\lambda_{a,nc}+\lambda_{b,nc}
\end{align}
\subsection{Gaussian mixture implementation}
Let us again consider agents $a$ and $b$ with Poisson posteriors.  As mentioned in Subsection IV-B,  before BIRD fusion, the marginal density with respect to the local FoV, i.e. $\pi_{i}(X\cap Fov_{i})$ is calculated. According to Proposition \ref{marginal density}, the resulting marginal posterior is  also Poisson  characterized by  parameter pairs $(\lambda_i,p_i)$, $i \in \{a,b\}$, with location densities
represented in bounded GM form as
\begin{equation}\label{GM-poisson}
p_i(x)=\sum_{j=1}^{N_{G,i}}\alpha_i^{(j)} \, \mathcal{N}(x;\hat x_i^{(j)},P_i^{(j)}) \, 1_{Fov_i}(x), ~~~i=a,b.
\end{equation}

Next, a step-by-step implementation of the proposed BIRD fusion is given as follows.

\textbf{STEP 1:}
Given the GM form of $p_i(x)$ in (\ref{GM-poisson}), the parameters of $\pi_{i,nc}$ and $\pi_{i,nc}$ are computed as follows.

\textit{\textbf{1) Calculation of $\lambda_{i,co}$}:}
First, the calculation of $K_{i,co}$ is considered.
    According to (\ref{decomposition-K-C}), we have
    \begin{equation}\label{GM-C_i1}
    \begin{split}
    K_{i,co}=&\int_{Fov_C}\sum_{j=1}^{N_{G,i}}\alpha^{(j)}_i\mathcal{N}(x;\hat x_i^{(j)},P_i^{(j)}) \, dx\\
    =&\sum_{j=1}^{N_{G,i}}\alpha^{(j)}_i c_i^{(j)}(Fov_C)
    \end{split}
    \end{equation}
where
\begin{equation}
c_i^{(j)}(Fov_C)=\int_{Fov_C}\mathcal{N}(x;\hat x_i^{(j)},P_i^{(j)}) \, dx .
\label{eq73}
\end{equation}
Even though (\ref{eq73}) is the integration of a Gaussian density, there is no closed-form expression for it.
Hence, we provide a Monte Carlo (MC) approximation of $c^{(j)}_i{(Fov_C)}$ as follows:
\textit{draw $M^{(j)}_i$ samples $x^{(j,m)}_i$, for $m=1,\cdots,M_i^{(j)}$, from the Gaussian density $\mathcal{N}(x;\hat x_i^{(j)},P_i^{(j)})$, then $c_i^{(j)}(Fov_C)$ can be approximated as
\begin{equation}
\begin{split}
c^{(j)}_i{(Fov_C)}
=& \dfrac{1}{M_i^{(j)}}  \, \sum_{m=1}^{M^{(j)}_i} 1_{Fov_C}(x^{(j,m)}_i).
\end{split}
\end{equation}}
Then, according to (\ref{GM-C_i1}), the quantity $K_{i,co}$ can be approximated as
\begin{equation}\label{GM-G_i1-evaluated}
K_{i,co}=\sum_{j=1}^{N_{G,i}} \,  \dfrac{\alpha_i^{(j)}}{M_i^{(j)}}  \,\sum_{m=1}^{M^{(j)}_i} 1_{Fov_C}(x^{(j,m)}_i).
\end{equation}
As a result, according to (\ref{decomposition-lambda-C}), the parameter $\lambda_{i,co}$ can be computed by
\begin{equation}
\lambda_{i,co}=\lambda_{i}K_{i,co}=\lambda_{i}\sum_{j=1}^{N_{G,i}} \dfrac{\alpha^{(j)}_i}{M_i^{(j)}}  \,  \sum_{m=1}^{M^{(j)}_i} 1_{Fov_C}(x^{(j,m)}_i).
\end{equation}

\textit{\textbf{2) Calculation of $p_{i,co}(x)$}:}
According to (\ref{decomposition-Poisson-C}) and the evaluated $K_{i,C}$ given in (\ref{GM-G_i1-evaluated}), the location density over $Fov_C$ is given by
\begin{equation}\label{GM-pi1}
p_{i,co}(x)=\sum_{j=1}^{N_{G,i}}\frac{\alpha_i^{(j)}}{K_{i,co}} \, \mathcal{N}(x;\hat x_i^{(j)},P_i^{(j)}) \, 1_{Fov_C}(x).
\end{equation}

\textit{\textbf{3) Calculation of  $\lambda_{i,nc}$ and  $p_{i,nc}(x)$:}}
Given  the evaluated $K_{i,c}$ in (\ref{GM-G_i1-evaluated}), the quantity $K_{i,nc}$ can be easily computed by
\begin{equation}
K_{i,nc}=1-K_{i,co}.
\end{equation}
Further, similarly as for $p_{i,co}$,  according to (\ref{decomposition-Poisson-NC}), $p_{i,nc}(x)$ is given by,
\begin{equation}\label{GM-p-i2}
p_{i,nc}(x)\!\!=\!\!\sum_{j=1}^{N_{G,i}}\frac{\alpha_i^{(j)}}{K_{i,nc}} \, \mathcal{N}(x;\hat x_i^{(j)},P_i^{(j)})\, 1_{Fov_i\backslash Fov_C}(x).
\end{equation}

\textbf{STEP 2:}
We first compute the numerator of the fused location density $\overline p_{co}(x)$.
By substituting the GM forms of $p_{a,co}(x)$ and $p_{b,co}(x)$ in (\ref{GM-pi1}) into the numerator of (\ref{fusion-p-w1}), we get
\begin{equation}\label{numerator-pw1}
 \begin{split}
 &1_{Fov_C}(x)p_{a,co}(x)^{\omega_{a}}p_{b,co}(x)^{\omega_{b}}\\
 =&1_{Fov_C}(x)\left[\sum_{j=1}^{N_{G,a}}\frac{\alpha_a^{(j)}}{K_{a,co}}\mathcal{N}(x;\hat x_a^{(j)},P_a^{(j)})1_{Fov_C}(x)\right]^{\omega_a}\times\\
    &\left[\sum_{j=1}^{N_{G,b}}\frac{\alpha_b^{(j)}}{K_{b,co}}\mathcal{N}(x;\hat x_b^{(j)},P_b^{(j)})1_{Fov_C}(x)\right]^{\omega_b}.
 \end{split}
 \end{equation}
Notice that (\ref{numerator-pw1}) involves exponentiation of a  GM which, unfortunately,  does not provide a GM in general.
As suggested in \cite{Battistelli}, the following approximation is reasonable
 \begin{align}
\begin{split}\label{GMs2}
&\left[\sum_{j=1}^{N_{G,i}}\frac{\alpha_i^{(j)}}{K_{i,co}}\mathcal{N}(x;\hat x_i^{(j)},P_i^{(j)})1_{Fov_C}(x)\right]^{\omega_i}\\
\cong &\sum_{j=1}^{N_{G,i}}\left(\frac{\alpha_i^{(j)}}{K_{i,co}}\right)^{\omega_i}\left[\mathcal{N}(x;\hat x_i^{(j)},P_i^{(j)})1_{Fov_C}(x)\right]^{\omega_i}
\end{split}
\end{align}
as long as the Gaussian components of $p_{i,co}(x)$ are well separated relatively to their corresponding covariances.
In case the components are not well separated, one can either perform merging before fusion (this is possible since, for a Poisson RFS, each location density $p_{i,co}(x)$ refers to a hypothetic single object) or use different approximations, e.g., by replacing the GM representation by a sigma-point approximation \cite{fractional-power}.

As a result, (\ref{numerator-pw1}) can be further re-expressed as
    \begin{equation}\label{GM-C_ww1}
    \begin{split}
    &1_{Fov_C}(x)p_{a,co}(x)^{\omega_{a}}p_{b,co}(x)^{\omega_{b}}\\
    =&\sum_{j=1}^{N_{G,a}}\!\sum_{j'=1}^{N_{G,b}}\left(\frac{\alpha_a^{(j)}}{K_{a,co}}\right)^{\!\!\omega_a}\!\!\!\left(\frac{\alpha_b^{(j)}}{K_{b,co}}\right)^{\!\!\omega_b}\!\!\!
\left[\mathcal{N}(x;\hat x_a^{(j)},P_a^{(j)})1_{Fov_C}(x)\right]^{\!\omega_a}\\
    &\times\left[\mathcal{N}(x;\hat x_b^{(j')},P_b^{(j')})1_{Fov_C}(x)\right]^{\omega_b}1_{Fov_C}(x).
   \end{split}
    \end{equation}
    Then, by exploiting  some elementary operations on GMs \cite{Battistelli}, the numerator (\ref{GM-C_ww1}) can be rewritten as
   \begin{equation}\label{quantity-before-Kc}
   \begin{split}
   &1_{Fov_C}(x)p_{a,co}(x)^{\omega_{a}}p_{b,co}(x)^{\omega_{b}}\\
   =&\sum_{j=1}^{N_{G,a}}\sum_{j'=1}^{N_{G,b}}\widetilde\alpha^{(j,j')}\mathcal{N}(x;\hat {x}^{(j,j')},P^{(j,j')}) \,1_{Fov_C}(x)
   \end{split}
   \end{equation}
   where
   \begin{align}
    P^{(j,j')}=&\left[\omega_a(P_{a}^{(j)})^{-1}+\omega_b(P_{b}^{(j')})^{-1}\right]^{-1}\\
   \hat{ x}^{(j,j')}=&P^{(j,j')}\left[\omega_a(P_{a}^{(j)})^{-1}\hat x_a^{(j)}+\omega_b(P_{b}^{(j')})^{-1}\hat x_b^{(j')}\right] \\
   \alpha^{(j,j')}=&\left(\frac{\alpha_a^{(j)}}{K_{a,co}}\right)^{\omega_a}\left(\frac{\alpha_b^{(j')}}{K_{b,co}}\right)^{\omega_b}\kappa(P_{a}^{(j)}, \omega_a) \,
   \kappa(P_{b}^{(j')}, \omega_b)\notag\\
   &\times\mathcal{N}\!\!\left(\!\!\hat x_{a}^{(j)}\!-\!\hat x_{b}^{(j')};0,\!\frac{P_{a}^{(j)}}{\omega_a}\!+\!\frac{P_{b}^{(j')}}{\omega_b}\!\!\right)\\
\kappa(P,\omega)&=\sqrt{\det[2\pi P\omega^{-1}](\det[2\pi P])^{-\omega}}.
\end{align}

\textit{\textbf{1) Calculation of $\overline \lambda_{co}$:}} First, calculation of $\overline K_{co}$ is discussed.
Given the GM form of $p_{i,co}(x)$ in (\ref{GM-pi1}), according to (\ref{fusion-C-w1}) and (\ref{quantity-before-Kc}), we have
    \begin{equation}\label{GM-C_w1}
    \begin{split}
    \overline K_{co}=&\int 1_{Fov_C}(x)p_{a,co}(x)^{\omega_{a}}p_{b,co}(x)^{\omega_{b}}dx\\
    =&\sum_{j=1}^{N_{G,a}}\sum_{j'=1}^{N_{G,b}}\widetilde\alpha^{(j,j')}\int_{Fov_C}\mathcal{N}(x;\hat x^{(j,j')},P^{(j,j')})dx.
   \end{split}
    \end{equation}
Then, based on (\ref{fusion-lambda-w1}), the evaluation of the parameter $\overline \lambda_{co}$ can be obtained.

\textit{\textbf{2) Calculation of $\overline p_{co}(x)$:}} Given the  numerator of $\overline p_{co}(x)$ in (\ref{quantity-before-Kc}) and $\overline K_{co}$ in (\ref{GM-C_w1}), the GM form of the location density $\overline p_{co}(x)$ is given by
    \begin{equation}\label{GM-p-w1}
   \!\overline p_{co}(x)\!=\!\!\sum_{j=1}^{N_{G,a}}\!\sum_{j'=1}^{N_{G,b}} \frac{\alpha^{(j,j')}}{\overline K_{co}} \mathcal{N}(x;\hat x^{(j,j')},P^{(j,j')}) \, 1_{Fov_C}(x).
    \end{equation}

\textbf{STEP 3:} Given the GM forms of $p_{co}$, $p_{a,nc}$, $p_{b,nc}$, the parameters of the global fused density are computed as follows.

\textit{\textbf{1) Calculation of $\overline \lambda$}:}
Replacing the evaluated $\lambda_{a,nc}$, $\lambda_{b,nc}$ and $\overline \lambda_{co}$ into (\ref{fused-cardinality}), the parameter $\overline \lambda$ can be obtained.

\textit{\textbf{2) Calculation of $\overline p(x)$}:}
 Replacing (\ref{GM-p-i2}) and (\ref{GM-p-w1}) into (\ref{fused-location-density}), the GM form of $\overline p(x)$ can be obtained.

\section{Simulation results}
This section considers two typical tracking scenarios in order  to assess the effectiveness of the proposed marginal posterior outside the FoV and the proposed BIRD fusion, respectively.

The object state is a vector consisting of planar position and velocity, i.e. $x_k=[p_{x,k},p_{y,k},\dot{p}_{x,k},\dot{p}_{y,k}]^{\top}$, where ``$^\top$'' denotes  transpose.
The single-object transition model $x_{k+1} = F_k x_k + w_k$ is linear Gaussian with matrices
\[
\displaystyle{
F_k=\left[
\begin{array}{cc}
I_2& \Delta I_2 \\ 0_2 & I_2
\end{array}
\right],\,\,\,\,Q_k=\sigma_w^{2}\left[
\begin{array}{cc}
\frac{\Delta^2}{4} I_2& \frac{\Delta^2}{3}\Delta I_2 \\ \frac{\Delta^2}{3} I_2 & \Delta^2 I_2
\end{array}
\right]
}
\]
where: $I_n$ and $0_n$ denote the $n\times n$ identity and zero matrices; $\Delta=1 ~[s]$  is the sampling interval; $Q_k$ is the covariance of the process disturbance $w_k$,
$\sigma_w=5~ [m/s^{2}]$ being the standard deviation
of object acceleration.
The probability of object survival is $P_{S,k}=0.98$.
The single-object observation model $z_{i,k} = H_k x_k + v_{i,k}$ is also linear Gaussian, for each sensor $i$, with matrices
\[
\displaystyle{
H_k=\left[
\begin{array}{cc}
I_2  & 0_2
\end{array}
\right],\,\,\,\,
R_k=\sigma_v^{2} I_2,
}
\]
where $R_k$ is the covariance of the measurement noise $v_{i,k}$, $\sigma_v=10 ~[m]$ being the standard deviation of position measurement on both axes.
The probability of object detection in each sensor is set to $P_D=0.98$.
The number of clutter reports in each scan is Poisson distributed with $\lambda=10$, where each clutter report is sampled uniformly over the local surveillance region.

PHD  filters, implemented in GM form, adopting the observation-based birth process \cite{partial-uniform-birth,diffuse-birth} have been adopted for local filtering in all sensor nodes.
The GM implementation parameters have been chosen as follows: the truncation threshold is  $\gamma_t=10^{-4}$; the pruning  threshold $\gamma_p=10^{-5}$; the merging threshold $\gamma_m=4$; the maximum number
of Gaussian components $N_{max}=150$.
\subsection{Comparison of three forms of posteriors outside the FoV}
In section \ref{section3}, the misbehaviour of standard GCI fusion has been analysed from the perspective of two inconsistent forms (Forms I and II) of posteriors outside the FoV, and their negative impact on the fused density.
Then, in section IV, the multi-object uninformative density has been introduced and exploited as  posterior outside the FoV in order to design the proposed BIRD fusion.
To show the negative impact of Forms I and II on GCI fusion and, on the other hand, the reasonability of BIRD fusion based on the multi-object uninformative density, in this subsection we carry out simulation experiments on a simple tracking scenario with just two agents and a single object. 

In order to get posteriors outside FoV of Form I, the value of $P_{D,k}$  over the global FoV  of each local sensor is set to $0.98$, and then standard GCI fusion is directly performed.
Conversely, in order to get posteriors of Form II, for each local filter $P_{D}$ is set to $0.98$  within the local FoV and to zero outside; $P_{S,k}$ is set to 0.98 over the global FoV,  and then standard
GCI fusion is directly carried out.
As for the posteriors of Form III, local filters have the same detection probability setting as with form I, but apply BIRD instead of standard GCI fusion.
\begin{figure}[tb]
\begin{minipage}[htbp]{0.49\linewidth}
  \centering
  \centerline{\includegraphics[width=4.95cm]{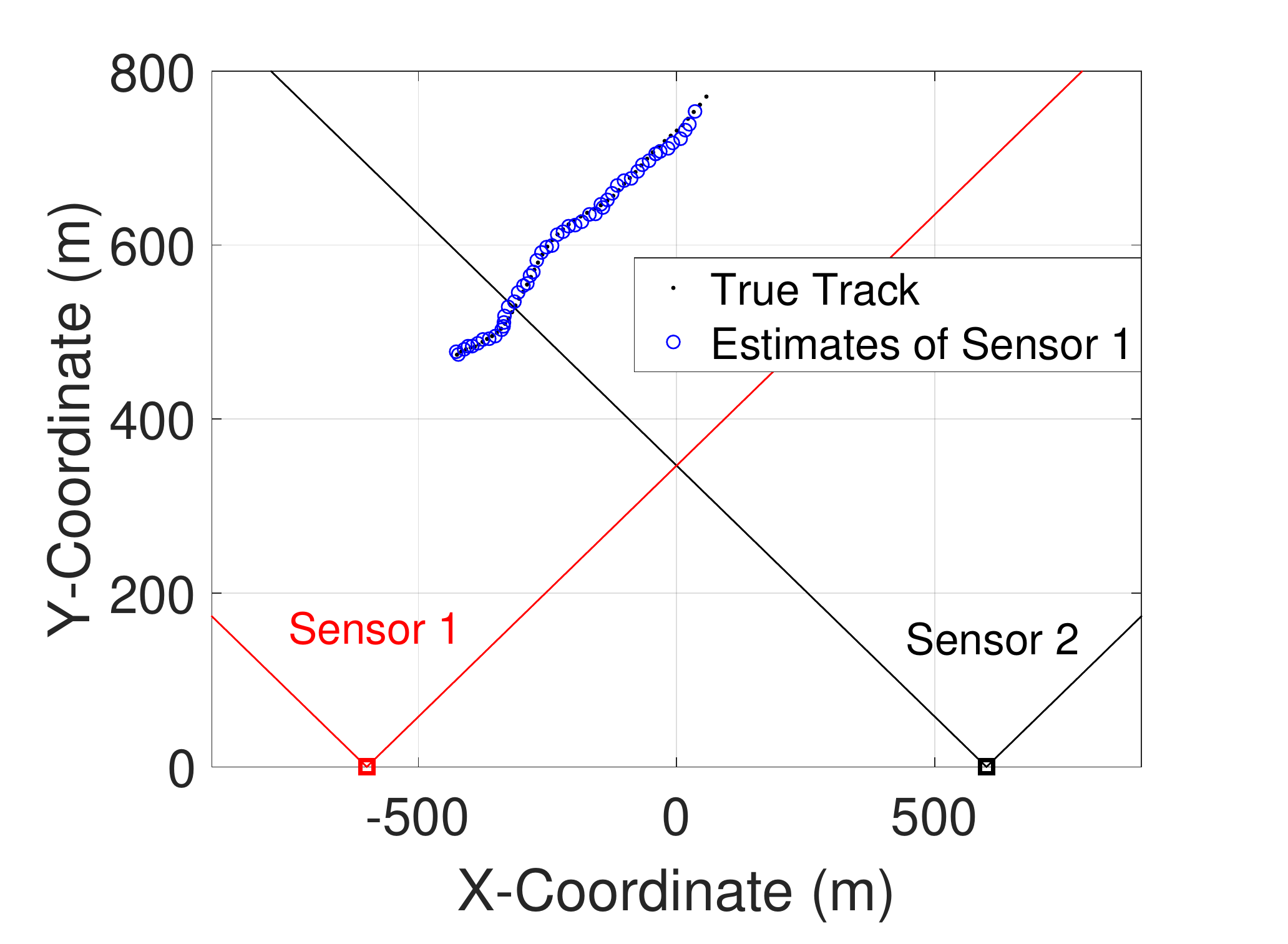}}

  \centerline{\small{\small{(a)}}}\medskip
\end{minipage}
  \hfill
\begin{minipage}[htbp]{0.49\linewidth}
  \centering
  \centerline{\includegraphics[width=4.95cm]{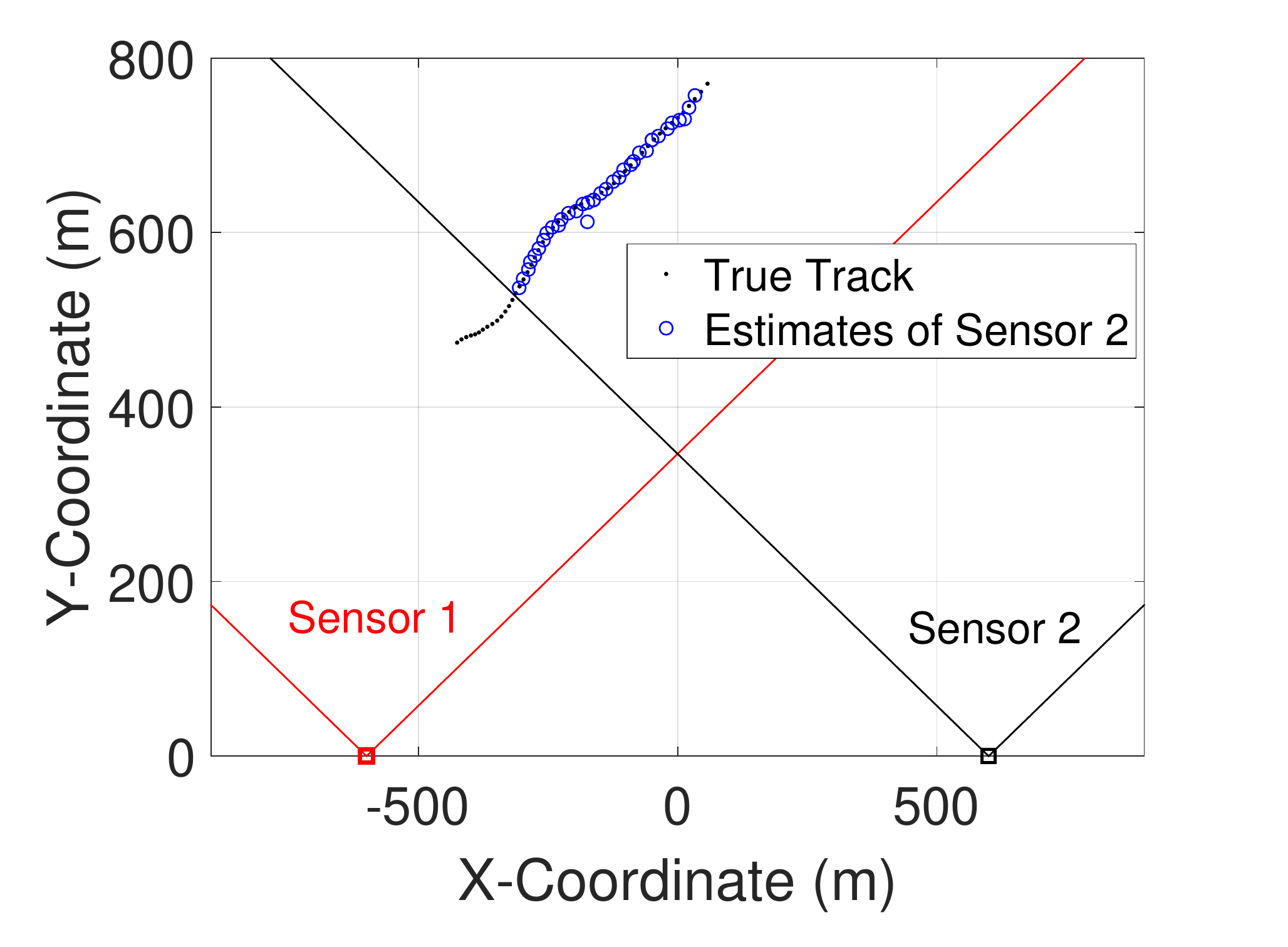}}
  \centerline{\small{\small{(b)}}}\medskip
\end{minipage}
\vfill
\centering{\begin{minipage}[htbp]{0.49\linewidth}
  \centerline{\includegraphics[width=4.95cm]{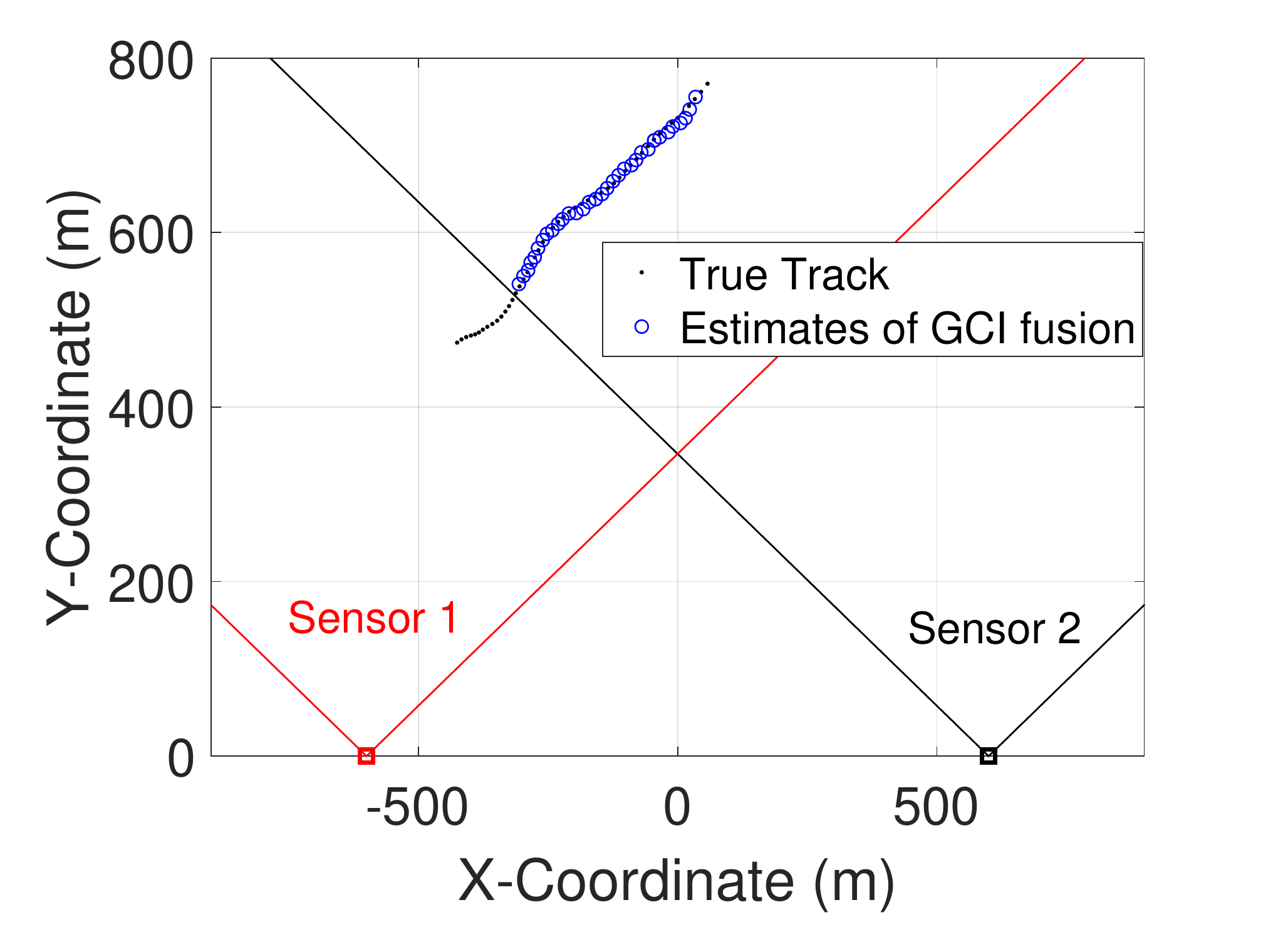}}
  \centerline{\small{\small{(c)}}}\medskip
\end{minipage}}
\caption{Posteriors outside FoV of Form I: (a) estimate of agent 1; (b) estimate of agent 2; (c) estimate after standard GCI fusion.}
\label{Form1}
\end{figure}
\begin{figure}[tb]
\begin{minipage}[htbp]{0.49\linewidth}
  \centering
  \centerline{\includegraphics[width=4.95cm]{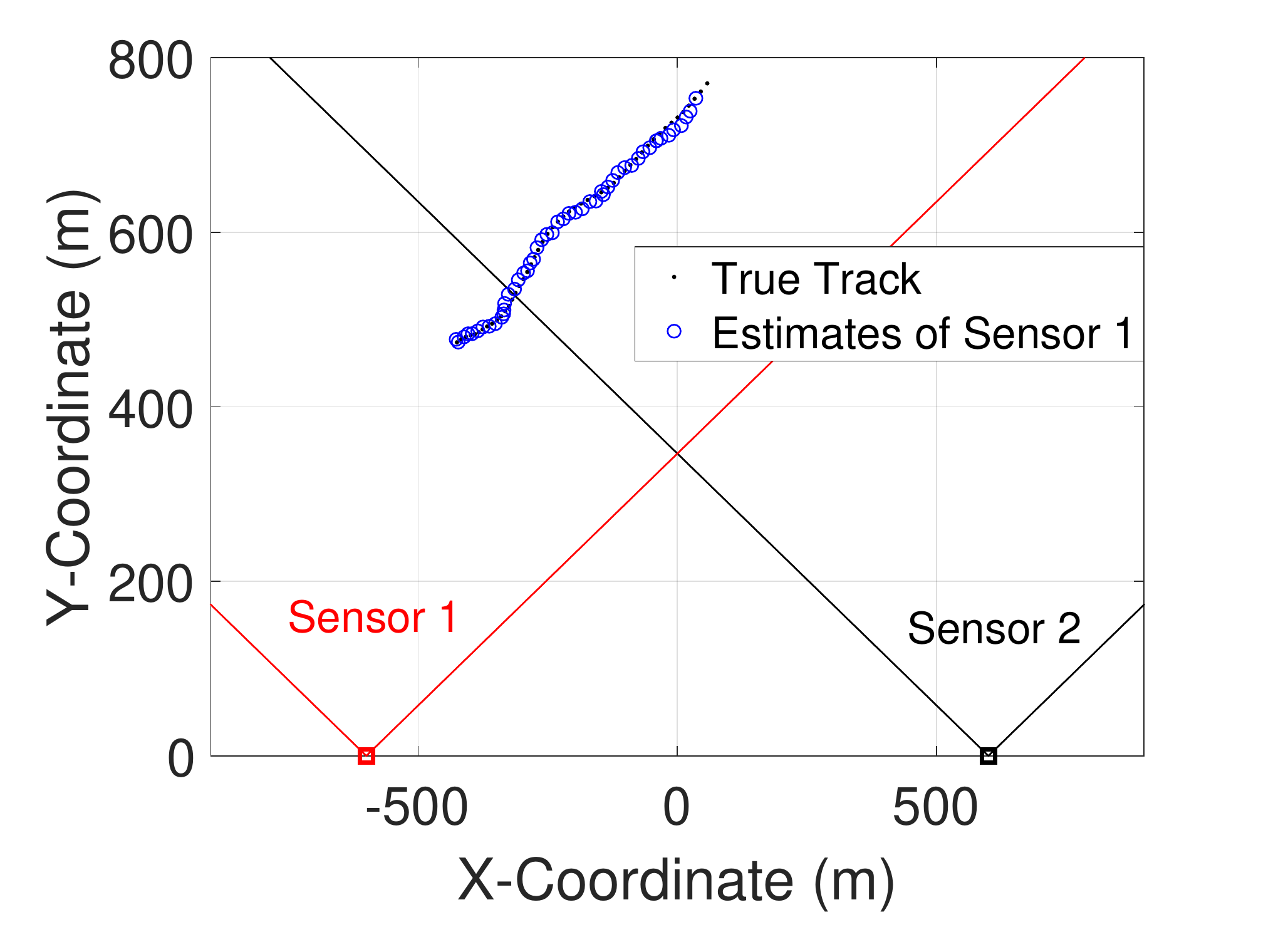}}

  \centerline{\small{\small{(a)}}}\medskip
\end{minipage}
  \hfill
\begin{minipage}[htbp]{0.49\linewidth}
  \centering
  \centerline{\includegraphics[width=4.95cm]{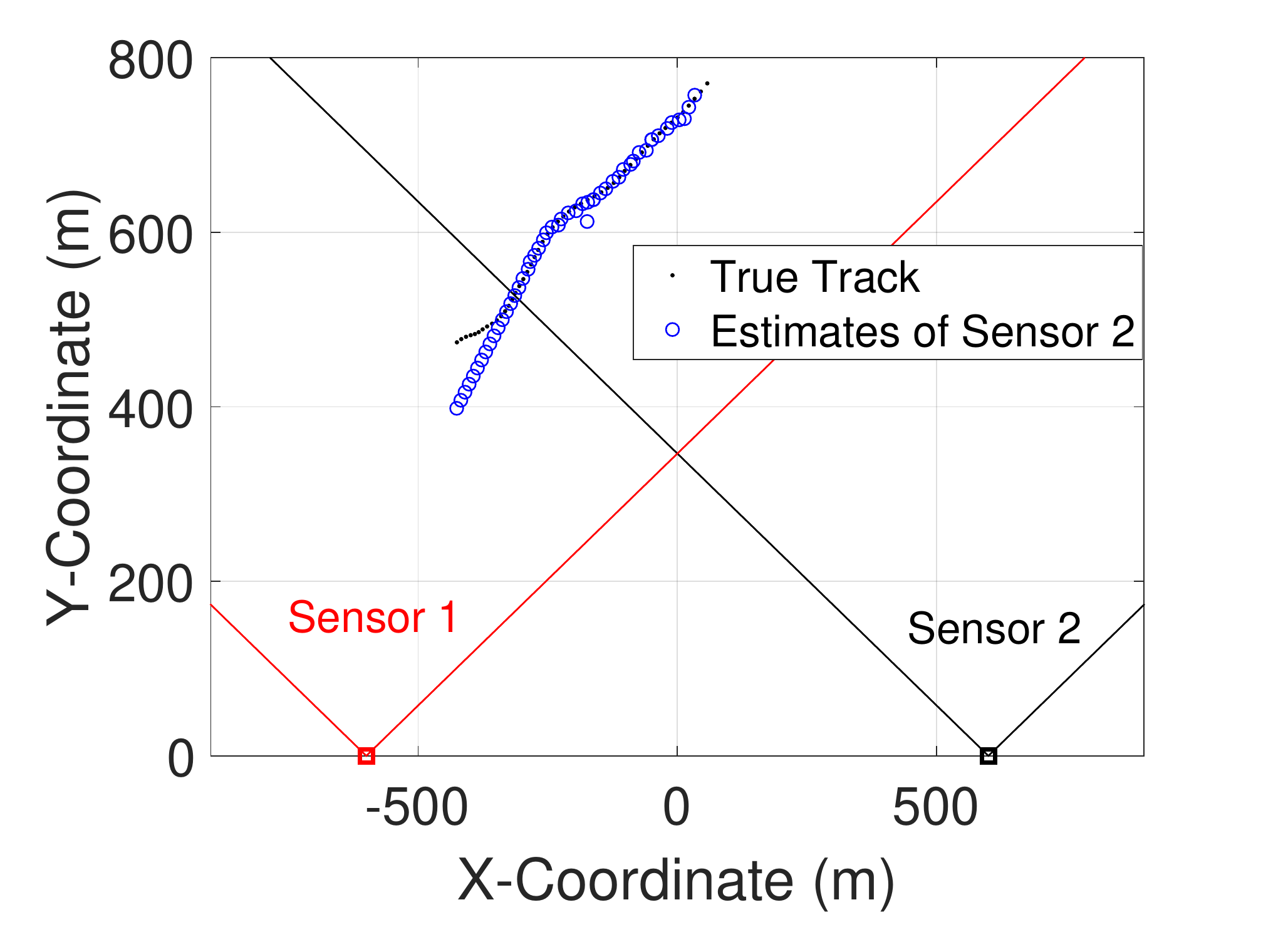}}
  \centerline{\small{\small{(b)}}}\medskip
\end{minipage}
\vfill
\centering{\begin{minipage}[htbp]{0.49\linewidth}
  \centerline{\includegraphics[width=4.95cm]{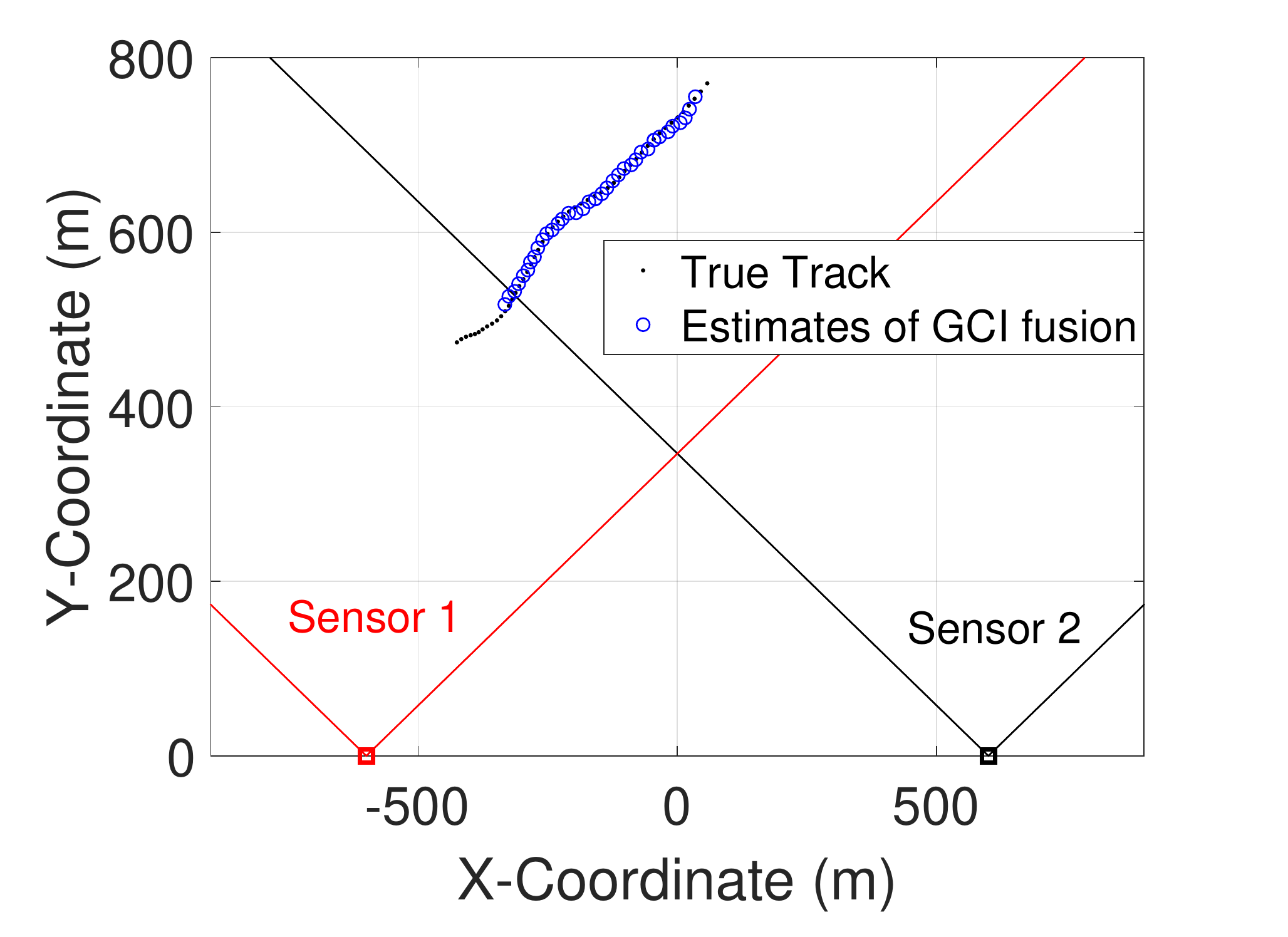}}
  \centerline{\small{\small{(c)}}}\medskip
\end{minipage}}
\caption{Posteriors outside FoV of Form II: (a) estimate of agent 1; (b) estimate of agent 2; (c) estimate after standard GCI fusion.}
\label{Form1}
\end{figure}
\begin{figure}[tb]
\begin{minipage}[htbp]{0.49\linewidth}
  \centering
  \centerline{\includegraphics[width=4.95cm]{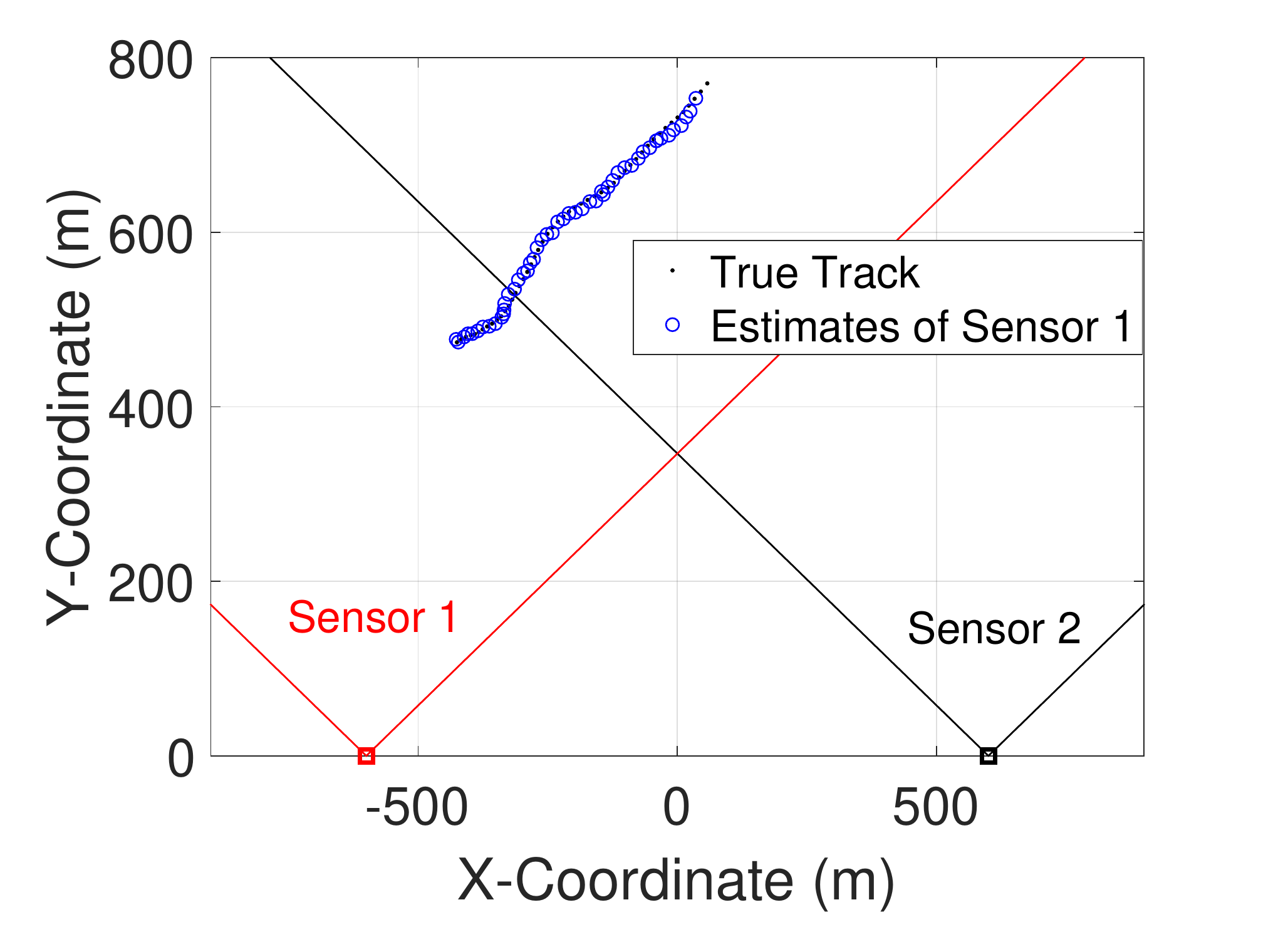}}

  \centerline{\small{\small{(a)}}}\medskip
\end{minipage}
  \hfill
\begin{minipage}[htbp]{0.49\linewidth}
  \centering
  \centerline{\includegraphics[width=4.95cm]{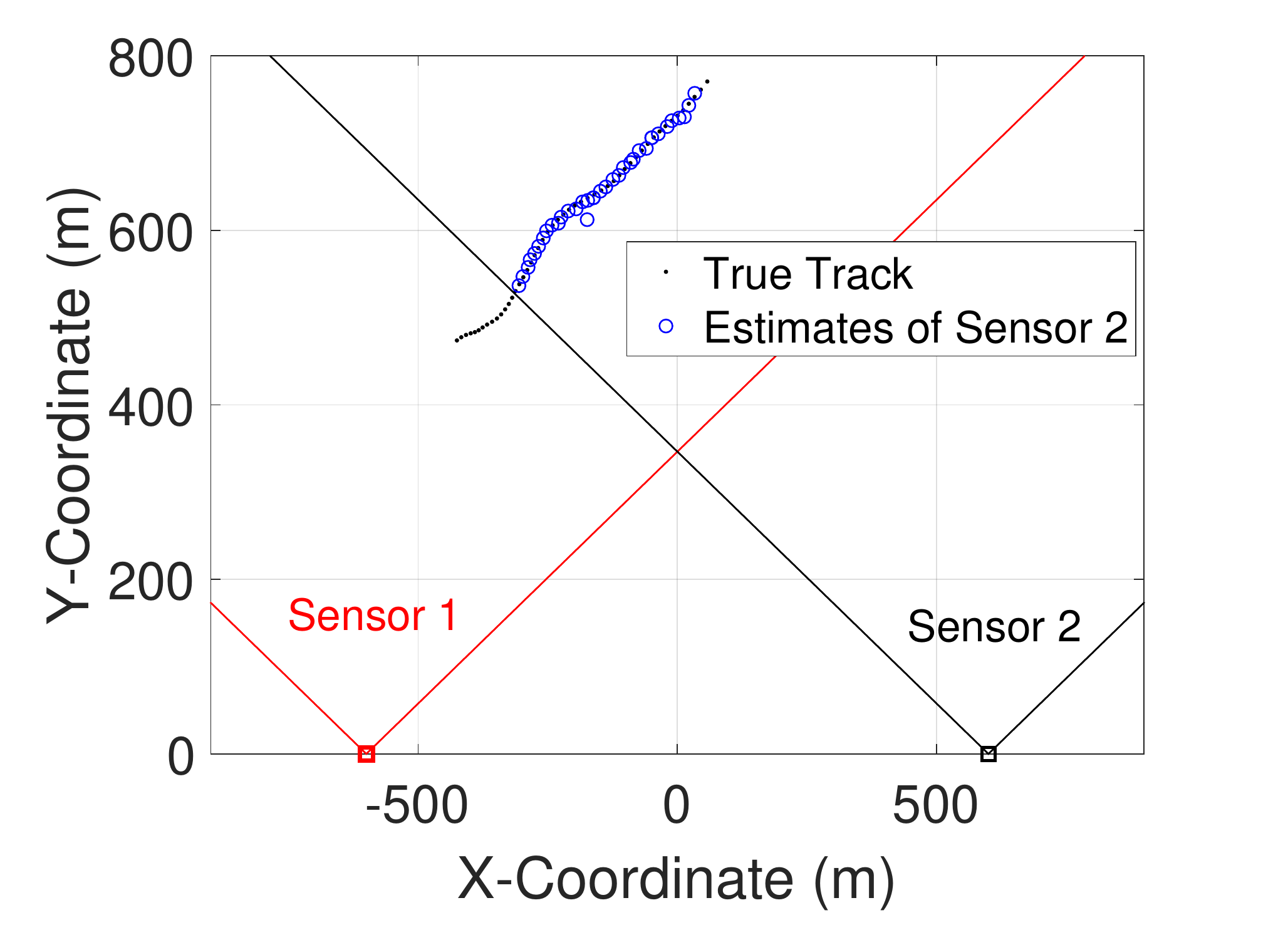}}
  \centerline{\small{\small{(b)}}}\medskip
\end{minipage}
\vfill
\centering{\begin{minipage}[htbp]{0.49\linewidth}
  \centerline{\includegraphics[width=4.95cm]{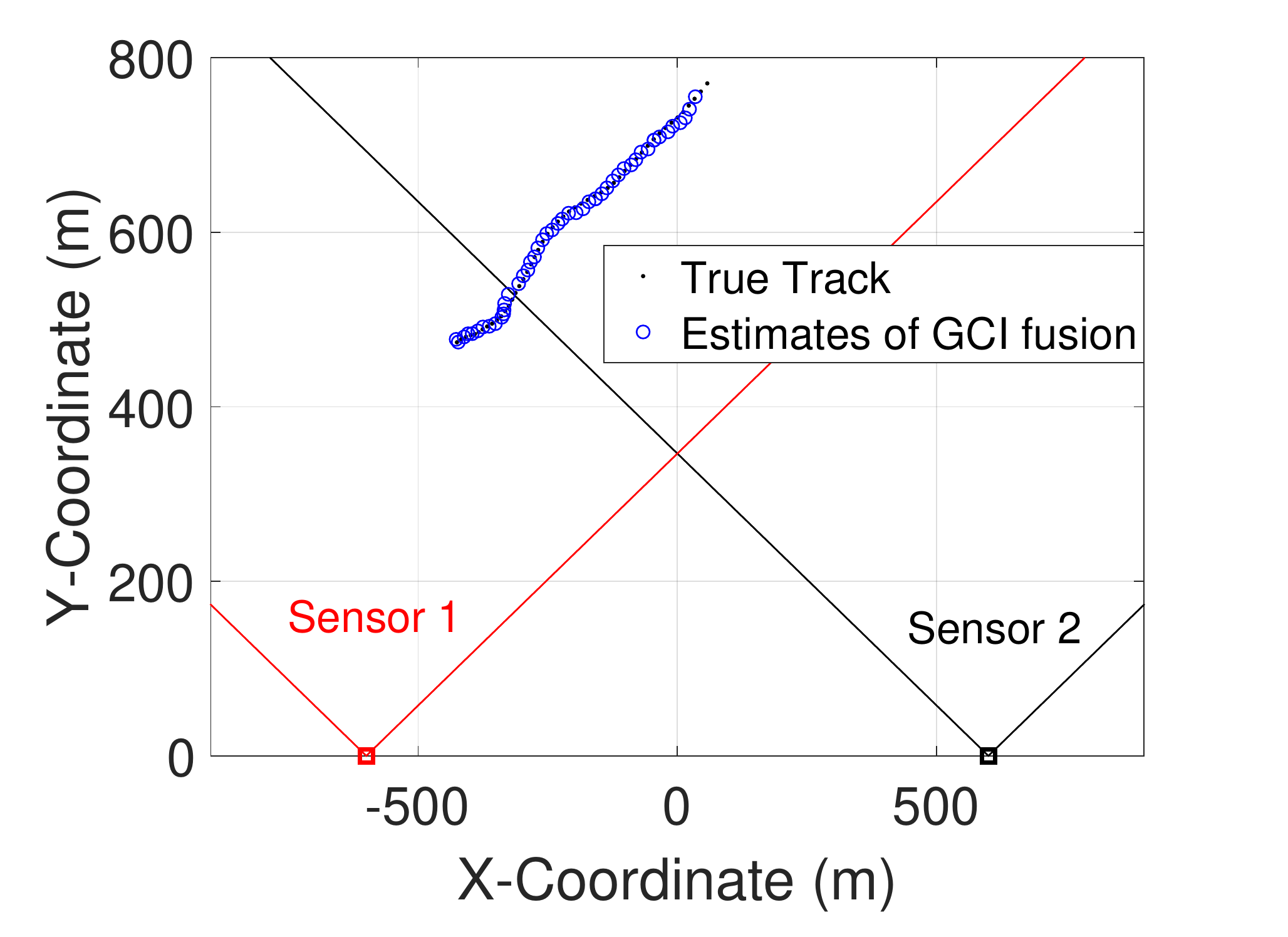}}
  \centerline{\small{\small{(c)}}}\medskip
\end{minipage}}
\caption{Posteriors outside FoV of Form III: (a) estimate of agent 1; (b) estimate of agent 2; (c) estimate after BIRD fusion.}
\label{Form1}
\end{figure}

Figs. 3-5 (a)-(c) respectively show the estimates of the two agents and the fusion results, for Forms I-III of posteriors outside FoVs.
From Fig. 3 (a)-(c),  it can be seen that adopting posteriors of Form I, GCI fusion can only preserve objects within the common FoV.
Then, it can be seen from Fig. 4 (b), for agent 2,  that when the object moves outside the FoV, the posterior actually becomes the prediction of the prior density,
thus making estimation more and more inaccurate. Accordingly, the corresponding fusion result in Fig. 4 (c) becomes worse with time, since the discrepancy between posteriors of the two agents increases with time.
Conversely, as shown in Fig. 5 (a)- (c), by  adopting the uninformative density, BIRD fusion provides satisfactory performance for the whole track over the global FoV.
\subsection{Performance assessment of BIRD fusion}
\begin{figure}[tb]
  \centering
\includegraphics[width=5cm]{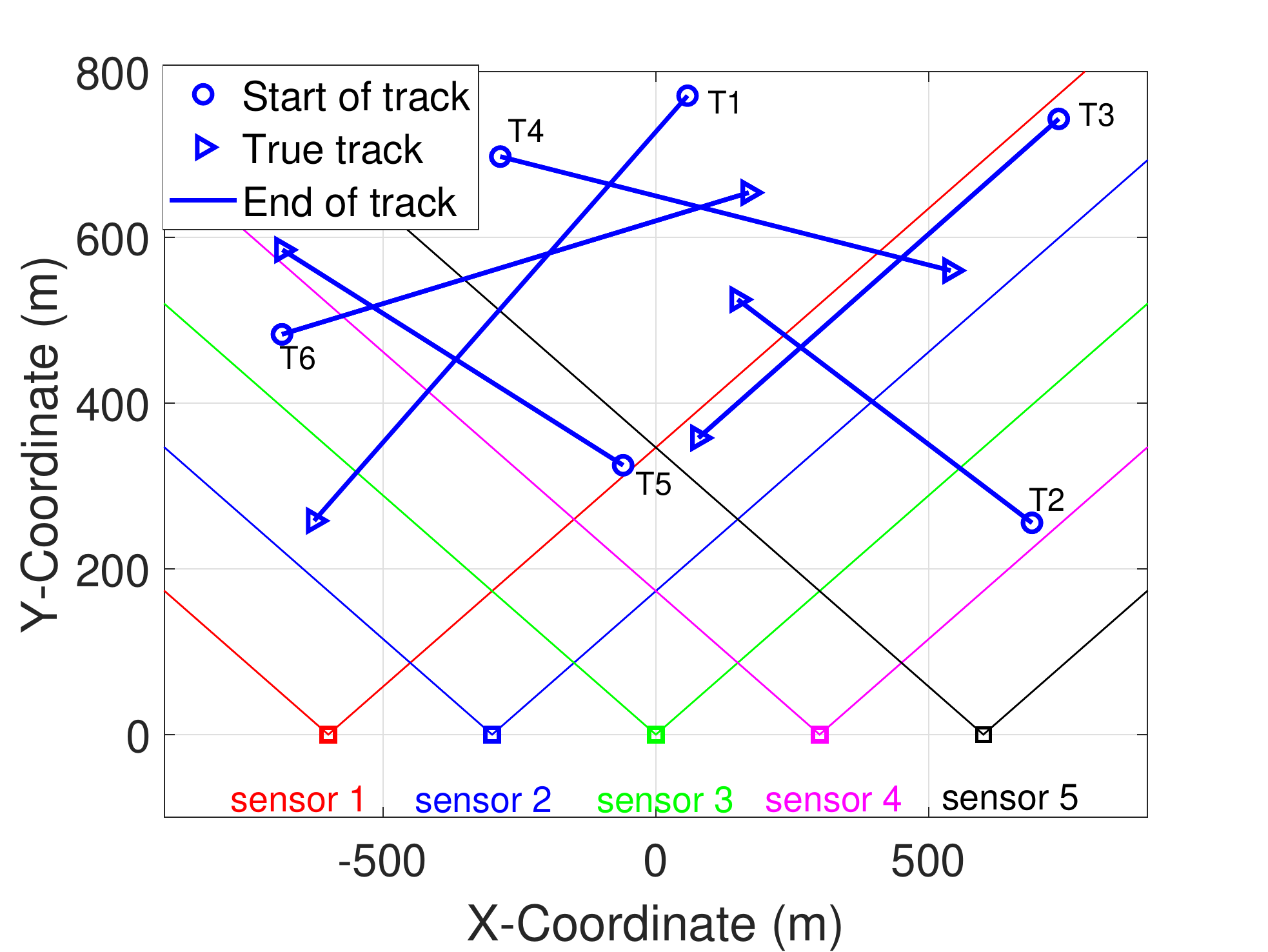}
\caption{{\bf } Simulation scenario: distributed sensor network with five sensors tracking six objects.}
\label{fig_scenario}
\end{figure}

This section aims to assess performance of the proposed BIRD fusion over distributed sensor networks,
also making a comparison with centralized fusion and local (no fusion) tracking.
The main adopted performance metric is the optimal sub-pattern assignment (OSPA) error \cite{MeMber_Vo1}.
A sensor network scenario with five sensors tracking six objects is considered, as depicted in Fig.~\ref{fig_scenario}  wherein
objects appear and disappear at different times, during the $100 \, [s]$ long simulations, as reported in Table~\ref{tab1}.  The number of consensus steps for the distributed fusion is set to $L=3$.

\begin{table}[tb]
	\caption{object birth and death times in the considered scenario.\label{tab1}}
	\centerline
	{
		\normalsize
	\begin{tabular}{cccccc}
		\hline
		object & Birth & Death & object & Birth & Death \\
		\hline
		T1 & 0\,s & 58\,s   & T4  & 15\,s  & 70\,s \\
		T2 & 0\,s & 50\,s    & T5 & 33\,s & 85\,s \\
		T3 & 15\,s & 70\,s    & T6 & 43\,s & 100\,s \\
		\hline
	\end{tabular}
	\normalsize
	}
\end{table}


The five sensor nodes of the considered network can work in the following three modes:
 \begin{enumerate}
 \item [\textbf{M}1:] (\textit{stand-alone mode}) they just perform local filtering without any information exchange;
 \item [\textbf{M}2:] (\textit{distributed mode}) they form a distributed network  exchanging information and performing consensus-based fusion with neighbours;
 \item [\textbf{M}3:] (\textit{centralized mode}) they form a centralized network sending posteriors to a fusion center which, in turn, feeds back the fused density to the sensor nodes.
 \end{enumerate}

\begin{figure}[tb]
\begin{minipage}[htbp]{0.49\linewidth}
  \centering
  \centerline{\includegraphics[width=4.95cm]{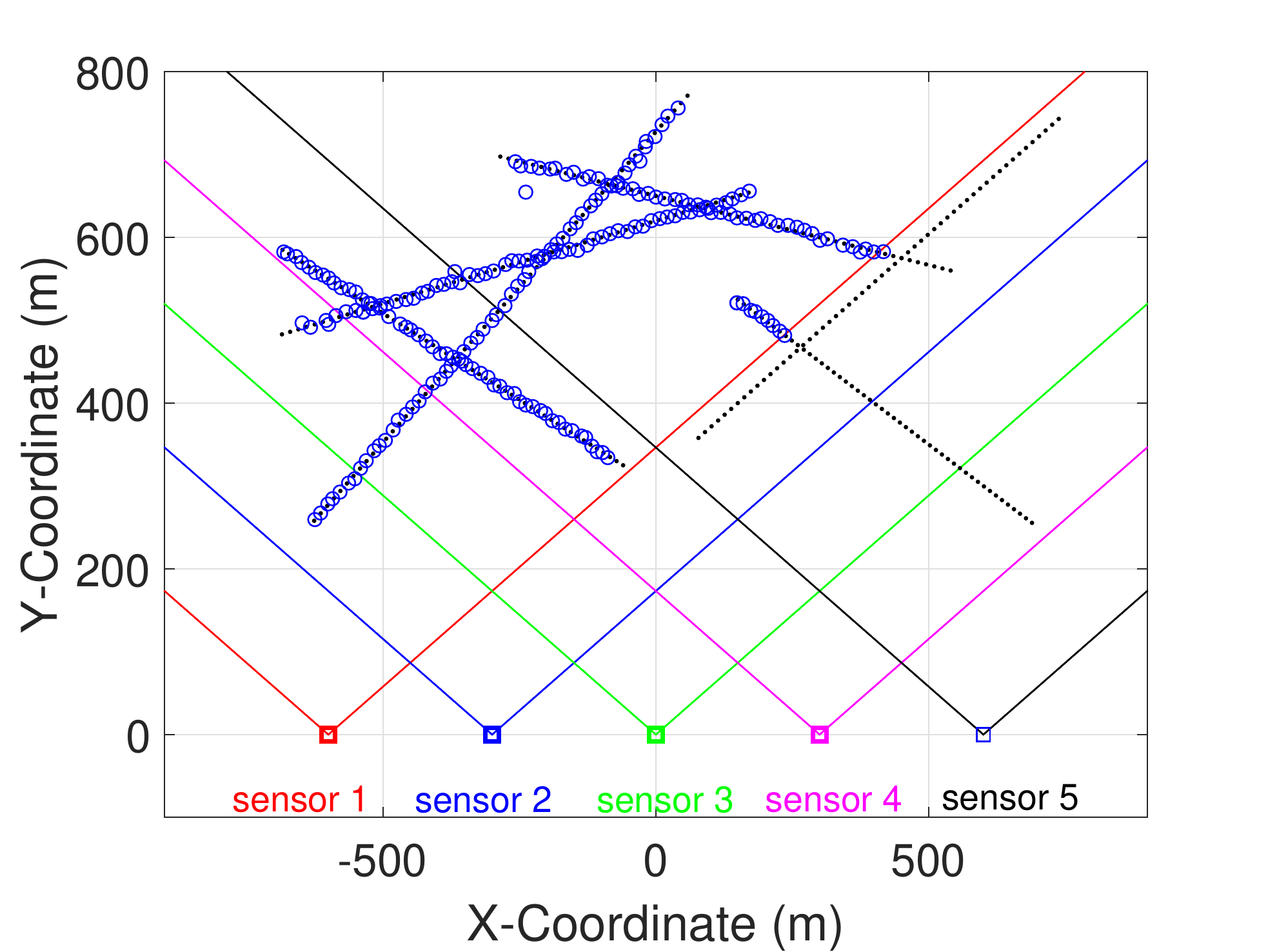}}
  \centerline{\small{\small{(a)}}}\medskip
\end{minipage}
  \hfill
\begin{minipage}[htbp]{0.49\linewidth}
  \centering
  \centerline{\includegraphics[width=4.95cm]{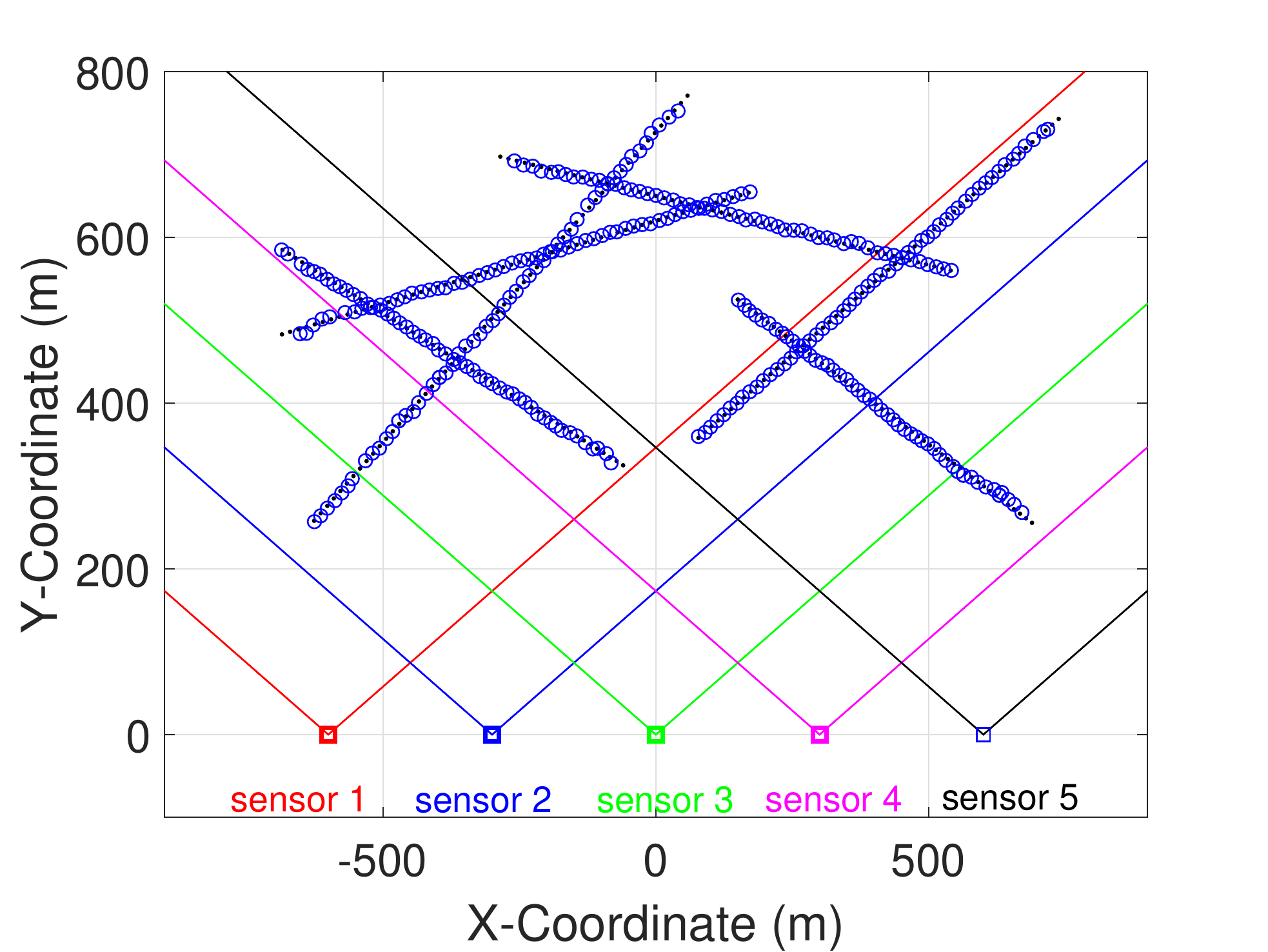}}
  \centerline{\small{\small{(b)}}}\medskip
\end{minipage}
\vfill
  \centering{\begin{minipage}[htbp]{0.49\linewidth}
  \centerline{\includegraphics[width=4.95cm]{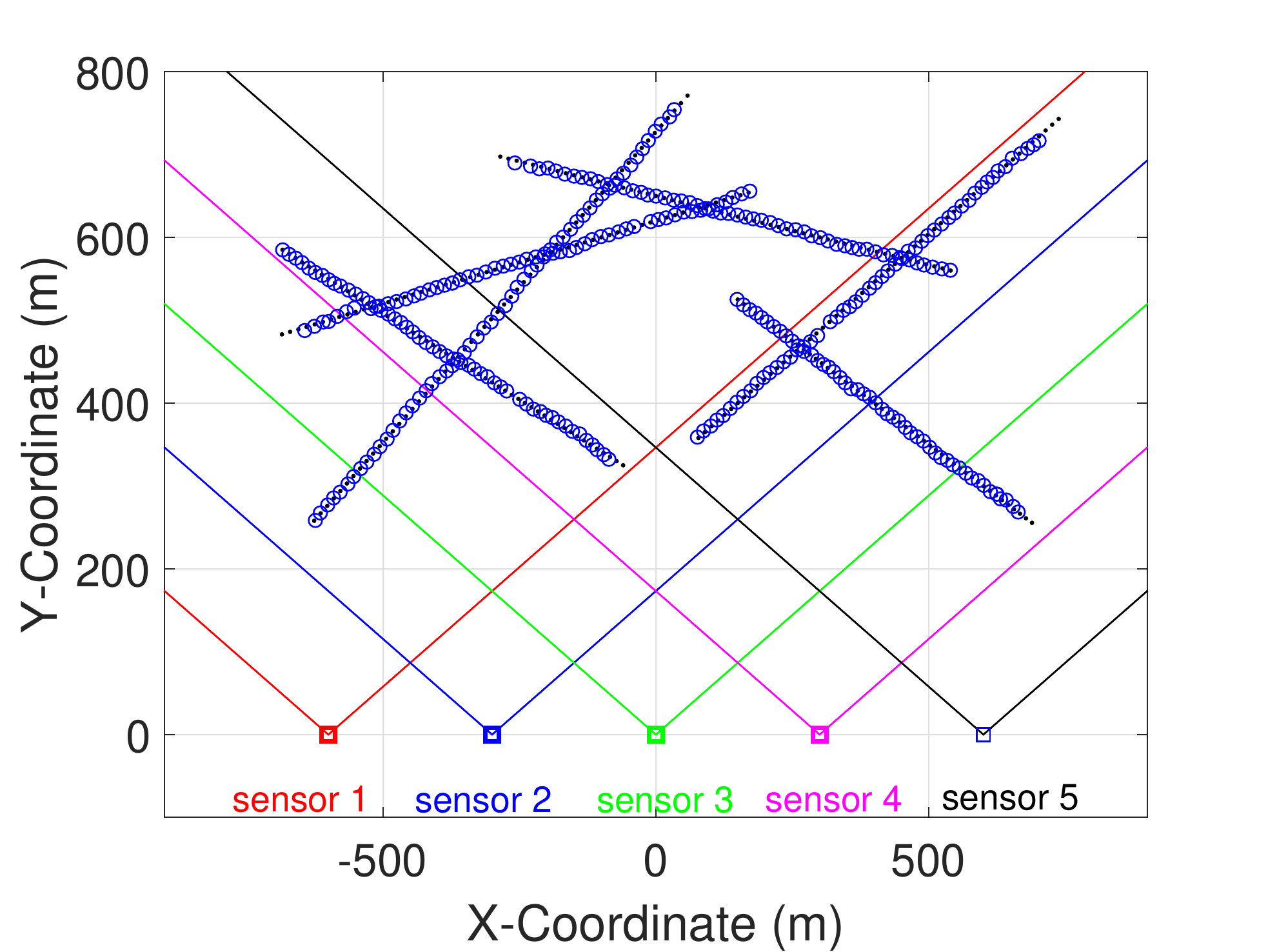}}
  \centerline{\small{\small{(b)}}}\medskip
\end{minipage}}
\caption{Single-run object tracks: (a) sensor 1 in mode M1; (b) sensor 1 in mode M2 (with 3 consensus steps); (c) centralized fusion (M3).}
\label{fig_single_run_fov}
\end{figure}

Fig. \ref{fig_single_run_fov} (a)-(c)  shows the respective outputs of a single sensor (sensor 5) in M1, M2 and M3  modes for a single Monte Carlo run;
similar bahaviours have been found for the other sensor nodes.
Hence, the proposed  fusion performs accurately and consistently in each node for the entire surveillance region,
in the sense that each node maintains locking on all tracks and correctly estimates object positions in both distributed and centralized cases (modes  M2 and M3).
On the other hand, individual sensors working independently (mode M1)  perform considerably worse in that each node is only able to track objects in its own FoV.
In fact, whenever  an object moves outside a node FoV, the node itself loses the object track very quickly as it cannot receive observations of it outside the FoV so that such track turns out to be terminated soon.

\begin{figure}[tb]
\begin{minipage}[htbp]{0.49\linewidth}
  \centering
\centerline{\includegraphics[width=4.9cm]{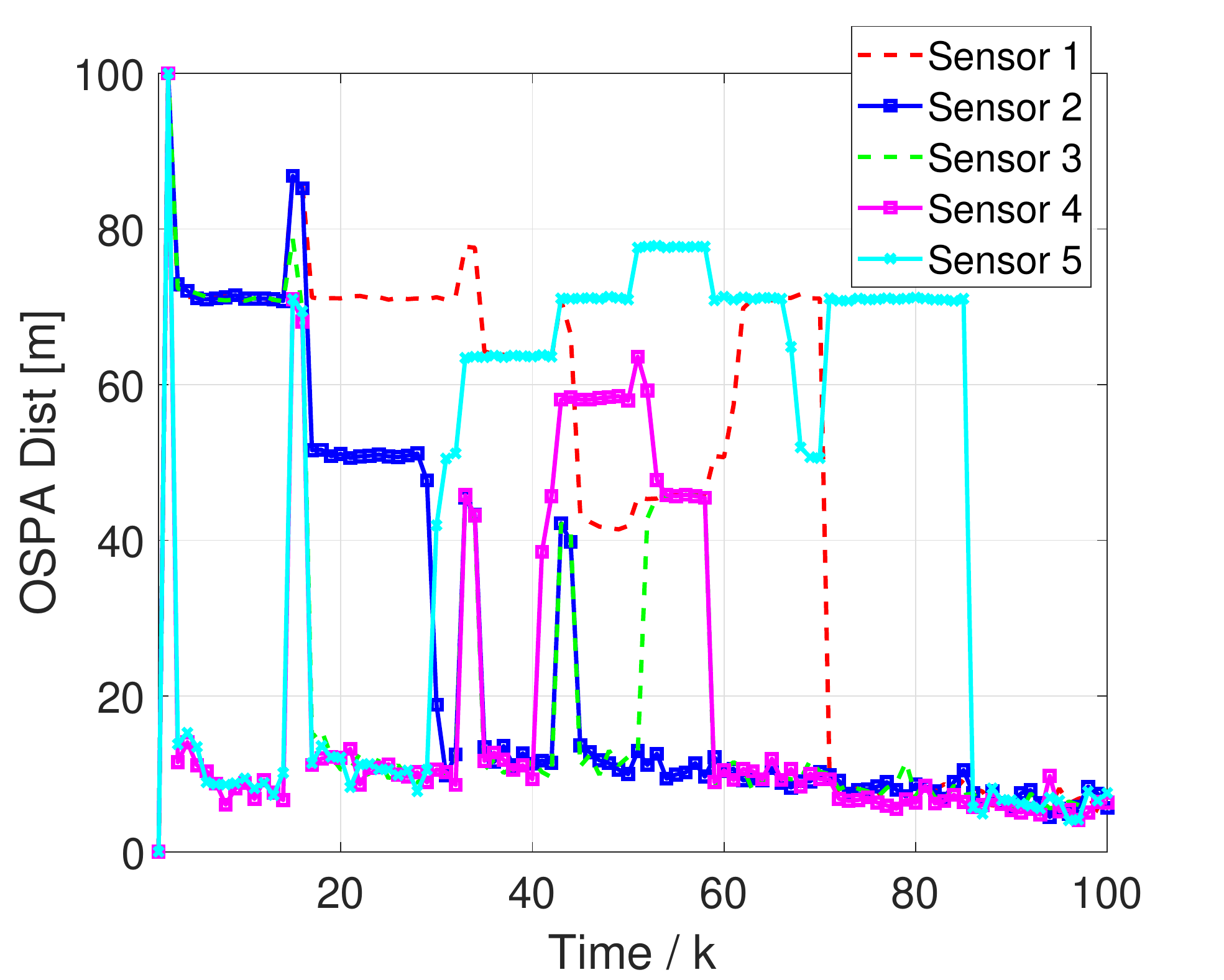}}
  \centerline{\small{\small{(a)}}}\medskip
  \end{minipage}
  \hfill
\begin{minipage}[htbp]{0.49\linewidth}
  \centering
  \centerline{\includegraphics[width=4.95cm]{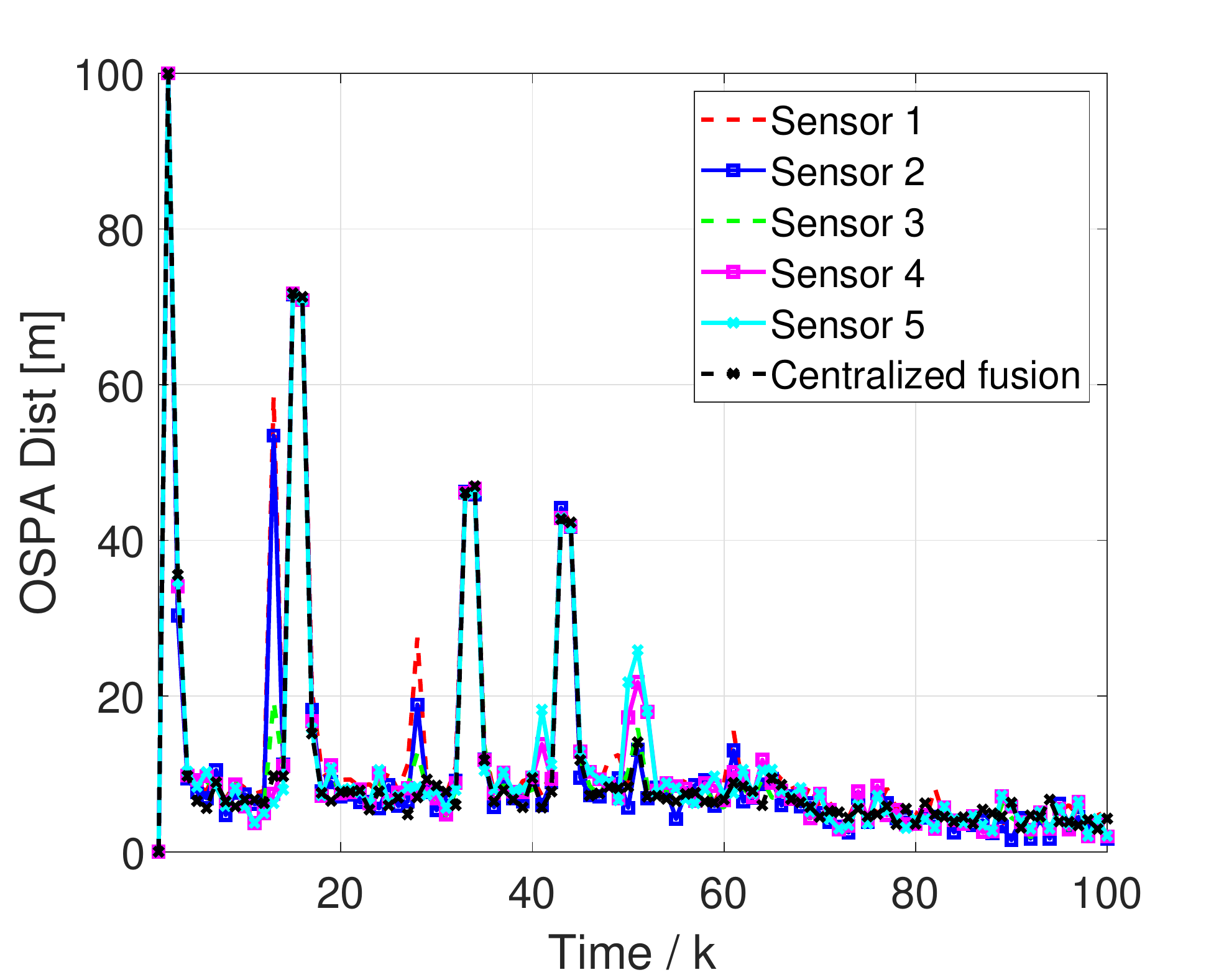}}

  \centerline{\small{\small{(b)}}}\medskip
\end{minipage}
\caption{OSPA errors (averaged over $100$ Monte Carlo trials) at sensors $1-5$: (a) in mode M1; (b) in modes M2 (with 3 consensus steps) and M3 (centralized fusion).}
\label{fig_ospa_fov}
\end{figure}

\begin{figure}[tb]
\begin{minipage}[htbp]{0.465\linewidth}
  \centering
\centerline{\includegraphics[width=4.95cm]{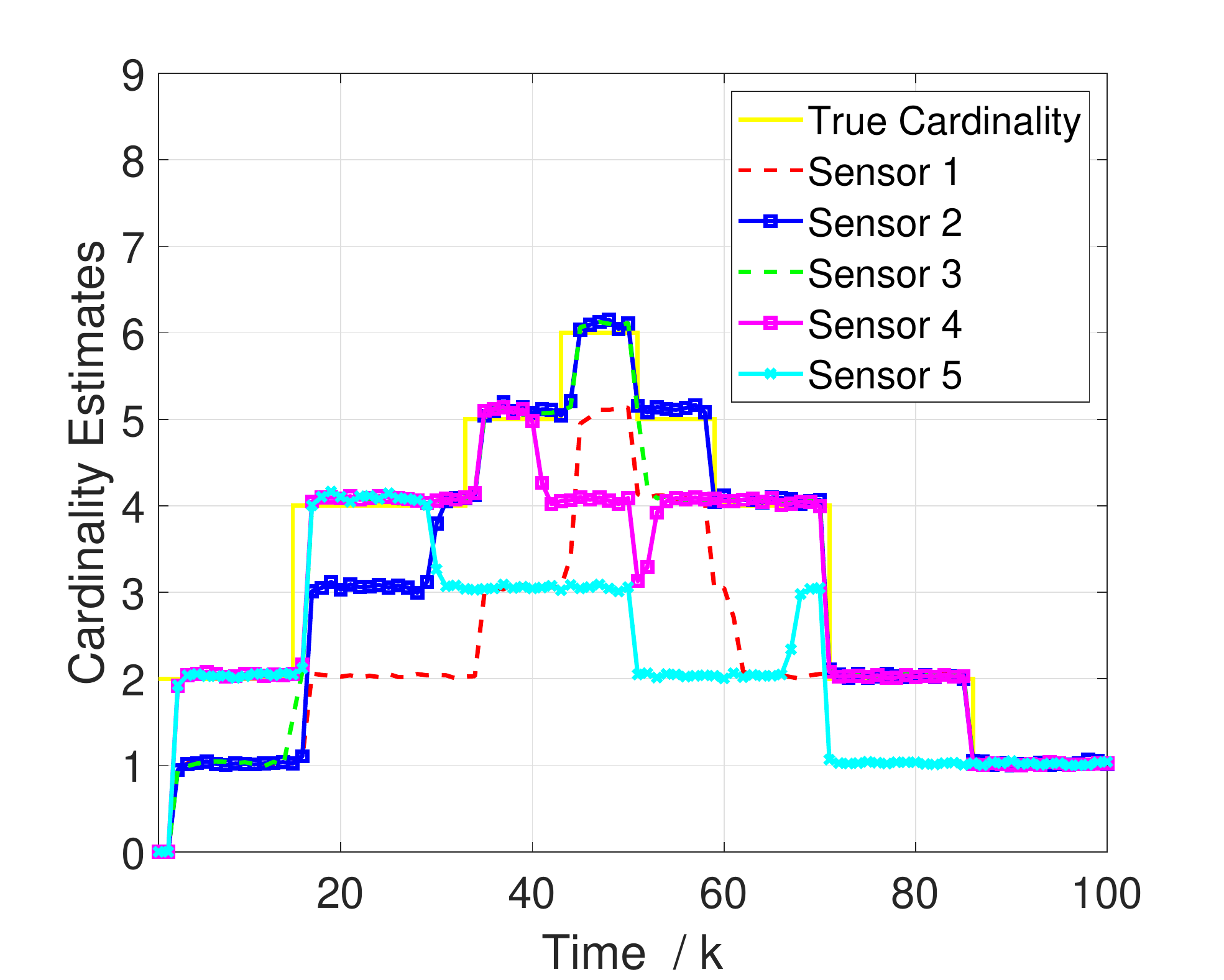}}
  \centerline{\small{\small{(a)}}}\medskip
  \end{minipage}
  \hfill
\begin{minipage}[htbp]{0.465\linewidth}
  \centering
  \centerline{\includegraphics[width=4.95cm]{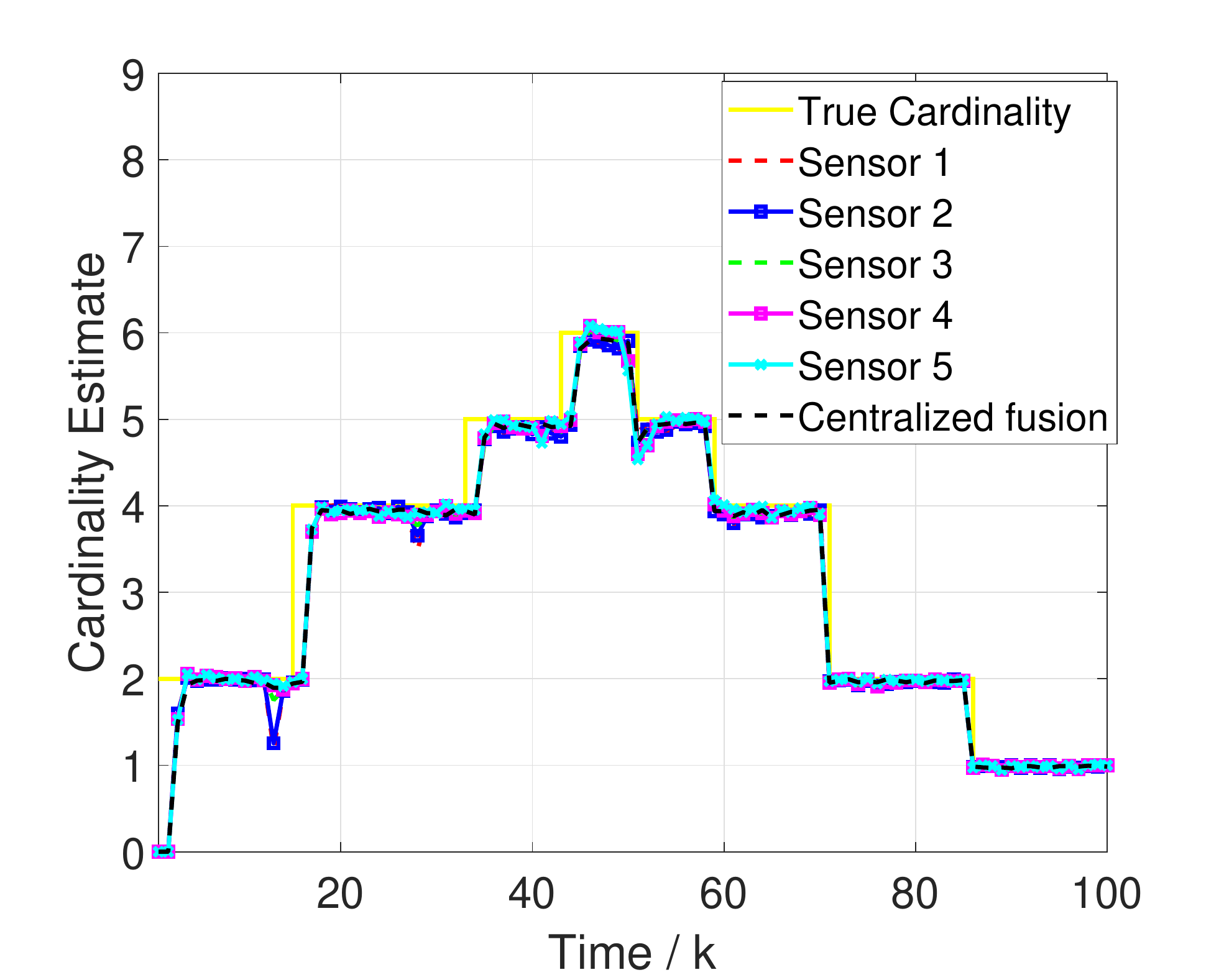}}

  \centerline{\small{\small{(b)}}}\medskip
\end{minipage}
\caption{Cardinality estimates (averaged over $100$ Monte Carlo trials) at sensors $1-5$  (a) in mode M1; (b) in modes M2 (with 3 consensus steps) and M3 (centralized fusion)}
\label{fig_card_fov}
\end{figure}
Figs.~\ref{fig_ospa_fov} and  \ref{fig_card_fov} show the average (over $100$ Monte Carlo runs) OSPA errors and cardinality estimates of  sensors 1-5 in M1, M2 and M3  modes.
It can be observed that  the five sensors working in M2 mode almost achieve the  same  performance in terms of OSPA error, approaching to the performance of the centralized case (M3)
and can cope well with the situation that objects leave the sensor FoV.
 Conversely, there are remarkable performance differences among the five sensors working in M1 mode due to their different FoVs.
 Specifically, whenever an object moves outside a sensor FoV, performance of the latter in
  M1 mode is highly deteriorated (increased OSPA and cardinality errors).
  Notice that the Monte Carlo results of Figs. \ref{fig_ospa_fov} and  \ref{fig_card_fov} are consistent  with the single run results of Fig. \ref{fig_single_run_fov}.
  Further, it can be seen that when the performance of tracking/fusion
  algorithms (in mode M2) reaches a steady-state level, the OSPA error for each sensor node in the cooperative (M2 and M3) modes is significantly lower than in the stand-alone (M1) mode.
  Figs.~\ref{fig_card_fov} (a) and (b) present the cardinality estimates  at sensors 1-5 under modes M1 and M2-M3, respectively.
  It is shown that cardinality estimates given by the sensors working in M2-M3 modes are more accurate than in M1 mode.

\section{Conclusion}
This paper addresses the problem of information fusion for multi-view
multi-agent surveillance systems. To counteract the limitation of standard GCI fusion, we present a novel principled fusion rule, called Bayesian-operator InvaRiance on Difference-sets (BIRD). The proposed BIRD fusion provides a general and principled framework  to combine the redundant information within the common FoV,  and merge the complementary information within the exclusive FoVs, leading to a coverage of the global FoV.
In particular, the proposed BIRD fusion can be performed
on both a centralized and a distributed peer-to-peer sensor network.
Simulation experiments on realistic multi-object tracking scenarios demonstrate effectiveness of the proposed solution.

\bibliographystyle{IEEEtran}


\begin{thebibliography}{99}

\bibitem{Battistelli}
G.~Battistelli, L.~Chisci, C.~Fantacci, A.~Farina, and A.~Graziano, ``Consensus
  {CPHD} filter for distributed multitarget tracking.'' \emph{IEEE J. Sel.
  Topics Signal Process.}, vol.~7, no.~3, pp. 508--520, 2013.

\bibitem{CY-Chong}
C.-Y. Chong, S.~Mori, and K.-C. Chang, ``Distributed multitarget multisensor
  tracking,'' \emph{Multitarget-multisensor tracking: Advanced applications},
  vol.~1, pp. 247--295, 1990.

\bibitem{Mahler-1}
R.~P. Mahler, ``Optimal/robust distributed data fusion: a unified approach,''
  in \emph{Proc. SPIE Defense and Security Symp.}, 2000, pp. 128--138.

\bibitem{Hurley}
M.~Hurley, ``An information-theoretic justification for covariance intersection
  and its generalization,'' in \emph{Proc. IEEE Int. Fusion Conf.}, July 2002,
  pp. 7--11.

\bibitem{EMD-Julier}
S.~J. Julier, T.~Bailey, and J.~K. Uhlmann, ``Using exponential mixture models
  for suboptimal distributed data fusion,'' in \emph{Proc. IEEE Nonlinear
  Statist. Signal Process. Workshop (NSSPW'6), Cambridge, U. K.}, 2006, pp.
  160--163.

\bibitem{Clark}
D.~E. Clark, S.~J. Julier, R.~Mahler, and B.~Ristic, ``Robust multi-object
  sensor fusion with unknown correlations,'' in \emph{Sens. Signal Process.
  Defence (SSPD'10)}, Sep. 2010, pp. 1--5.

\bibitem{double-counting}
G.~Battistelli, L.~Chisci, C.~Fantacci, A.~Farina, and R.~Mahler, ``Distributed
  fusion of multitarget densities and consensus {PHD}/{CPHD} filters,'' in
  \emph{Proc. SPIE Defense, Security and Sensing}, vol. 9474, Baltimore, MD,
  2015.
%

\bibitem{GCI-MB}
B.~L. Wang, W.~Yi, R.~Hoseinnezhad, S.~Q. Li, L.~J. Kong, and X.~B. Yang,
  ``Distributed fusion with multi-{B}ernoulli filter based on generalized
  covariance intersection,'' \emph{IEEE Trans. Signal Process.}, vol.~65,
  no.~1, pp. 242--255, Jan. 2017.

\bibitem{Fantacci-BT}
C.~Fantacci, B.-N. Vo, B.-T. Vo, G.~Battistelli, and L.~Chisci, ``Robust fusion
  for multisensor multiobject tracking,''
  \emph{IEEE Signal Processing Letters}, vol. 25, no. 5, pp. 640-644, 2015.

\bibitem{GCI-GMB}
S.~Q. Li, W.~Yi, R.~Hoseinnezhad, G.~Battistelli, B.~L. Wang, and L.~J. Kong,
  ``Robust distributed fusion with labeled random finite sets,'' \emph{IEEE
  Trans. on Signal Process.}, vol.~66, no.~2, pp. 278--293, Jan. 2018.




\bibitem{PHD-DFOV}
L.~Gao and G.~Battistelli and L.~Chisci, ``Random-finite-set-based distributed multi-robot {SLAM},''
 \emph{IEEE Transactions on Robotics}, vol. 36, No. 6, pp. 1758-1777, 2020.

\bibitem{DFoV-LMB}
S.~Li, G.~Battistelli, L.~Chisci, W.~Yi, B.~Wang, and L.~Kong, ``Multi-sensor
  multi-object tracking with different fields-of-view using the {LMB} filter,''
  in \emph{21st International Conference on Information Fusion (FUSION)}, 2018.

    \bibitem{GCI-PHD-guchong}
  G.~Li, G.~Battistelli, W.~Yi, and L.~Kong,  ``Distributed multi-sensor multi-view fusion based on generalized covariance intersection,'' \emph{Signal Processing}, vol. 166, pp. 107--246, Jan. 2020.


   \bibitem{DFOV-LMB-Reza}
X.~Wang, A.~K.~Gostar, T.~Rathnayake, and B.~Xu, A. B.~Hadiashar and R.~Hoseinnezhad, ``Centralized multiple-view sensor fusion using labeled multi-Bernoulli filters,''
  \emph{Signal Processing}, vol. 150, pp. 75--84, Sep. 2018.



  \bibitem{DFOV-PHD-WeiYi}
W.~Yi, G.~Li, G.~Battistelli,
  ``Distributed multi-sensor fusion of PHD filters with different sensor fields of view,'' \emph{IEEE
  Trans. on Signal Process.}, vol.~68, pp. 5204--5218, Oct.. 2020.



   \bibitem{DFOV-LMB-Gostar}
K.~Gostar, T.~Rathnayake, R.~Tennakoon, A. B.~Hadiashar, G.~Battistelli, L.~Chisci and R.~Hoseinnezhad,
  ``Centralized cooperative censor fusion for dynamic
sensor network with limited field-of-view via
labeled multi-Bernoulli filter,'' \emph{IEEE
  Trans. on Signal Process.}, vol.~69, pp. 878--891, Feb. 2021.



   \bibitem{DFOV-CPHD-TianchengLi}
K.~Da, T.~Li, Y.~Zhu and Q.~Fu,
  ``Gaussian mixture particle Jump-Markov-CPHD
fusion for multitarget tracking using sensors
with limited views,'' \emph{IEEE
  Trans. on Signal Process.}, vol.~6, pp. 605--616, Aug. 2020.




\bibitem{Fusion-CS}
A.~K. Gostar, T.~Rathnayake, R.~Tennakoon, A.~Bab-Hadiashar, G.~Battistelli,
  L.~Chisci, and R.~Hoseinnezhad, ``Cooperative sensor fusion in centralized
  sensor networks using {C}auchy--{S}chwarz divergence,'' \emph{Signal
  Process.}, vol. 167, pp. 107--278, Feb. 2020.

\bibitem{Fusion-LKLA}
L.~Gao, G.~Battistelli, and L.~Chisci, ``Multiobject fusion with minimum
  information loss,'' \emph{IEEE Signal Processing Letters}, vol. 27, pp. 201-205, 2020.

\bibitem{AA-li}
T.~Li and F.~Hlawatsch, ``A distributed particle-PHD filter using
  arithmetic-average fusion of Gaussian mixture parameters,'' \emph{Information
  Fusion}, 2021.

\bibitem{mahler_book}
R.~Mahler, \emph{Statistical multisource-multitarget information fusion}.\hskip
  1em plus 0.5em minus 0.4em\relax Norwood, MA, USA: Artech House, 2007.

\bibitem{Uhlmann}
J.~K. Uhlmann, ``Dynamic map building and localization for autonomous
  vehicles,'' \emph{Unpublished doctoral dissertation, Oxford University},
  vol.~36, 1995.

\bibitem{Heskes}
T.~Heskes, ``Selecting weighting factors in logarithmic opinion pools,'' in
  \emph{Advances in Neural Information Processing Systems}, Cambridge, MA, USA:
  MIT Press, 1998, pp. 266--272.

\bibitem{uninformative-1}
H.~Jeffreys, \emph{Theory of probablity}.\hskip 1em plus 0.5em minus 0.4em\relax
  Oxford University Press, 1961.

\bibitem{uninformative-2}
E.~T. Jaynes, ``Prior probabilities,'' \emph{IEEE Trans. on Systems
  Science and Cybernetics}, vol.~4, no.~3, Sep. 1968.

\bibitem{CU-Fusion}
J.~K. Uhlmann, ``Covariance consistency methods for fault-tolerant distributed
  data fusion,'' \emph{Information Fusion}, vol.~4, pp. 201--215, 2003.

\bibitem{Xiao}
L.~Xiao, S.~Boyd, and S.~Lall, ``A scheme for robust distributed sensor fusion
  based on average consensus,'' in \emph{Proc. 4th Int. Symposium on
  Information Processing in Sensor Networks, \emph{Los Angeles, USA}}, 2005,
  pp. 63--70.

 \bibitem{stone-streit}
 L.~D.~Stone, R.~L. Streit, T.~L. Corwin, and K.L. Bell, ``Bayesian multiple target tracking'', \hskip
  1em plus 0.5em minus 0.4em\relax Norwood, MA, USA: Artech House, 2014.

\bibitem{PHD-Vo}
B.-N. Vo and W.-K. Ma, ``The {G}aussian mixture probability hypothesis density
  filter,'' \emph{IEEE Trans. Signal Process.}, vol.~54, no.~11, pp.
  4091--4104, 2006.

\bibitem{fractional-power}
S.~Julier and J.~Uhlmann, ``Unscented filtering and nonlinear estimation,''
  \emph{Proceedings of the IEEE}, vol.~92, no.~3, pp. 401--422, 2004.

\bibitem{MeMber_Vo1}
D.~Schuhmacher, B.-T. Vo, and B.-N. Vo, ``A consistent metric for performance
  evaluation of multi-object filters,'' \emph{IEEE Trans. Signal Process.},
  vol.~56, no.~8, pp. 3447--3457, 2008.

\bibitem{partial-uniform-birth}
M.~Beard, B.-T. Vo, B.-V. Vo, and S.~Arulampalam, ``A partially uniform target
  birth model for Gaussian mixture PHD/CPHD filtering,'' \emph{IEEE Trans.
  Aerosp. Electron. Syst.}, vol.~49, no.~4, pp. 2835--2844, 2013.

\bibitem{diffuse-birth}
J.~Houssineau and D.~Laneuville, ``PHD filter with diffuse spatial prior on the
  birth process with applications to GM-PHD filter,'' in \emph{Proc. IEEE Int.
  Fusion Conf.}, Jul. 2010.
\end{thebibliography}
\end{document}


\onecolumn{\title{Mathematical proofs of propositions}
\author{ Bailu Wang, Suqi Li*, Giorgio Battistelli, Luigi Chisci, Wei Yi
\thanks{

B. Wang and S. Li  are with  Chongqing University, Chongqing, 400044, China (Email: w\_b\_l3020@163.com; qi\_qi\_zhu1210@163.com;).
G. Battistelli and L. Chisci are  with Dipartimento di Ingegneria dell'Informazione, Universit$\grave{\mbox{a}}$ degli Studi di Firenze, Via Santa Marta 3, 50139, Firenze, Italy (Email: giorgio.battistelli@unifi.it; luigi.chisci@unifi.it).

W. Yi is with University of Electronic Science and Technology of China, Chengdu, Sichuan, 611731, China (Email: kussoyi@gmail.com)

Corresponding author: Suqi Li.

}}
\maketitle

 This document provides proofs of Propositions 1-5, and 7 of  the paper ``Principled information fusion  for multi-view multi-agent surveillance system''.  The cited references and equations are  in the main-body. The equations in this document are enumerated by following the equations in the main-body.

\section{Proof of Proposition 1}

\begin{IEEEproof}
Based on the definition [18], the belief-mass function of $X_W$ is given by
\begin{equation}\label{eq:first-extension}
\begin{split}
\beta_W(S)=&\Pr\left(X\cap \mathbb{W}\subseteq S,X\backslash(X\cap \mathbb{W})\subseteq \mathbb{X}\right)\\
=&\Pr\left(X\cap \mathbb{W}\subseteq S,X\cap \mathbb{V}\subseteq \mathbb{X}\right)\\
=&\sum_{m=0}^{+\infty}\Pr(X\cap \mathbb{W}\subseteq S,X\cap \mathbb{V}\subseteq \mathbb{X}\big||X|=m)\Pr(|X|=m)
\end{split}
\end{equation}
where $S$ can be any non-empty subset of $\mathbb{X}$;  $|\cdot|$ denotes the cardinality of a set; the third equality is obtained by applying \textit{the law of total probability}.
\begin{figure}[htp]
  \centering
\includegraphics[width=16cm]{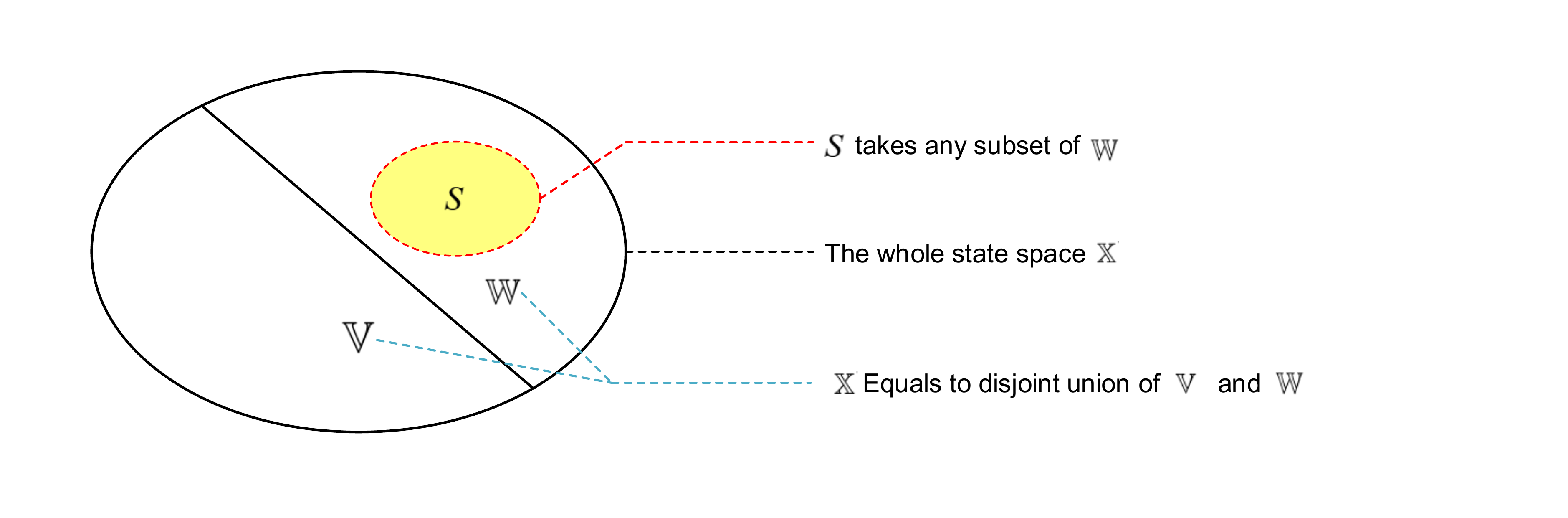}
\caption{The relationship between  sets involved in (\ref{eq:first-extension}).}
\label{fig:Division of the global FoV}
\end{figure}

Further, again exploiting  the \textit{law of total probability},  $\Pr(X\cap \mathbb{W}\in S,X\cap \mathbb{V}\in \mathbb{X} | |X|=m)$ in (\ref{eq:first-extension}) can be computed by
\begin{equation}\label{eq:second-extension}
\begin{split}
&\Pr\left(X\cap \mathbb{W}\subseteq S,X\cap \mathbb{V}\subseteq \mathbb{X}\,\big|\,\,|X|=m\right)\\
=&\sum_{n=0}^{m}\sum_{\{x_1,\cdots,x_n\}\uplus\{x'_{1},\cdots,x'_{n'}\}=X}\!\!\!\Pr\left(X\cap \mathbb{W}\subseteq S,X\cap \mathbb{V}\subseteq \mathbb{X},x_1\in\mathbb{W},\cdots,x_n\in\mathbb{W},x'_{1}\in\mathbb{V},\cdots,x'_{n'}\in\mathbb{V}\big||X|=m\right)\\
=&\!\!\sum_{n=0}^{m}\sum_{\{x_1,\cdots,x_n\}\uplus\{x'_{1},\cdots,x'_{n'}\}=X}\!\!\Pr\left(x_1\in S,\!\cdots\!,x_n\in S,\cdots,x_1\in\mathbb{W},\!\cdots\!,x_n\in\mathbb{W},x'_{1}\in\mathbb{V},\cdots,x'_{n'}\in\mathbb{V}\big|
|X|=m\right).\\
\end{split}
\end{equation}

Without loss of generality, the density $\pi(X)$ can be represented as
\begin{equation}
\pi(\{x_{1},\cdots,x_{n}\})\triangleq  n! \, \rho(n) \,p(\{x_{1},\cdots,x_{n}\}), \text{if\,\,} n>0;
\end{equation}
otherwise, if $n=0$, $\pi(\{x_{1},\cdots,x_{n}\})\triangleq \rho(n)$, where $\rho(n)$ denotes the cardinality distribution, and $p(\{x_{1},\cdots,x_{n}\})$ is a symmetric density defined on $\mathbb{X}^{n}$.

Then, we have
\begin{equation}\label{eq:third-extension}
\begin{split}
&\Pr\left(X\cap \mathbb{W}\subseteq S,X\cap \mathbb{V}\subseteq \mathbb{X}\big||X|=m\right)=\\
&\sum_{n=0}^{m}\sum_{\{x_1,\cdots,x_n\}\uplus\{x'_{1},\cdots,x'_{n'}\}\!=\!X}\int \prod_{i=1}^n 1_S(x_i)\!\prod_{i=1}^n 1_\mathbb{W}(x_i)\!\prod_{j=1}^{n'}  1_\mathbb{V}(x_{j})p(\{x_1,\cdots,x_n\}\!\cup\!\{x'_{1},\!\cdots\!,x'_{n'}\}) d(x_1,\!\cdots\!,x_n,x'_{1},\cdots,x'_{n'}) . \\
\end{split}
\end{equation}
As a result, the belief-mass function $\beta_W(S)$ can be further written by
{\small{
\begin{equation}\label{mass-belief-marginal-1}
\begin{split}
&\beta_{W}(S)\\
=&\sum_{m=0}^{+\infty}\rho(m)\!\sum_{n=0}^{m}\!\sum_{\{x_1,\cdots,x_n\}\uplus\{x'_{1},\cdots,x'_{n'}\}\!=\!X}\int \!\prod_{i=1}^n 1_S(x_i)\!\prod_{i=1}^n 1_\mathbb{W}(x_i)\!\prod_{j=1}^{n'}  1_\mathbb{V}(x_{j})p(\{x_1,\cdots,x_n\}\!\cup\!\{x'_{1},\!\cdots\!,x'_{n'}\}) d(x_1,\!\cdots\!,x_n,x'_{1},\cdots,x'_{n'})\\
=&\sum_{m=0}^{+\infty}\frac{1}{m!}\!\sum_{n=0}^{m}\!\sum_{\{x_1,\cdots,x_n\}\uplus\{x'_{1},\cdots,x'_{n'}\}\!=\!X}\int \!\prod_{i=1}^n 1_S(x_i)\!\prod_{i=1}^n 1_\mathbb{W}(x_i)\!\prod_{j=1}^{n'}  1_\mathbb{V}(x_{j})m!\rho(m)p(\{x_1,\cdots,x_n\}\!\cup\!\{x'_{1},\!\cdots\!,x'_{n'}\}) d(x_1,\!\cdots\!,x_n,x'_{1},\cdots,x'_{n'}).\\
\end{split}
\end{equation}}}
Let us define
\begin{align}
X_1\triangleq&\{x_1,\cdots,x_n\}\\
X_2\triangleq&\{x'_1,\cdots,x'_{n'}\},
\end{align}
then (\ref{mass-belief-marginal-1}) yields:
\begin{equation}\label{mass-belief-marginal-2}
\beta_{W}(S)=\int \int \sum_{X_1\uplus X_2=X}1_{S}(X_1)1_{\mathbb{W}}(X_1)1_{\mathbb{V}}(X_2)\pi(X_1\uplus X_2)\delta X_1\delta X_2
\end{equation}
where $\int\cdot\delta X$ denotes the set integral.

Observing (\ref{mass-belief-marginal-2}), only the summation terms with $X_1\subseteq \mathbb{W}$ and $X_2\subseteq \mathbb{V}$ can survive; other summation terms are equal to zero. Hence, (\ref{mass-belief-marginal-2}) can be further simplified as
\begin{equation}
\begin{split}
\beta_W(S)
=& \int_S 1_{\mathbb{W}}(X_1)\left(\int_\mathbb{V} \pi(X_1\cup X_2)\delta X_2\right)\delta X_1 .
\end{split}
\end{equation}

According to \textit{Radon-Nikodym theorem}, taking the set derivative of $\beta_{W}(S)$ with $S=\emptyset$, we can get the  density of $X_{W}$, i.e.,
\begin{equation}
\pi_{W}(X)=[1_{\mathbb{W}}]^{X}\int_{\mathbb{V}}\pi(X\cup X') \, \delta X'  . \end{equation}
Since  we have
\begin{equation}
\begin{split}
\int_{\mathbb{X}}\pi_{W}(X)\delta X=&\int_{\mathbb{X}}[1_{\mathbb{W}}]^{X}\int_{\mathbb{V}}\pi(X\cup X')\delta X'\\
=&\int_{\mathbb{W}}\int_{\mathbb{V}}\pi(X\cup X')\delta X'\delta X\\
=&\Pr(X\in\mathbb{W},X'\in\mathbb{V})\\
=&\Pr(X\cup X'\in\mathbb{X})\\
=&\int \pi(X)\delta X=1,
\end{split}\end{equation}
i.e. $\pi_{W}(X)$ is a density.
\end{IEEEproof}

\section{Proof of Proposition 2}
\begin{IEEEproof}
When $\pi_{W}(X\cap\mathbb{W})\neq 0$,  according to (11) we have
\begin{equation}
\begin{split}
&\pi_{V}(X\cap \mathbb{V}|X\cap\mathbb{W})\pi_{W}(X\cap\mathbb{W})\\
=&\,\,\frac{\pi((X\cap\mathbb{W})\cup(X\cap\mathbb{V}))}{\pi_{W}(X\cap\mathbb{W})}\pi_{W}(X\cap\mathbb{W})\\
=&\,\, \pi(X).
\end{split}
\end{equation}
As for the case $\pi_{W}(X\cap\mathbb{W})=0$, we have $\Pr(X\cap\mathbb{W}\subseteq S)=0, \forall S\subseteq\mathbb{X}$, from which
\begin{equation}
\begin{split}
\pi(X)=&\frac{\delta\Pr(X\subseteq S)}{\delta X}\bigg|_{S=\emptyset}\\
=&\frac{\delta\Pr(X\cap\mathbb{W}\subseteq S,X\cap\mathbb{V}\subseteq S )}{\delta X}\bigg|_{S=\emptyset}=0 .
\end{split}
\end{equation}
Then, according to the definition in (13), it trivially holds that
\begin{equation}
\pi(X)=\pi_{V}(X\cap \mathbb{V}|X\cap\mathbb{W}) \, \pi_{W}(X\cap\mathbb{W}).
\end{equation}
\end{IEEEproof}

\section{Proof of Proposition 3}

 \subsection{Lemma 1 and its proof}

  \begin{Lem}
 Suppose that $\pi_a(X)$ and $\pi_b(X)$ are the posteriors of agents $a$ and $b$. If $\pi_a(\emptyset)\rightarrow 1$, then the GCI fusion of $\pi_a(X)$ and $\pi_b(X)$ yields:
 \begin{equation}
 \overline\pi(\emptyset)=\frac{\pi_a(\emptyset)^{\omega_a}\pi_b(\emptyset)^{\omega_b}}{\int \pi_a(X)^{\omega_a}\pi_b(X)^{\omega_b}\delta X}\rightarrow 1.
 \end{equation}
 \end{Lem}
 \begin{IEEEproof}
The denominator of the GCI fusion of $\pi_a$ and $\pi_b$ is represented as
\begin{equation}\label{zero-forcing-denominator}
\begin{split}
&\int \pi_a(X)^{\omega_a}\pi_b(X)^{\omega_b}\delta X\\
=&\pi_a(\emptyset)^{\omega_a}\pi_b(\emptyset)^{\omega_b}+\sum_{n=1}^{+\infty}p_a(n)^{\omega_a}p_b(n)^{\omega_b}
\int \pi_a(\{x_1,\cdots,x_n\}|n)^{\omega_a}\, \pi_b(\{x_1,\cdots,x_n\}|n)^{\omega_b}d(x_1,\cdots,x_n).
\end{split}
\end{equation}

Since the following holds, for any $n\geq 1$,
\begin{equation}
\int \pi_a(\{x_1,\cdots,x_n\}|n)^{\omega_a} \pi_b(\{x_1,\cdots,x_n\}|n)^{\omega_b}d(x_1,\cdots,x_n)\leq 1,
\end{equation}
we have
\begin{equation}
\int \pi_a(X)^{\omega_a}\pi_b(X)^{\omega_b}\delta X \leq \pi_a(\emptyset)^{\omega_a}\pi_b(\emptyset)^{\omega_b}+\sum_{n=1}^{+\infty}p_a(n)^{\omega_a}  p_b(n)^{\omega_b}.
\end{equation}


Hence, the GCI fusion of $\pi_a(X)$ and $\pi_b(X)$ when $X=\emptyset$ can be bounded as
\begin{equation}
\frac{\pi_a(\emptyset)^{\omega_a}\pi_b(\emptyset)^{\omega_b}}{\pi_a(\emptyset)^{\omega_a}\pi_b(\emptyset)^{\omega_b}+\sum_{n=1}^{+\infty}p_a(n)^{\omega_a} p_b(n)^{\omega_b} } \leq\overline\pi(\emptyset)\leq 1 . \end{equation}


 In order to consider the limit situation in which the no-object probability $\pi_a(\emptyset)$ goes to $1$,
 consider a scalar $\varepsilon \in [0,1)$ and let $\pi_a (\emptyset) = 1-\varepsilon$. Accordingly,  consider a parametric family of cardinality distributions $p_a(n,\varepsilon)$ such that
 \begin{equation}
  \sum_{n=1}^{+\infty}p_a(n, \varepsilon)=\varepsilon
 \end{equation}
 in which each $p_a(n, \varepsilon)$, for $n \ge 1$, is a continuous non-decreasing function of $\varepsilon$ in some
 interval $[0, \bar \varepsilon)$ with $\bar \varepsilon \in (0,1)$.

 Notice now that the following upper bound holds
 \begin{equation}
 p_a(n, \varepsilon)^{\omega_a } p_b(n)^{\omega_b} \le \omega_a  p_a(n, \bar \varepsilon) + \omega_b \, p_b (n)
 \end{equation}
Since the series
\begin{equation}
\sum_{n=1}^{+ \infty}  \left (  \omega_a  p_a(n, \bar \varepsilon) + \omega_b \, p_b (n) \right )
\end{equation}
 is convergent, by virtue of the Weierstrass M-test the series
 \begin{equation}\label{eq:other-cardinalities}
 \sum_{n=1}^{+ \infty}   p_a(n, \varepsilon)^{\omega_a } p_b(n)^{\omega_b}
 \end{equation}
 converges uniformly in $[0, \bar \varepsilon)$. Hence, the uniform convergence theorem ensures that
 (\ref{eq:other-cardinalities}) is a continuous function of $\varepsilon$ in $[0, \bar \varepsilon)$.
 As a result, noting that $p_a(n, 0) = 0$ for any $n \ge 1$, we have
 \begin{equation}
 \lim_{\varepsilon \rightarrow 0}  \sum_{n=1}^{+ \infty}   p_a(n, \varepsilon)^{\omega_a } p_b(n)^{\omega_b}  = 0
 \end{equation}
 which in turn implies
 \begin{equation}
 \lim\limits_{\varepsilon\to 0} \frac{\pi_a(\emptyset)^{\omega_a}\pi_b(\emptyset)^{\omega_b}}{\pi_a(\emptyset)^{\omega_a}\pi_b(\emptyset)^{\omega_b}+ \sum_{n=1}^{+ \infty}   p_a(n, \varepsilon)^{\omega_a } p_b(n)^{\omega_b} }=1.
 \end{equation}

%

Finally, according to the Sandwich Theorem, Lemma 1 holds, i.e.
\begin{equation}
\lim\limits_{\varepsilon\to 0}\overline\pi(\emptyset)=1 .
\end{equation}
\end{IEEEproof}

\subsection{Proof of Proposition 3}
\begin{IEEEproof}
First, for better understanding of  the proof, the several sets involved in the decompositions are illustrated in Fig. 2 in the main body of the paper.

According to the decomposition in (14), the posterior of sensor $i=a,b$ can be decomposed as
\begin{equation}
\pi_i(X)=\pi_{i}(X\cap Fov_i) \, \pi_{i}(X\cap \mathbb{X}\backslash Fov_i|X\cap Fov_i), i=a,b.
\end{equation}
Again using (\ref{decomposition-pi}), $\pi_{i}(X\cap Fov_i)$ can be further decomposed into
\begin{equation}
\begin{split}
&\pi_{i}(X\!\cap\!Fov_i)\!=\pi_{i,co}(X\!\cap\!Fov_C)\pi_{i,nc}(X\cap(Fov_{i}\!\backslash Fov_C)|X\!\cap\!Fov_C).
\end{split}
\end{equation}
According to the GCI rule, the fusion of $\pi_a(X)$ and $\pi_b(X)$ yields:
\begin{equation}
\begin{split}
\overline\pi(X)=\overline\pi_{co}(X\cap Fov_C)\overline\pi_{nc}(X\cap(\mathbb{X}\backslash Fov_C)|X\cap Fov_C)
\end{split}
\end{equation}
where
\begin{align}
\overline\pi_{co}(X\cap Fov_C)
\propto&\pi_{a,co}(X\!\cap\!Fov_C)^{\omega_a}\pi_{b,co}(X\!\cap\!Fov_C)^{\omega_b},\\
\begin{split}\label{fused-NC}
\overline\pi_{nc}\left(X\cap\left(\mathbb{X}\backslash Fov_C\right)|X\cap Fov_C\right)
\propto &\pi_{a,nc}(X\cap(Fov_{a}\!\backslash Fov_C)|X\!\cap\!Fov_C)^{\omega_a}\pi_{b}(X\cap (\mathbb{X}\backslash Fov_b)|X\cap Fov_b)^{\omega_b}\times\\
&\pi_{a}(X\cap (\mathbb{X}\backslash Fov_a)|X\cap Fov_a)^{\omega_a}\pi_{b,nc}(X\cap(Fov_{b}\!\backslash Fov_C)|X\!\cap\!Fov_C)^{\omega_b} .
\end{split}
\end{align}
Since  $\pi_{a}(X\cap(\mathbb{X}\backslash Fov_a)|X\cap Fov_a)$ and $\pi_{b}(X\cap(\mathbb{X}\backslash Fov_b)|X\cap Fov_b)$ can be respectively decomposed into
\begin{equation}\label{decomposition-out_1}
\begin{split}
\pi_{a}(X\!\cap\!(\mathbb{X}\backslash Fov_a)|X\!\cap\!Fov_a)=&\pi_{a}(X\!\cap\!(Fov_b\backslash Fov_C)|X\!\cap\!Fov_a)\pi_{a}(X\!\cap\!(\mathbb{X}\backslash Fov_G)|X\!\cap\!Fov_G)\\
\end{split}
\end{equation}
\begin{equation}\label{decomposition-out_2}
\begin{split}
\pi_{b}(X\!\cap\!(\mathbb{X}\backslash Fov_b)|X\!\cap\!Fov_b)\!
=&\pi_{b}(X\!\cap\!(Fov_a\backslash Fov_C)|X\!\cap\!Fov_b)\pi_{b}(X\!\cap\!(\mathbb{X}\backslash Fov_G)|X\!\cap\!Fov_G),
\end{split}
\end{equation}
(\ref{fused-NC}) can be further represented by
\begin{equation}
\begin{split}
&\overline \pi_{nc}(X\cap(\mathbb{X}\backslash Fov_C)|X\cap Fov_C)\\
 \propto&\underbrace{\pi_{a,nc}(X\cap(Fov_{a}\!\backslash Fov_C)|X\!\cap\!Fov_C)^{\omega_a}\pi_{b}(X\!\cap\!Fov_a\backslash Fov_C|X\!\cap\!Fov_b)^{\omega_b}}_{E_{1}(\cdot)}\\
&\times\underbrace{\pi_{a}(X\!\cap\!(Fov_b\backslash FoV_C)|X\!\cap\!Fov_a)^{\omega_a}\pi_{b,nc}(X\cap(Fov_{b}\!\backslash Fov_C)|X\!\cap\!Fov_C)^{\omega_b}}_{E_{2}(\cdot)}\\
&\times\underbrace{\pi_{a}(X\!\cap\!(\mathbb{X}\backslash Fov_G)|X\!\cap\!Fov_G)^{\omega_a}\pi_{b}(X\!\cap\!(\mathbb{X}\backslash Fov_G)|X\!\cap\!Fov_G)^{\omega_b}}_{E_{3}(\cdot)}.
\end{split}
\end{equation}
Then the normalization of $\overline \pi_{nc}(X\cap(\mathbb{X}\backslash Fov_C)|X\cap Fov_C)$ can be computed by
\begin{equation}
\begin{split}
&\int \overline \pi_{nc}(X\cap(\mathbb{X}\backslash Fov_C)|X\cap Fov_C)\delta(X\cap(\mathbb{X}\backslash Fov_C))\\
=&\underbrace{\int\pi_{a,nc}(X\cap(Fov_{a}\!\backslash Fov_C)|X\!\cap\!Fov_C)^{\omega_a}\pi_{b}(X\!\cap\!(Fov_a\backslash Fov_C)|X\!\cap\!Fov_b)^{\omega_b}\delta(X\!\cap\!(Fov_a\backslash Fov_C))}_{F_1}\\
&\times\underbrace{\int\pi_{a}(X\!\cap\!(Fov_b\backslash Fov_C)|X\!\cap\!Fov_a)^{\omega_a}\pi_{b,N\!C}(X\cap(Fov_{b}\!\backslash Fov_C)|X\!\cap\!Fov_C)^{\omega_b}\delta(X\!\cap\!(Fov_b\backslash Fov_C))}_{F_2}\\
&\times\underbrace{\int \pi_{a}(X\!\cap\!(\mathbb{X}\backslash Fov_G)|X\!\cap\! Fov_G)^{\omega_a}\pi_{b}(X\!\cap\!(\mathbb{X}\backslash Fov_G)|X\!\cap\!Fov_G)^{\omega_b}\delta(X\! \cap\!Fov_G)}_{F_3} .
\end{split}
\end{equation}
Suppose that $\pi_{i,ou}(X\cap \left(\mathbb{X}\backslash Fov_i\right)=\emptyset|X\cap Fov_i)=1-\varepsilon_i, i=a,b$, where $\varepsilon_i$'s are any non-negative real-values.  According to (\ref{decomposition-out_1}) and (\ref{decomposition-out_2}), the following holds:
\begin{align}
&0\geq\pi_{a}(X\cap(Fov_b\backslash Fov_C)=\emptyset|X\!\cap\!Fov_a)\geq1-\varepsilon_a\\
&0\geq\pi_{a}(X\cap(\mathbb{X}\backslash Fov_G)=\emptyset|X\cap Fov_G)\geq1-\varepsilon_a\\
&0\geq\pi_{b}(X\cap(Fov_a\backslash Fov_C)=\emptyset|X\!\cap\!Fov_b)\geq 1-\varepsilon_b\\
&0\geq\pi_{b}(X\cap(\mathbb{X}\backslash Fov_G)=\emptyset|X\cap Fov_G)\geq 1-\varepsilon_b.
\end{align}
Again, based on the Sandwich Theorem, when $\varepsilon_i\to 0$ ($\pi_{i}(X\cap \left(\mathbb{X}\backslash Fov_i\right)=\emptyset|X\cap Fov_i)\to 1$), $i=a,b$, we have
\begin{align}
&\pi_{a}(X\cap(Fov_b\backslash Fov_C)=\emptyset|X\!\cap\!Fov_a)\to 1\\
&\pi_{a}(X\cap(\mathbb{X}\backslash Fov_G)=\emptyset|X\cap Fov_G)\to 1\\
&\pi_{b}(X\cap(Fov_a\backslash Fov_C)=\emptyset|X\!\cap\!Fov_b)\to 1\\
&\pi_{b}(X\cap(\mathbb{X}\backslash Fov_G)=\emptyset|X\cap Fov_G)\to 1.
\end{align}
Then, the following holds, according to Lemma 1,
\begin{align}
&\frac{E_1(\emptyset)}{F_1}\to 1\\
&\frac{E_2(\emptyset)}{F_2}\to 1\\
&\frac{E_3(\emptyset)}{F_3}\to 1 . \\
\end{align}
Finally, the proposition holds, i.e.,
\begin{equation}
\begin{split}
&\lim\limits_{\varepsilon\to 0} \overline \pi_{N\!C}(X\cap(\mathbb{X}\backslash Fov_C)=\emptyset|X\cap Fov_C)\\
=&\lim\limits_{\varepsilon\to 0}\frac{E_1(\emptyset)}{F_1}\frac{E_2(\emptyset)}{F_2}\frac{E_3(\emptyset)}{F_3}\\
=&1 .
\end{split}
\end{equation}
\end{IEEEproof}

\section{Proof of Proposition 4}
\begin{IEEEproof}
To prove that (24) provides an uninformative density, it is required to check  both (7)  and (8) of Definition 1.
\begin{itemize}
\item First, we verify (7). Substituting (24) into (7), for any $\pi\in\digamma$, the normalization factor yields:
\begin{equation}\label{normalization-factor-1}
\begin{split}
&\int \pi(X)\pi_{ui}(X) \delta X\\
=&\sum_{n=0}^{+\infty} \frac{\frac{[V(\mathbb{X})]^n}{n!}}{\sum_{m=0}^{n_{\text{max}}} \frac{[V(\zeta)]^m}{m!}}\int \pi(\{x_1,\cdots,x_n\}) \frac{1}{[V(\zeta)]^n} d(x_1,\cdots,x_2)\\
=& \frac{1}{\sum_{m=0}^{n_{\text{max}}} \frac{[V(\zeta)]^m}{m!}}\sum_{n=0}^{+\infty} \frac{1}{n!}\int \pi(\{x_1,\cdots,x_n\}) d(x_1,\cdots,x_2)\\
=&\frac{1}{\sum_{m=0}^{n_{\text{max}}} \frac{[V(\zeta)]^m}{m!}}.
\end{split}
\end{equation}
As for the denominator, we get:
\begin{equation}\label{denominator-1}
\pi(X)\pi_{ui}(X)=\frac{1}{\sum_{m=0}^{n_{\text{max}}} \frac{[V(\zeta)]^m}{m!}}\pi(\{x_1,\cdots,x_n\}) .
\end{equation}
Combining (\ref{normalization-factor-1}) and (\ref{denominator-1}), we have
\begin{equation}
\pi\oplus \pi_{ui}=\frac{\pi(X)\pi_{ui}(X)}{\int \pi(X)\pi_{ui}(X) \delta X}=\pi(X).
\end{equation}
\item Then, we verify (8). Substituting (24) into (8), the normalization factor yields:
\begin{equation}\label{normalization-factor-2}
\begin{split}
&\int [\pi_{ui}(X)]^\omega \delta X\\
=&\sum_{n=0}^{+\infty} \frac{1}{n!}\left[\sum_{m=0}^{n_{\text{max}}} \frac{[V(\zeta)]^m}{m!}\right]^\omega d(x_1,\cdots,x_2)\\
=&\frac{1}{\left[\sum_{m=0}^{n_{\text{max}}} \frac{[V(\zeta)]^m}{m!}\right]^\omega} \sum_{n=0}^{n_{\text{max}}} \frac{[V(\zeta)]^n}{n!}\\
=&\left[\sum_{m=0}^{n_{\text{max}}} \frac{[V(\zeta)]^m}{m!}\right]^{(1-\omega)}.
\end{split}
\end{equation}
As a result, using (\ref{normalization-factor-2}), we have
\begin{equation}
\begin{split}
&\omega \odot \pi_{ui}(X)\\
=&\frac{\pi_{ui}(X)}{\int [\pi_{ui}(X)]^\omega \delta X}\\
=&\frac{\frac{1}{\left[\sum_{m=0}^{n_{\text{max}}} \frac{[V(\zeta)]^m}{m!}\right]^{\omega}}}{\left[\sum_{m=0}^{n_{\text{max}}} \frac{[V(\zeta)]^m}{m!}\right]^{(1-\omega)}}\\
=&\frac{1}{\sum_{m=0}^{n_{\text{max}}} \frac{[V(\zeta)]^m}{m!}}\\
=&\pi_{ui}(X) .
\end{split}
\end{equation}
\end{itemize}
\end{IEEEproof}

\section{Proof of Proposition 5}
\begin{IEEEproof}
First, we compute  the marginal density of $X\cap\mathbb{W}$, $\pi_{W}(X)$.  Based on (9), $\pi_{W}(X)$ is given by
\begin{equation}\label{marginal-density-a-1}
\begin{split}
\pi_{W}(X)=&[1_{\mathbb{W}}]^{X}\int_{\mathbb{V}} \pi(X\cup X')\delta X'\\
=&[1_{\mathbb{W}}]^{X}\int_{\mathbb{V}} [\lambda]^{|X|+|X'|} \exp(-\lambda)p^{X}p^{X'}\delta X'\\
=&[1_{\mathbb{W}}]^{X}[\lambda]^{|X|} \exp(-\lambda)p^{X}\int_{\mathbb{V}} [\lambda]^{|X'|}p^{X'}\delta X' .
\end{split}
\end{equation}
Define
\begin{align}
p_{W}(x)\triangleq&\frac{1_{\mathbb{W}}(x)p(x)}{K_{W}}\\
p_{V}(x)\triangleq&\frac{1_{\mathbb{V}}(x)p(x)}{K_{V}}\\
K_{W}\triangleq&\int_{\mathbb{W}}p(x) dx\\
K_{V}\triangleq&\int_{\mathbb{V}}p(x) dx .
\end{align}
Since $\mathbb{X}=\mathbb{W}\uplus\mathbb{V}$, it holds that:
\begin{equation}
K_{W}=1-K_{V}.
\end{equation}
Hence, (\ref{marginal-density-a-1}) can be further expressed as
\begin{equation}
\begin{split}
\pi_{W}(X)=&[1_{\mathbb{W}}]^{X}[\lambda]^{|X|} \exp(-\lambda)\exp(\lambda K_{V})p^{X}\int_{\mathbb{V}} \exp(-\lambda K_{V})[\lambda K_{V}]^{|X'|}p_{V}^{X'}\delta X'\\
=&[1_{\mathbb{W}}]^{X}[\lambda K_{W}]^{|X|} \exp(-\lambda(1-K_{V}))p_{W}^{X}\\
=&[1_{\mathbb{W}}]^{X}[\lambda K_{W}]^{|X|} \exp(-\lambda K_{W})p_{W}^{X}.
\end{split}
\end{equation}
As a result, $\pi_{W}$ is Poisson on $\mathbb{W}$ with parameter pair $\left( \lambda K_{W}, p_{W} \right)$.

Next, we compute the density of $X\cap\mathbb{V}$ conditioned on $X\cap\mathbb{W}$.
Since $\pi_W(X)$ is Poisson, $\pi_W(X)\neq 0$ for any $X\in \mathcal{F}(\mathbb{X})$. Then we have
\begin{equation}\label{conditioned-density-B}
\begin{split}
\pi_{V}(X'|X)= &\frac{[1_{\mathbb{W}}]^{X}[1_{\mathbb{V}}]^{X'}\pi(X\cup X')}{\pi_{W}(X)}\\
=&\frac{[1_{\mathbb{W}}]^{X}[1_{\mathbb{V}}]^{X'}[\lambda]^{|X|+|X'|} \exp(-\lambda)p^{X}p^{X'}}{[1_{\mathbb{W}}]^{X}[\lambda K_{W}]^{|X|} \exp(-\lambda K_{W})p_{W}^{X}}\\
=&\frac{[1_{\mathbb{V}}]^{X'}[\lambda]^{|X'|} p^{X'}[1_{\mathbb{W}}]^{X}[\lambda]^{|X|}\exp(-\lambda)p^{X}}{[1_{\mathbb{W}}]^{X}[\lambda K_{W}]^{|X|} \exp(-\lambda K_{W})p_{W}^{X}}\\
=&\frac{[1_{\mathbb{V}}]^{X'}[\lambda]^{|X'|} \exp(-\lambda(1-K_{W}))p^{X'}[1_{\mathbb{W}}]^{X}[\lambda K_{W}]^{|X|}\exp(-\lambda K_{W})p_{W}^{X}}{[1_{\mathbb{A}}]^{X}[\lambda K_{W}]^{|X|} \exp(-\lambda K_{W})p_{W}^{X}}\\
=&[1_{\mathbb{W}}]^{X'}[\lambda K_{V}]^{|X'|} \exp(-\lambda K_{V})p_{V}^{X'} .
\end{split}
\end{equation}

It can be seen from (\ref{conditioned-density-B}) that $X\cap\mathbb{V}$ is independent of $X\cap\mathbb{W}$. Hence  without loss of generality, we use the following notation:
\begin{equation}
\pi_{V}(X'|X)=\pi_{V}(X')=[1_{\mathbb{V}}]^{X'}[\lambda K_{V}]^{|X'|} \exp(-\lambda K_{V})p_{V}^{X'}.
\end{equation}
Hence, $\pi_{V}$ is Poisson on $\mathbb{V}$ with parameter pair $\left( \lambda K_{V}, p_{V} \right)$.
\end{IEEEproof}

\section{Proof of Proposition 7}
\begin{IEEEproof}
Since $X_W$ and $X_V$ are Poisson RFSs defined on disjoint subspaces, then according to Proposition 4, $X_W$ and $X_V$ are statistically independent.
Further, according to [17], the density of the union  $X=X_W\cup X_V$ of disjoint sets  can be computed by
\begin{equation}\label{density-x}
\begin{split}
\pi(X)=&\sum_{X'\subseteq X} \pi_W(X')\ \, \pi_V(X\backslash X')\\
=&\sum_{X'\subseteq X} 1^{\mathbb{W}}(X') \, \pi_W(X') \, 1^{\mathbb{V}}(X\backslash X') \, \pi_V(X\backslash X')\\
=&\pi_W(X\cap\mathbb{W}) \, \pi_V(X\cap\mathbb{V}).
\end{split}
\end{equation}
Then, substitution of (49) and (50) into (\ref{density-x}) yields
\begin{equation}
\pi(X)=\exp(-\lambda_W)\left[\lambda_W p_W\right]^{X\cap\mathbb{W}}\exp(-\lambda_V)\left[\lambda_V p_V\right]^{X\cap\mathbb{V}}.
\end{equation}
Furthermore, since $p_W(x)$ and $p_V(x)$ are defined on disjoint spaces $\mathbb{W}$ and $\mathbb{V}$, it holds that:
\begin{align}
\lambda_W p_W(x)=&\lambda_W p_W(x)+\lambda_V p_V(x), \forall x\in\mathbb{W},\\
\lambda_V p_V(x)=&\lambda_W p_W(x)+\lambda_V p_V(x), \forall x\in\mathbb{V}.
\end{align}
Hence, we have
\begin{equation}
\begin{split}
\pi(X)=&\exp(-\lambda_W)\left[\lambda_W p_W(x)+\lambda_V p_V(x)\right]^{X\cap\mathbb{W}}\exp(-\lambda_V)\left[\lambda_W p_W(x)+\lambda_V p_V(x)\right]^{X\cap\mathbb{V}}\\
=&\exp(-(\lambda_W+\lambda_V))\left[\lambda_W p_W(x)+\lambda_V p_V(x)\right]^{X}.
\end{split}
\end{equation}
To normalize $\lambda_W p_W(x)+\lambda_V p_V(x)$, we compute the corresponding integral as
\begin{equation}\label{density-x-2}
\int_{\mathbb{W}\cup\mathbb{V}}\lambda_W p_W(x)+\lambda_V p_V(x)dx=\lambda_W+\lambda_V.
\end{equation}
Next, define
\begin{align}
\label{union-space}\mathbb{X}&\triangleq\mathbb{W}\cup\mathbb{V}\\
\label{union-lambda}\lambda&\triangleq\lambda_W+\lambda_V\\
\label{union-p} p(x)&\triangleq\frac{\lambda_W p_W(x)+\lambda_V p_V(x)}{\lambda}, x\in \mathbb{X},
\end{align}
where $p(x)$ is a density defined on $\mathbb{X}$.
Substitution of (\ref{union-space})--(\ref{union-p}) into (\ref{density-x-2}) yields
\begin{equation}
\pi=\lambda^{|X|} \, \exp(-\lambda) \, p^{X},
\end{equation}
which is a Poisson density parameterized by $(\lambda, p)$.
\end{IEEEproof}
